\documentclass[a4paper,table,11pt]{article}
\usepackage{jheppub} 
\usepackage{mathtools}
\usepackage{mathrsfs}
\usepackage{bbold}
\usepackage{cancel}
\usepackage{float}
\usepackage{pgfplots}
\pgfplotsset{compat=1.18}
\usetikzlibrary{hobby}
\usetikzlibrary{decorations.markings}

\usepackage{makecell}
\usepackage{circuitikz}
\usepackage{subcaption}
\usepackage{ifthen}
\usepackage{braket}
\usepackage[bbgreekl]{mathbbol}

\newcommand{\bbG}{\mathbb{G}}
\newcommand{\bbGA}{\mathbb{G}_{\text{adv}}}
\newcommand{\bbGR}{\mathbb{G}_{\text{ret}}}

\newcommand{\Gin}{G^{\text{in}}}
\newcommand{\Gout}{G^{\text{out}}}
\newcommand{\Kin}{K^{\text{in}}}

\newcommand{\Psmall}{\scalebox{.9}{$\scriptscriptstyle P$}}
\newcommand{\F}{\scalebox{.9}{$\scriptscriptstyle F$}}
\newcommand{\Pb}{\bar{\Psmall}}
\newcommand{\Fb}{\bar{\F}}

\newcommand{\sR}{{}_{\text{R}}}
\newcommand{\sL}{{}_{\text{L}}}

\newcommand{\z}{\zeta}
\renewcommand{\b}{\beta}

\newcommand{\Ja}{J_{a}}
\newcommand{\Jd}{J_{d}}
\newcommand{\Jba}{\bar{J}_{a}}
\newcommand{\Jbd}{\bar{J}_{d}}

\newcommand{\nbe}{n}

\newcommand{\Phib}{\bar{\Phi}}
\newcommand{\Jb}{\bar{J}}
\newcommand{\Ginb}{\bar{G}^{\text{in}}}
\newcommand{\Goutb}{\bar{G}^{\text{out}}}
\newcommand{\gb}{\bar{g}}

\newcommand{\nbeb}{\bar{\nbe}}

\newcommand{\nq}{n}
\newcommand{\nqb}{\bar{\nq}}

\newcommand{\dt}{\partial_{t}}
\newcommand{\mG}{\mathcal{G}}

\newcommand{\bTh}{\mathbb{\Theta}}
\newcommand{\muq}{\mu_{q}}

\newcommand*\diff{\mathop{}\!\mathrm{d}}

\begin{document}

\title{Holographic Fluctuation-Dissipation Relations in Finite Density Systems }
\author{Shivam K. Sharma}
\affiliation{International Centre for Theoretical Sciences (ICTS-TIFR)\\ 
Tata Institute of Fundamental Research, Shivakote, Hesaraghatta Hobli, Bengaluru 560089, India.}
\emailAdd{shivam.sharma@icts.res.in}  
\abstract{ Using real-time holography, we investigate fluctuation-dissipation relations in holographic systems at finite density. In the bulk, it corresponds to the study of scattering of a charged scalar field against a charged black hole background, leading to an exterior field theory outside the AdS Reissner-Nordström black hole. This theory captures dissipation (from infalling Quasi-normal modes) and fluctuations (from Hawking radiation). Here, we also develop a Witten diagrammatic framework for computing $n$-point functions for interacting charged scalars. From the boundary perspective, it gives real-time correlations of the dual CFT held at finite temperature and finite chemical potential. Our results yield (non)-linear holographic fluctuation-dissipation relations at small but finite density, which are shown to recover the correct behaviour in the zero-density limit.}


\newcommand{\Gsemicap}[4]{
\draw[dashed,{|[scale=1.8]}-] (#1 +  #3,#2 + #4)--(#3,#4);
}

\newcommand{\Gdiodearrow}[4]{
\draw[dashed,-{Triangle[scale=1.8,open]}] (#1,#2)--(#1 +  #3 , #2 + #4);
}

\newcommand{\Ssemicap}[4]{
\draw[thick,{|[scale=1.8]}-] (#1 +  #3,#2 + #4)--(#3,#4);
}

\newcommand{\Sdiodearrow}[4]{
\draw[thick,-{Triangle[scale=1.8,open]}] (#1,#2)--(#1 +  #3 , #2 + #4);
}

\newcommand{\Gdiode}[4]{
	\draw[dashed,-{Triangle[scale=1.8,open]}] (#1,#2)--(#1/2 +  #3/2 , #2/2 + #4/2);
	\draw[dashed,{|[scale=1.8]}-] (#1/2 +  #3/2,#2/2 + #4/2)--(#3,#4);
}

\newcommand{\Sdiode}[4]{
	\draw[-{Triangle[scale=1.8,open]}] (#1,#2)--(#1/2 +  #3/2 , #2/2 + #4/2);
	\draw[{|[scale=1.8]}-] (#1/2 +  #3/2,#2/2 + #4/2)--(#3,#4);
}

\newcommand{\threeptPFYuk}[3]{
\begin{tikzpicture}[scale=1]
    \coordinate (z) at (0,-0.5);
    \draw[blue] (-1.4,1)--(1.4,1);
    \ifthenelse{#1 > 0}{\Sdiode{-1.2}{1}{0}{-0.5}}{\Sdiode{0}{-0.5}{-1.2}{1}};
    \ifthenelse{#2 > 0}{\Gdiode{0}{1}{0}{-0.5}}{\Gdiode{0}{-0.5}{0}{1}};
    \ifthenelse{#3 > 0}{\Sdiode{1.2}{1}{0}{-0.5}}{\Sdiode{0}{-0.5}{1.2}{1}};
    \node at (z) {$\bullet$};
\end{tikzpicture}
}

\newcommand{\fourptPFcmplxphifour}[4]{
\begin{tikzpicture}[scale=1.4]
    \coordinate (z) at (0,-0.6);
    \draw[blue] (-1.75,1)--(1.75,1);
    \ifthenelse{#1 > 0}{\Sdiode{-1.5}{1}{0}{-0.6}}{\Sdiode{0}{-0.6}{-1.5}{1}};
    \ifthenelse{#2 > 0}{\Sdiode{-0.5}{1}{0}{-0.6}}{\Sdiode{0}{-0.6}{-0.5}{1}};
    \ifthenelse{#3 > 0}{\Sdiode{0.5}{1}{0}{-0.6}}{\Sdiode{0}{-0.6}{0.5}{1}};
    \ifthenelse{#4 > 0}{\Sdiode{1.5}{1}{0}{-0.6}}{\Sdiode{0}{-0.6}{1.5}{1}};
    \node at (z) {$\bullet$};
    \draw[red, arrows={->[scale=1.2,red]}] (-1.55,0.8) -- (-1.2,0.4);
    \draw[red, arrows={->[scale=1.2,red]}] (-0.50,0.4) -- (-0.6,0.8);
    \draw[red, arrows={->[scale=1.2,red]}] (0.55,0.8) -- (0.4,0.4);
    \draw[red, arrows={->[scale=1.2,red]}] (1.2,0.4) -- (1.55,0.8);
\end{tikzpicture}
}

\newcommand{\sixptPFcmplxphifour}[7]{
\begin{tikzpicture}[scale=1.4]
    \coordinate (z) at (0,-0.6);
    \coordinate (z') at (2,-0.6);
    \draw[blue] (-1.75,1)--(3.75,1);
    \ifthenelse{#1 > 0}{\Sdiode{-1.5}{1}{0}{-0.6}}{\Sdiode{0}{-0.6}{-1.5}{1}};
    \ifthenelse{#2 > 0}{\Sdiode{-0.5}{1}{0}{-0.6}}{\Sdiode{0}{-0.6}{-0.5}{1}};
    \ifthenelse{#3 > 0}{\Sdiode{0.5}{1}{0}{-0.6}}{\Sdiode{0}{-0.6}{0.5}{1}};
    \ifthenelse{#4 > 0}{\Sdiode{0}{-0.6}{2}{-0.6}}{\Sdiode{2}{-0.6}{0}{-0.6}}
    \ifthenelse{#5 > 0}{\Sdiode{1.5}{1}{2}{-0.6}}{\Sdiode{2}{-0.6}{1.5}{1}};
    \ifthenelse{#6 > 0}{\Sdiode{2.5}{1}{2}{-0.6}}{\Sdiode{2}{-0.6}{2.5}{1}};
    \ifthenelse{#7 > 0}{\Sdiode{3.5}{1}{2}{-0.6}}{\Sdiode{2}{-0.6}{3.5}{1}};
    \node at (z) {$\bullet$};
    \node at (z') {$\bullet$};
    \draw[red, arrows={->[scale=1.2,red]}] (-1.55,0.8) -- (-1.2,0.4);
    \draw[red, arrows={->[scale=1.2,red]}] (-0.50,0.4) -- (-0.6,0.8);
    \draw[red, arrows={->[scale=1.2,red]}] (0.55,0.8) -- (0.4,0.4);
    \draw[red, arrows={->[scale=1.2,red]}] (0.7,-0.4) -- (1.3,-0.4);
    \draw[red, arrows={->[scale=1.2,red]}] (1.55,0.4) -- (1.4,0.8);
    \draw[red, arrows={->[scale=1.2,red]}] (2.35,0.8) -- (2.2,0.4);
    \draw[red, arrows={->[scale=1.2,red]}] (3.05,0.4) -- (3.45,0.8);
\end{tikzpicture}
}

\newcommand{\fourptScalExPFYuk}[5]{
\begin{tikzpicture}[scale=1.4]
	\draw[blue] (-1.75,0)--(1.75,0);
	\ifthenelse{#1 > 0}{\Sdiode{-1.5}{0}{-1}{-1}}{\Sdiode{-1}{-1}{-1.5}{0}}
	\ifthenelse{#2 > 0}{\Sdiode{-0.5}{0}{-1}{-1}}{\Sdiode{-1}{-1}{-0.5}{0}}
	\ifthenelse{#3 > 0}{\Sdiode{0.5}{0}{1}{-1}}{\Sdiode{1}{-1}{0.5}{0}}
	\ifthenelse{#4 > 0}{\Sdiode{1.5}{0}{1}{-1}}{\Sdiode{1}{-1}{1.5}{0}}
	\ifthenelse{#5 > 0}{\Gdiode{-1}{-1}{1}{-1}}{\Gdiode{1}{-1}{-1}{-1}}
	\node at (-1,-1) {$\bullet$};
	\node at (1,-1) {$\bullet$};
\end{tikzpicture}
}

\newcommand{\fourptFermExPFYuk}[5]{
\begin{tikzpicture}[scale=1.4]
	\draw[blue] (-1.75,0)--(1.75,0);
	\ifthenelse{#1 > 0}{\Gdiode{-1.5}{0}{-1}{-1}}{\Gdiode{-1}{-1}{-1.5}{0}}
	\ifthenelse{#2 > 0}{\Sdiode{-0.5}{0}{-1}{-1}}{\Sdiode{-1}{-1}{-0.5}{0}}
	\ifthenelse{#3 > 0}{\Gdiode{0.5}{0}{1}{-1}}{\Gdiode{1}{-1}{0.5}{0}}
	\ifthenelse{#4 > 0}{\Sdiode{1.5}{0}{1}{-1}}{\Sdiode{1}{-1}{1.5}{0}}
	\ifthenelse{#5 > 0}{\Sdiode{-1}{-1}{1}{-1}}{\Sdiode{1}{-1}{-1}{-1}}
	\node at (-1,-1) {$\bullet$};
	\node at (1,-1) {$\bullet$};
\end{tikzpicture}
}

\maketitle

\section{Introduction}\label{sec:Intro}

Thermal systems are present throughout physics, from everyday objects like a cup of coffee to exotic objects like black holes. The behaviour of these systems is governed by various constraints on observable quantities. A key example is the \textit{Fluctuation-Dissipation Theorem} (FDT), which fundamentally links the strength of fluctuations to the dissipation coefficient. In other words, the way a system responds to small external disturbances (dissipation) is directly related to the fluctuations it exhibits in thermal equilibrium.

\medskip

A simple illustration of the FDT can be found in classical Brownian motion. It is a random, erratic movement of tiny particles suspended in a fluid (like pollen in air), caused by continuous collisions with the molecules of the fluid. The dynamics of a Brownian particle is described by Langevin's equation:
\begin{equation}\label{eq:linLangevin}
    \ddot{q}+\gamma \dot{q}  = \mathfrak{f}\, \eta(t) \ ,
\end{equation}
where $\gamma$ is the damping (or dissipation) coefficient, $\eta$ represents the thermal noise\footnote{\emph{Thermal noise} refers to the random fluctuations that arise due to the thermal motion of particles.} and $\mathfrak{f}$ measures the fluctuation strength. The relationship between these coefficients, as given by the following linear FDT:
\begin{equation}\label{eq:linFDT}
    \frac{2}{\b} \gamma =  \mathfrak{f} \ ,
\end{equation}
where $\beta$ denotes the inverse temperature.

\medskip 

The origin of FDT lies in the thermal nature of the bath \cite{Kubo:1966fyg}. This theorem, though powerful, is also very intuitive because the mechanisms causing dissipation are inherently linked to the fluctuations observed. A common classical argument is that the thermal fluctuations causing noise in the bath are also responsible for dissipation. In Brownian motion, for instance, random kicks from the environment not only hinder the particle’s motion but also simultaneously generate noise. Then the FDT precisely asserts that, on average, the random kicks from the fluctuating force balance out the energy lost through dissipation.

\medskip

Another way to understand FDTs is by considering the role of Kubo-Martin-Schwinger (KMS) conditions \cite{Kubo:1957mj, Martin:1959jp} governing bath correlators. These conditions emerge from the thermal density matrix ($\hat{\rho} = e^{-\beta H}$) and define the relationships between different thermal correlators. The two-point KMS condition, for instance, naturally gives rise to linear FDT \eqref{eq:linFDT} in the high-temperature limit.

\medskip

To see, we first write the two-point KMS condition in the time domain, as given by,
\begin{equation}\label{eq:2KMStime}
    \braket{A(t -i \beta)\  B(0)}_{\beta} = \braket{B(0) \   A(t) }_{\beta} \ ,
\end{equation}
where the subscript $\beta$ denotes the thermal state. In the frequency domain, this becomes:
\begin{equation}\label{eq:2KMSfreq}
    e^{- \beta \omega } \braket{A(\omega)\  B(0)}_{\beta} =  \braket{B(0) \   A(\omega) }_{\beta}  \ ,
\end{equation}
where $\omega$ is the Fourier conjugate variable to time $t$.

To understand it better, let us start by defining two different types of thermal Wightman functions, as given below: 
\begin{equation}\label{eq:defWightman}
    G^{>}(t) :=  \braket{A(t)\  B(0)}_{\beta} \ , \qquad \qquad G^{<}(t) :=  \braket{B(0) \ A(t)}_{\beta} \ .
\end{equation}
Then, as seen from the Eqs.~\eqref{eq:2KMStime} and \eqref{eq:2KMSfreq}, the KMS condition in terms of the two-point Wightman function is given by,
\begin{equation}
    G^>(t) = G^<(t - i\beta) \ , \qquad \qquad  G^>(\omega) = e^{\beta \omega} \, G^<(\omega) \ . 
\end{equation}
Now, writing the commutator and anti-commutator in terms of thermal Wightman functions, and using the above-mentioned KMS conditions, we find:
\begin{equation}
\begin{split}
    \braket{[A(\omega), B(-\omega)]} &= G^>(\omega) - G^<(\omega) = G^>(\omega)(1 - e^{-\beta \omega}) \ , \\
    \braket{\{A(\omega), B(-\omega)\}} &= G^>(\omega) + G^<(\omega) = G^>(\omega)(1 + e^{-\beta \omega}) \ .
\end{split}
\end{equation}
Dividing the above equations gives the following fluctuation-dissipation relation:
\begin{equation}
\boxed{
\braket{\{A(\omega), B(-\omega)\}} = \coth\left( \frac{\beta \omega}{2} \right) \braket{[A(\omega), B(-\omega))]} \ .
}    
\end{equation}
Here,
\begin{itemize}
    \item The \textbf{anti-commutator} gives the thermal fluctuations,
    \item The \textbf{commutator} gives the dissipative response,
    \item The factor \( \coth(\beta \omega/2) \) encodes thermal weighting.
\end{itemize}
In the high-temperature limit $\beta \to 0$, this reduces to the familiar classical form of the linear FDT as given in Eq.~(\ref{eq:linFDT}) before.\footnote{For textbook treatments of FDT via linear response theory, see \cite{Rammer:2007zz, kamenev_2011}.}

\medskip

On the other hand, \emph{non-linear} FDTs \cite{PhysRevD.66.025008} emerge from higher-point KMS conditions, often supplemented by principles such as time-reversibility. These relations are typically formulated using effective descriptions such as non-linear Langevin equations and can involve \emph{out-of-time-ordered} (OTO) correlators. These correlators are sensitive to microscopic reversibility and quantum chaos, and are useful for diagnosing non-linear response. For detailed studies on this front, see \cite{Tsuji:2016kep, Haehl:2017eob, Chakrabarty:2018dov, Chakrabarty:2019qcp}.

\medskip

At the opposite end of the spectrum of examples given above, black holes also behave as thermal systems characterised by well-defined temperatures. As such, they are expected to satisfy fluctuation-dissipation theorems. In strongly coupled systems — including those with a holographic dual with a black hole — directly accessing FDTs can be challenging. However, the AdS/CFT correspondence \cite{Maldacena:1997re, Gubser:1998bc, Witten:1998qj} provides a powerful tool: it maps strongly coupled quantum field theories to weakly coupled classical gravitational systems, where FDTs can be extracted using gravitational methods.

\medskip

Since Brownian motion provides an intuitive realization of the FDT, it is natural to ask: Can we construct a Langevin-like effective theory for Brownian motion in a holographic setting, and thereby derive FDTs from gravity? This question finds motivation in the fluid/gravity correspondence \cite{Rangamani:2009xk, Hubeny:2011hd} and the first part of the question has been explored in a variety of works on holographic Brownian motion \cite{Son:2009vu, deBoer:2008gu, Giecold:2009cg, Casalderrey-Solana:2007ahi, Chakrabortty:2013kra, Banerjee:2013rca}. In particular, a detailed discussion of non-linear Langevin dynamics and corresponding FDTs using real-time holography is given in \cite{Chakrabarty:2019aeu}.

\medskip

The essential idea is to model the Brownian particle (e.g., a heavy quark) as the endpoint of a fundamental string in a black hole spacetime. One end of the string is anchored at the AdS boundary, representing the particle, while the other end stretches into the bulk, interacting with the horizon. Solving the string equations of motion and computing the on-shell action yields a generating functional for the boundary theory. This functional serves as an effective action for the particle, from which one can extract a stochastic equation of motion using a Hubbard-Stratonovich transformation \cite{Hubbard:1959ub, 1957SPhD....2..416S, 2011arXiv1104.5161K}. By comparing the noise and dissipation terms in this effective theory, one directly obtains the corresponding FDT \cite{Chakrabarty:2019aeu}. While this analysis applies to particles, a natural next step is to extend it to field-theoretic degrees of freedom.

\medskip

Deriving non-linear FDTs in field theory remains technically challenging due to the limitations of perturbation theory. Holography, however, provides a non-perturbative framework well-suited to such problems. A complete analysis of fluctuations and dissipation in this setting requires a fully real-time formulation of holography. Earlier developments by Son-Starinets \cite{Son:2002sd, Son:2009vu} and Skenderis–van Rees \cite{Skenderis:2008dg, Skenderis:2008dh} laid the groundwork for this, but did not fully account for thermal fluctuations or Hawking radiation in the bulk. This limitation was addressed in \cite{Glorioso:2018mmw}, which proposed the \emph{grSK geometry} (See figure~\ref{fig:grSKpic} below), a gravitational dual to the Schwinger-Keldysh (SK) formalism in QFT. Although a complete derivation of this prescription remains elusive, the grSK framework has proven remarkably effective for capturing fluctuation phenomena in holographic systems.
\begin{figure}[H]
\begin{center}
\begin{tikzpicture}

\begin{scope}[shift={(0,0)},scale=1]
\draw  (2,-2) -- (2,2);

\draw[decoration = {zigzag,segment length = 2mm, amplitude = 0.5mm}, decorate,shift={(0,2)}] (-2,0) .. controls (0,-0.5) .. (2,0);

\draw[color=blue!80, fill=orange!60, thick] (2,1.8) -- (0.2,0) -- (2,-1.8);
\draw[dashed,thick] (2,-1.2) -- ({0.25+0.3*cos(45)},0.3);
\draw[densely dashed,thick] ({0.25+0.3*cos(45)},0.3) arc (-40:-310:0.13);

\draw[dashed,thick] (2,-0.6) -- ({0.25+0.3*cos(45)+0.6*cos(45)*cos(45)},{0.3+0.6*sin(45)*sin(45)});
\draw[densely dashed,thick] ({0.25+0.3*cos(45)+0.6*cos(45)*sin(45)},{0.3+0.6*sin(45)*sin(45)}) arc (-40:-310:0.13);

\draw[dashed,thick] (2,0) -- ({0.25+0.3*cos(45)+1.2*cos(45)*cos(45)},{0.3+1.2*sin(45)*sin(45)});
\draw[densely dashed,thick] ({0.25+0.3*cos(45)+1.2*cos(45)*sin(45)},{0.3+1.2*sin(45)*sin(45)}) arc (-40:-310:0.13);

\draw (0,0) -- (2,2);
\draw (-2,2) -- (2,-2);
\node[rotate=-45] at ({0.15+1.2*cos(135)},{0.15+1.2*sin(135)-1}) {{\scriptsize{$\leftarrow \mathcal{H}^{-} \rightarrow$}}};
\node[rotate=45] at ({-0.15+1.2*cos(225)+1.4},{0.15+1.2*sin(225)+1.6}) {{\scriptsize{$\leftarrow \mathcal{H}^{+}\rightarrow $}}};

\node at (2.6,0) {{\scriptsize{$M_R$}}$\uparrow$};

\node at (0,-2.8) {$(a)$};
\end{scope}

\begin{scope}[shift={(5,0)},scale=1]

\draw  (2,-2) -- (2,2);

\draw[decoration = {zigzag,segment length = 2mm, amplitude = 0.5mm}, decorate,shift={(0,2)}] (-2,0) .. controls (0,-0.5) .. (2,0);

\draw[color=blue!80, fill=blue!40, thick] (2,1.8) -- (0.2,0) -- (2,-1.8);
\draw[dashed,thick] (2,-1.2) -- ({0.25+0.3*cos(45)},0.3);
\draw[densely dashed,thick] ({0.25+0.3*cos(45)},0.3) arc (-40:-310:0.13);

\draw[dashed,thick] (2,-0.6) -- ({0.25+0.3*cos(45)+0.6*cos(45)*cos(45)},{0.3+0.6*sin(45)*sin(45)});
\draw[densely dashed,thick] ({0.25+0.3*cos(45)+0.6*cos(45)*sin(45)},{0.3+0.6*sin(45)*sin(45)}) arc (-40:-310:0.13);

\draw[dashed,thick] (2,0) -- ({0.25+0.3*cos(45)+1.2*cos(45)*cos(45)},{0.3+1.2*sin(45)*sin(45)});
\draw[densely dashed,thick] ({0.25+0.3*cos(45)+1.2*cos(45)*sin(45)},{0.3+1.2*sin(45)*sin(45)}) arc (-40:-310:0.13);

\draw (0,0) -- (2,2);
\draw (-2,2) -- (2,-2);
\node[rotate=-45] at ({0.15+1.2*cos(135)},{0.15+1.2*sin(135)-1}) {{\scriptsize{$\leftarrow \mathcal{H}^{-} \rightarrow$}}};
\node[rotate=45] at ({-0.15+1.2*cos(225)+1.4},{0.15+1.2*sin(225)+1.6}) {{\scriptsize{$\leftarrow \mathcal{H}^{+}\rightarrow $}}};

\node at (0,-2.8) {$(b)$};
\end{scope}

\begin{scope}[shift={(9,0)},scale=1.1]

\draw[color=blue!100, fill=blue!60, thick] (2,2) -- (2.4,-1.6) -- (0,0);

\draw[color=blue!80, fill=orange!60, thick] (2,2) -- (0,0) -- (2,-2);

\draw[dashed,thick] (2,-1.2) -- ({0.25+0.3*cos(45)},0.3);
\draw[densely dashed,thick] ({0.25+0.3*cos(45)},0.3) arc (-40:-330:0.13);

\draw[dashed,thick] (2,-0.6) -- ({0.25+0.3*cos(45)+0.6*cos(45)*cos(45)},{0.3+0.6*sin(45)*sin(45)});
\draw[densely dashed,thick] ({0.25+0.3*cos(45)+0.6*cos(45)*sin(45)},{0.3+0.6*sin(45)*sin(45)}) arc (-40:-330:0.13);

\draw[dashed,thick] (2,0) -- ({0.25+0.3*cos(45)+1.2*cos(45)*cos(45)},{0.3+1.2*sin(45)*sin(45)});
\draw[densely dashed,thick] ({0.25+0.3*cos(45)+1.2*cos(45)*sin(45)},{0.3+1.2*sin(45)*sin(45)}) arc (-40:-330:0.13);

\node[] at (2.5,1) {\scriptsize{\(M_L\)}};
\draw[->] (2.4,0.8)--(2.5,0); 

\node[] at (1.8,0.7) {\scriptsize{\(M_R\)}};
\draw[->] (1.8,0.9)--(1.8,1.6);

\node[rotate=45] at ({-0.15+1.2*cos(225)+1.4},{0.15+1.2*sin(225)+1.6}) {{\scriptsize{$\leftarrow \mathcal{H}^{+}\rightarrow $}}};

\node at (1,-2.7) {$(c)$};
\end{scope}

\end{tikzpicture}
\end{center}
\caption{
 Schematic construction of grSK geometry via Penrose diagrams: Figures  (a) and (b) show two copies of the AdS black hole, respectively. Figure (c) represents the grSK geometry.
}
\label{fig:grSKpic}
\end{figure}

Using this grSK prescription, the authors of \cite{Jana:2020vyx} provided the first derivation of non-linear FDTs entirely within a field-theoretic framework by integrating out holographic bath degrees of freedom. Their derivation relied on analyzing Hawking radiation and its interaction with ingoing modes near the horizon. Before reviewing their gravitational analysis in the upcoming section~$\S$\ref{sec:realatmuzeror}, we briefly outline their setup and key steps involved in deriving the FDTs.

\medskip

To illustrate, they considered a non-linear generalisation of the Langevin equation for a Brownian field $\Phi$ of the form:
\begin{equation}\label{eq:nlinLangevin}
\left(-K \partial_{t}^2+ D \nabla^2 +\gamma \partial_t\right)\Phi+\sum_{k=1}^{n-1} \frac{\eta^k}{k!}\left( \theta_{k}+\bar{\theta}_{k}\partial_t \right)\Phi^{n-k} = \mathfrak{f} \, \eta \  ,
\end{equation}
where $K,D,\theta_{k},\bar{\theta}_{k}$ are some unknown coefficients. These coefficients were shown to satisfy the following non-linear FDTs,
\begin{equation}\label{eq:nlinFDT}
    \frac{2}{\b} \bar{\theta}_{k}+ \theta_{k+1}+\frac{1}{4}\theta_{k-1} = 0 \ ,
\end{equation}
where $\theta_{k}$ and $\bar{\theta}_{k}$ correspond to non-linear generalization of fluctuation and dissipation coefficients.\footnote{The explicit expressions for $\theta_{k}$ and $\bar{\theta}_{k}$ can be found in the section 6 of \cite{Jana:2020vyx}.} Importantly, Eq.~(\ref{eq:nlinFDT}) provides a non-linear FDT valid to arbitrary order, capturing the full structure of fluctuations and dissipation in the zero-density holographic system.

\medskip

The key steps behind deriving these holographic non-linear FDTs can be summarised as:
\begin{enumerate}
    \item One probes the grSK geometry with a self-interacting real scalar field and constructs the on-shell action in the average-difference basis (see section~$\S$\ref{sec:realatmuzeror} for details).

    \item Applying inverse Martin-Siggia-Rose (MSR) trick \cite{Martin:1973zz} then yields a classical Langevin equation, from which fluctuation and dissipation coefficients ($\theta_k$, $\bar{\theta}_k$) and their relations can be extracted. These relations can also be directly read off from the on-shell gravitational action, as we will discuss in section~$\S$\ref{sec:realatmuzeror}.
\end{enumerate}

Here, it is important to highlight that obtaining non-linear FDTs crucially requires an understanding of the interactions in the bulk. From the bulk perspective, it amounts to studying the scattering processes against black holes in the presence of Hawking radiation. Recently, in \cite{Loganayagam:2024mnj, Martin:2024mdm}, an exterior field-theoretic description has been constructed to study these scattering processes against neutral black holes. By \textit{field-theoretic}, we refer to a diagrammatic understanding of scattering with Feynman rules to calculate the scattering amplitudes. Before discussing this field theory, let us outline some of its salient features:
\begin{itemize}
    \item \textit{Dissipation}: It accounts for the effects of particles falling into the black hole.
    \item \textit{Fluctuations}: It incorporates thermal fluctuations caused by Hawking radiation.
    \item \textit{Exterior nature}: The field theory is restricted to the region outside the black hole, as the interior remains inaccessible.
\end{itemize}

The approach involves treating the exterior field theory as an open system, with the black hole as the environment. Such systems are described by the grSK geometry ( See \cite{Haehl:2024pqu} for more details on the SK formalism and grSK geometry), which is constructed by connecting the two copies of the black hole exterior along their future horizons. The grSK prescription has been subjected to a range of such consistency checks and has proven successful in capturing the physics of Hawking radiation across multiple scenarios. These include systems involving Brownian particle \cite{Chakrabarty:2019aeu, Bu:2021jlp}, scalar fields \cite{Jana:2020vyx, Pantelidou:2022ftm, Loganayagam:2022zmq}, fermionic fields \cite{Loganayagam:2020eue}, gauge fields \cite{Bu:2020jfo, He:2022jnc}, and even linearized gravity \cite{Ghosh:2020lel} (also see \cite{Colin-Ellerin:2020mva, He:2021jna, He:2022deg, Caron-Huot:2022lff}). It has also been extended to cases involving chemical potentials \cite{Loganayagam:2020iol, Chakrabarty:2020ohe}.\footnote{For the gravitational dual of the deformed SK contour, see \cite{Sivakumar:2024iqs}.}

\medskip

Our objective in this note is to establish the relationship between fluctuations and dissipation in finite-density holographic systems. As a first step toward this goal, we address the problem of scattering in the background of a charged black hole. For instance, we ask:
\begin{center}
    \emph{Does the existence of exterior field theory extend to charged black holes?} 
\end{center}
\noindent
To answer this question, we consider an interacting complex scalar field in the Reissner-Nordstr\"{o}m black hole (RN-AdS) background. The answer suggests the existence of a field theory living outside of the outermost horizon of a charged black hole. This has been achieved by the gravitational Schwinger-Keldysh formalism for charged black holes, known as \emph{RNSK geometry} \cite{Loganayagam:2020iol}, which we will review in this note (see section~\S\ref{sec:RNSK}).

\medskip

This exterior field theory captures both dissipation (due to infalling modes) and fluctuations (corresponding to Hawking radiation). In here, we develop a diagrammatic framework for computing $n$-point functions for interacting charged scalars. From the boundary perspective, it gives real-time correlations of the dual CFT held at finite temperature and finite density. Finite-density systems exhibit less understood properties of matter, such as high-$T_c$ superconductors and non-Fermi liquids. Thus, our analysis would add to the progress that has been made using the AdS/CFT correspondence to study these systems \cite{Hartnoll:2016apf}. 

\medskip

Using this exterior field theory, we want to answer whether non-linear \textit{fluctuation-dissipation relations} (FDRs) exist for holographic systems at finite density. To do this, we will closely follow \cite{Jana:2020vyx} where authors have evaluated non-linear FDRs by probing a real scalar field in a neutral black hole at the level of contact interactions. The only major difference would be that our probe is charged along with a charged black hole. We will keep our focus on contact interaction in this note; however, the whole analysis can be generalised to the arbitrary order in coupling constant using Feynman diagrammatics.

\medskip

We conclude this introduction by summarising the key findings of this article:

\begin{itemize}
\item We constructed an \emph{exterior field theory} for interacting Hawking radiation against charged black holes and developed Witten diagrammatics to analyze scattering processes.
\item Using this exterior field theory, we derived the (non)-linear \emph{holographic fluctuation-dissipation relations} at small but finite density and verified its consistency in the zero-density limit.
\end{itemize}

\bigskip

\subsection*{Outline of the article}

\medskip

In this article, we begin with a discussion of the motivation and a short review of the holographic fluctuation-dissipation theorem, including the role of exterior field theory in describing black hole scattering. In section~\S\ref{sec:RNSK}, we briefly review the RNSK geometry and examine \emph{CPT} action on fields. In section~\S\ref{sec:phiinRNSK}, we solve the dynamics of a self-interacting complex scalar in the RNSK geometry perturbatively. Section~\S\ref{sec:extdiagram} introduces the Feynman diagrammatics to construct an exterior field theory and lists relevant Witten diagrams for four-point and six-point functions. In section~\S\ref{sec:gradexp}, we construct the on-shell in a gradient expansion near the small charge regime. Section~\S\ref{sec:holFDR} presents (non)-linear fluctuation-dissipation relations at small but finite density, verifying their zero-density limit. Finally, section~\S\ref{sec:Discussion} summarises the results and suggests future directions, with appendices providing explicit calculations omitted from the main text.

\section{RNSK geometry}\label{sec:RNSK}

In this section, we briefly review the gravitational dual of the Schwinger-Keldysh (SK) formalism in the context of charged black holes, namely the \emph{RNSK geometry} \cite{Loganayagam:2020iol}. This bulk geometry describes the dynamics of a boundary conformal field theory (CFT) at finite temperature and finite density, formulated within the SK framework. After introducing the essential features of this geometry, we also discuss the role of the \emph{CPT (Charge-Parity-Time reversal)} transformation, which maps ingoing bulk solutions to outgoing ones.

Let us start with the Reissner--Nordstr\"{o}m (RN) black brane solution in AdS$_{d+1}$ in ingoing Eddington--Finkelstein coordinates and a \emph{mock tortoise coordinate} $\zeta $:
\begin{equation}\label{eq:gA}
    \begin{split}
        &\mathrm{d}s^{2} = - r^{2} \, f \, \mathrm{d}v^{2} + i \, \beta \, r^{2} \, f \, \mathrm{d}v \, \mathrm{d}\zeta + r^{2} \, \mathrm{d} \mathbf{x}_{d-1}^{2} \ , \\& \mathcal{A}_{M} \, \mathrm{d}x^{M} = - \mu \left(\frac{r_{+}}{r}\right)^{d-2} \, \mathrm{d}v \ , 
    \end{split}
\end{equation}
where $f$ is the emblackening factor of the RN--AdS black brane, $\beta $ is its inverse Hawking temperature, $r_+$ is its outer horizon radius, and $\mu$ is the chemical potential of boundary theory.\footnote{For an alternative description of the geometry in terms of orthonormal 1--forms, see \cite{Loganayagam:2020iol}.} Here, we have chosen the gauge so that the gauge field $\mathcal{A}_{M}$ vanishes at infinity (i.e., $ r \to \infty$), unlike in \cite{Liu:2009dm, Faulkner:2009wj, Faulkner:2013bna}, where it is set to vanish at the horizon. As a result, the chemical potential is determined by the gauge field at the horizon rather than at the boundary. 

The above-mentioned quantities can also be given in terms of the mass parameter $M$ and the charge parameter $Q$ of the black brane as,
\begin{equation}\label{temp}
\begin{split}
f(r) &= 1 +\frac{Q^2}{r^{2d-2}} -\frac{M}{r^d} \ , \\
\frac{1}{\beta} &= \frac{d\, r_+}{4 \pi}  \left(1- \frac{\left(d-2\right) Q^2}{d \, r_{+}^{2d-2}}\right) \ , \\
\mu  &= \frac{g_{_{\mathcal{F}}} \, Q}{r_{+}^{d-2}} \left[ \frac{d-1}{2 \left(d-2\right)} \right]^{\frac{1}{2}}\ ,
\end{split}
\end{equation}
where $g_{_{\mathcal{F}}}$ represents the gauge coupling constant in the Einstein-Maxwell bulk action, given by,
\begin{equation}
    \frac{1}{2 \kappa^2} \int d^{d+1} x \sqrt{-g} \left[ \mathcal{R}+d(d-1)-\frac{1}{g_{_{\mathcal{F}}}^2}\mathcal{F}_{MN} \mathcal{F}^{MN}\right] \ ,
\end{equation}
where $\kappa^2$ labels Newton's gravitational constant, and we have set the AdS radius to unity ($l_{\rm AdS} =1$).

The mock tortoise coordinate $\zeta$ is related to the standard radial coordinate $r$ via the following differential equation,

\begin{equation}\label{mock tortoise}
\begin{split}
\frac{\mathrm{d}r}{\mathrm{d}{\zeta }} = \frac{i \beta }{2} \, r^{2} f(r) \, .
\end{split}
\end{equation}
The differential equation \eqref{mock tortoise} above then normalises $\zeta$ such that it has a unit jump across the logarithmic branch cut that starts from $r_+$ and $\zeta (r=\infty + i \varepsilon)=0$ can be taken without any loss of generality (see figure~\ref{fig:fixedv}). This condition, along with the above differential equation, then uniquely defines $\zeta$ everywhere in the neighbourhood of $r \in [r_{+}, \infty)$. The real part of $\z$ starts at 0, remains unchanged as the contour moves toward $r_+$ above the real axis in the complex $r$-plane, and gains $+1$ when the contour passes below the real axis, maintaining this value at the other infinity. The imaginary part of $\z$ has no monodromy and parametrizes both legs of the contour in the complex $r$-plane identically.

\begin{figure}[H]
	\begin{center}
		\begin{tikzpicture}[scale=0.8]
		\draw[white!60!gray, line width = .7] (-4,2.3) -- (-4,-2) ;
		\draw[white!60!gray, line width = .7] (-5,0) --(10,0);
		\draw[teal, ultra thick] (1,1) -- (8,1);
		\draw[teal, ultra thick] (1,1) arc (45: 315 : 1.414);
		\draw[teal, ultra thick] (1,-1) --(8,-1);
		\draw[decoration={snake, segment length = 2.5 mm, amplitude=3}, decorate, red!80!black, thick] (0,0) -- (8,0);
		\draw[decoration={snake, segment length = 2.5 mm, amplitude=3}, decorate, red!80!black, thick] (-3,0) -- (-5,1);
		\node[circle,scale= 0.4,red!80!black,fill] at (-3,0) {} ;
		\node[circle,scale= 0.4,red!80!black,fill] at (0,0) {} ;
		\node[circle,scale= 0.4,red!80!black,fill] at (8,0) {} ;
		\node[circle, fill, cyan!40!teal, scale= 0.17] at (1,1) {} ;
		\node[circle, fill, cyan!40!teal, scale= 0.17] at (1,-1) {};
		\node[circle, fill, black, scale= 0.4] at (8,1) {} ;
		\node[circle, fill, black, scale= 0.4] at (8,-1) {} ;
		
		\node[white!60!gray] at (-4,2.3) {\scriptsize$\blacktriangle$};
		\node[white!60!gray] at (10,0) {\scriptsize $\blacktriangleright$};
		\node[blue!40!black] at (5,1) {$\blacktriangleleft$};
		\node[blue!40!black] at (5,-1) {$\blacktriangleright$};
		\node[blue!40!black] at (-1.41,0) {$\blacktriangledown$};

		\node[teal!40!black] at (4,1.5) {\scriptsize $\mathbf{Re} \,  \zeta = 0$};
		\node[teal!40!black] at (4,-1.5) { \scriptsize$\mathbf{Re} \, \zeta  = 1$};
		\node[red!65!black] at (0,-0.4) {\scriptsize $r_+$};
		\node[red!65!black] at (-3,-0.4) {\scriptsize $r_h^{\mathrm{int}}$};
		\node[red!65!black] at (8,-0.4) { \scriptsize$\infty$};
		\node at (9.7,1) {  \scriptsize $\infty+i\varepsilon \ (\z=0)$};
		\node at (9.7,-1) { \scriptsize $\infty-i\varepsilon \ (\z=1)$};
		\node[white!20!gray] at (-4, 2.75) {\scriptsize $\mathbf{Im} \, r$};
		\node[white!20!gray] at (10.7,0) {\scriptsize $\mathbf{Re} \, r$};
		\end{tikzpicture}
		\caption{ The radial contour drawn on the complex $r$ plane, at fixed $v$. The locations of the two boundaries and the two horizons have been indicated.}
        \label{fig:fixedv}
	\end{center}
\end{figure}
We define the \emph{Reissner-Nordstr\"{o}m Schwinger-Keldysh} (RNSK) geometry as one constructed by taking the RN--AdS exterior and replacing the radial interval extending from the outer horizon to infinity by a doubled contour, as indicated in Fig.~\ref{fig:fixedv}.

We then obtain a geometry with two copies of RN--AdS exteriors smoothly stitched together by an `outer-horizon cap' region. This spacetime requires one further identification --- each radially constant slice must meet at the future turning point. Using sections of the Penrose diagram, the following figure (Fig.~\ref{fig:RNSKpic}) schematically describes the construction of RNSK geometry.

\begin{figure}[H]
\begin{center}
\begin{tikzpicture}

\begin{scope}[shift={(0,0)},scale=0.8]
\draw (-2,-2) -- (-2,2) (2,-2) -- (2,2);

\draw[color=blue!80, fill=orange!60, thick] (2,1.8) -- (0.2,0) -- (2,-1.8);
\draw[dashed,thick] (2,-1.2) -- ({0.25+0.3*cos(45)},0.3);
\draw[densely dashed,thick] ({0.25+0.3*cos(45)},0.3) arc (-40:-310:0.13);

\draw[dashed,thick] (2,-0.6) -- ({0.25+0.3*cos(45)+0.6*cos(45)*cos(45)},{0.3+0.6*sin(45)*sin(45)});
\draw[densely dashed,thick] ({0.25+0.3*cos(45)+0.6*cos(45)*sin(45)},{0.3+0.6*sin(45)*sin(45)}) arc (-40:-310:0.13);

\draw[dashed,thick] (2,0) -- ({0.25+0.3*cos(45)+1.2*cos(45)*cos(45)},{0.3+1.2*sin(45)*sin(45)});
\draw[densely dashed,thick] ({0.25+0.3*cos(45)+1.2*cos(45)*sin(45)},{0.3+1.2*sin(45)*sin(45)}) arc (-40:-310:0.13);

\draw (-2,-2) -- (2,2);
\draw (-2,2) -- (2,-2);
\node[rotate=-45] at ({0.15+1.2*cos(135)},{0.15+1.2*sin(135)}) {{\scriptsize{$\leftarrow r_{+} \rightarrow$}}};
\node[rotate=45] at ({-0.15+1.2*cos(225)},{0.15+1.2*sin(225)}) {{\scriptsize{$\leftarrow r_{+}\rightarrow $}}};

\draw (-2,-2) -- (-1,-3);
\draw (-2,2) -- (-1,3);
\node[rotate=45] at ({-0.6+0.15+1.2*cos(135)},{0.35+1.15+1.2*sin(135)}) {{\scriptsize{$ r_{-} \rightarrow$}}};
\node[rotate=-45] at ({-0.4 -0.15+1.2*cos(225)},{0.3-2+0.15+1.2*sin(225)}) {{\scriptsize{$ r_{-} \rightarrow$}}};

\draw (2,-2) -- (1,-3);
\draw (2,2) -- (1,3);
\node[rotate=-45] at ({2.7-0.6+0.15+1.2*cos(135)},{-0.1+0.35+1.15+1.2*sin(135)}) {{\scriptsize{$\leftarrow r_{-} $}}};
\node[rotate=45] at ({2.8-0.4 -0.15+1.2*cos(225)},{0.3-2+0.15+1.2*sin(225)}) {{\scriptsize{$ \leftarrow r_{-} $}}};

\node at (2.6,0) {{\scriptsize{$M_R$}}$\uparrow$};

\node at (0,3.2) {$\vdots$};
\node at (0,-3.2) {$\vdots$};
\node at (0,-4) {$(a)$};
\end{scope}

\begin{scope}[shift={(5,0)},scale=0.8]

\draw (-2,-2) -- (-2,2) (2,-2) -- (2,2);

\draw[color=blue!80, fill=blue!40, thick] (2,1.8) -- (0.2,0) -- (2,-1.8);
\draw[dashed,thick] (2,-1.2) -- ({0.25+0.3*cos(45)},0.3);
\draw[densely dashed,thick] ({0.25+0.3*cos(45)},0.3) arc (-40:-310:0.13);

\draw[dashed,thick] (2,-0.6) -- ({0.25+0.3*cos(45)+0.6*cos(45)*cos(45)},{0.3+0.6*sin(45)*sin(45)});
\draw[densely dashed,thick] ({0.25+0.3*cos(45)+0.6*cos(45)*sin(45)},{0.3+0.6*sin(45)*sin(45)}) arc (-40:-310:0.13);

\draw[dashed,thick] (2,0) -- ({0.25+0.3*cos(45)+1.2*cos(45)*cos(45)},{0.3+1.2*sin(45)*sin(45)});
\draw[densely dashed,thick] ({0.25+0.3*cos(45)+1.2*cos(45)*sin(45)},{0.3+1.2*sin(45)*sin(45)}) arc (-40:-310:0.13);

\draw (-2,-2) -- (2,2);
\draw (-2,2) -- (2,-2);
\node[rotate=-45] at ({0.15+1.2*cos(135)},{0.15+1.2*sin(135)}) {{\scriptsize{$\leftarrow r_{+} \rightarrow$}}};
\node[rotate=45] at ({-0.15+1.2*cos(225)},{0.15+1.2*sin(225)}) {{\scriptsize{$\leftarrow r_{+}\rightarrow $}}};

\draw (-2,-2) -- (-1,-3);
\draw (-2,2) -- (-1,3);
\node[rotate=45] at ({-0.6+0.15+1.2*cos(135)},{0.35+1.15+1.2*sin(135)}) {{\scriptsize{$ r_{-} \rightarrow$}}};
\node[rotate=-45] at ({-0.4 -0.15+1.2*cos(225)},{0.3-2+0.15+1.2*sin(225)}) {{\scriptsize{$ r_{-} \rightarrow$}}};

\draw (2,-2) -- (1,-3);
\draw (2,2) -- (1,3);
\node[rotate=-45] at ({2.7-0.6+0.15+1.2*cos(135)},{-0.1+0.35+1.15+1.2*sin(135)}) {{\scriptsize{$\leftarrow r_{-} $}}};
\node[rotate=45] at ({2.8-0.4 -0.15+1.2*cos(225)},{0.3-2+0.15+1.2*sin(225)}) {{\scriptsize{$ \leftarrow r_{-} $}}};

\node at (2.6,0) {{\scriptsize{$M_L$}}$\downarrow$};

\node at (0,3.2) {$\vdots$};
\node at (0,-3.2) {$\vdots$};

\node at (0,-4) {$(b)$};

\end{scope}

\begin{scope}[shift={(9,0)},scale=1.1]

\draw[color=blue!100, fill=blue!60, thick] (2,2) -- (2.4,-1.6) -- (0,0);

\draw[color=blue!80, fill=orange!60, thick] (2,2) -- (0,0) -- (2,-2);

\draw[dashed,thick] (2,-1.2) -- ({0.25+0.3*cos(45)},0.3);
\draw[densely dashed,thick] ({0.25+0.3*cos(45)},0.3) arc (-40:-330:0.13);

\draw[dashed,thick] (2,-0.6) -- ({0.25+0.3*cos(45)+0.6*cos(45)*cos(45)},{0.3+0.6*sin(45)*sin(45)});
\draw[densely dashed,thick] ({0.25+0.3*cos(45)+0.6*cos(45)*sin(45)},{0.3+0.6*sin(45)*sin(45)}) arc (-40:-330:0.13);

\draw[dashed,thick] (2,0) -- ({0.25+0.3*cos(45)+1.2*cos(45)*cos(45)},{0.3+1.2*sin(45)*sin(45)});
\draw[densely dashed,thick] ({0.25+0.3*cos(45)+1.2*cos(45)*sin(45)},{0.3+1.2*sin(45)*sin(45)}) arc (-40:-330:0.13);

\node[] at (2.5,1) {\scriptsize{\(M_L\)}};
\draw[->] (2.4,0.8)--(2.5,0); 

\node[] at (1.8,0.7) {\scriptsize{\(M_R\)}};
\draw[->] (1.8,0.9)--(1.8,1.6);

\node at (1.5,-2.9) {$(c)$};

\end{scope}

\end{tikzpicture}
\end{center}
\caption{
 Schematic construction of RNSK geometry: Figures  (a) and (b) show two copies of RN-AdS, respectively. Figure (c) represents the RNSK geometry with the black dashed lines denoting the fixed \(v\) contours shown in Fig.~\ref{fig:fixedv}.
}
\label{fig:RNSKpic}
\end{figure}

Now that we have elaborated on the RNSK geometry and its associated notation, let us answer the following question:
\begin{center}
  \emph{Given the ingoing modes, how do we construct the outgoing Hawking modes?}  
\end{center}
The idea behind answering this question is to make use of \emph{CPT} symmetry of the boundary CFT, which acts as,
\begin{equation}\label{CFT_CPT}
\begin{split}
v \longmapsto - v \, , \ \qquad \mathbf{x} \longmapsto - \mathbf{x} \, ,
\end{split}
\end{equation}
in addition to a charge conjugation. The implementation of this boundary \emph{CPT} symmetry on the bulk fields can be done by using standard AdS/CFT rules. The key point that is used is that the boundary symmetries can always be composed with bulk gauge symmetries (since they act trivially on the Hilbert space of states of the boundary theory). The detailed construction of the implementation of this boundary \emph{CPT} symmetry, onto the bulk, is given in \cite{Loganayagam:2020iol}. In particular, it was shown that the boundary \emph{CPT} transformation is an
automorphism of the RN--AdS principle bundle in the bulk. 

To see the action of \emph{CPT} explicitly, we begin by working in boundary momentum space, which proves to be convenient for our analysis. For example, a complex scalar field can be expressed as,
\begin{equation}
\Phi(\z, v,\textbf{x} ) = \int \frac{\mathrm{d}{k^0} \ \mathrm{d}^{d-1}{\textbf{k}}}{(2 \pi )^{d}} \ \Phi (\z, k^0 , \textbf{k}) , e^{- i k^0 v + i \textbf{k} \cdot \textbf{x}} \ .
\end{equation}

Since our focus in this work is restricted to complex scalar fields, we will not discuss the \emph{CPT} transformation for more general field content (e.g., see \cite{Loganayagam:2020iol} for spinors). Also, we only present the action of \emph{CPT} on a charged scalar field of charge $q$, showing how it converts an ingoing solution into an outgoing one, as:
\begin{equation}\label{eq:intooutmap}
\Phi^{\rm in}_{\rm q}(\z, k) \longmapsto e^{-\beta k^0 \zeta} e^{ \b \muq \, \bTh }\ \Phi^{\rm in}_{\rm -q}(\z, -k)  \equiv \Phi^{\rm out}_{\rm q}(\z, k)\ ,
\end{equation} 
where $k$ denotes the pair $\{k^0, \textbf{k}\}$ and we have defined\footnote{Here, we note that $\bTh$ is proportional to the gauge transformation parameter $\mathbb{\Lambda}$ of the bulk gauge field $\mathcal{A}_{M}$, introduced in \cite{Loganayagam:2020iol} to derive the action of \emph{CPT} in the bulk. Explicitly, $\bTh = \frac{\mathbb{\Lambda}}{i \beta \mu}$.}
\begin{equation}\label{eq:muqTheta}
\muq \equiv \  \mu q \ , \qquad \qquad \bTh(\z) \equiv \ -\frac{1}{\mu} \int _{0}^{\zeta } \mathrm{d}\zeta' \  \mathcal{A}_{v}(\zeta') = \int _{0}^{\zeta } \mathrm{d}\zeta' \left( \frac{r_+}{r}\right)^{d-2}   \ . 
\end{equation}

Note that we have chosen to indicate the charge $q$ as a subscript on the field, and we will continue to follow this convention throughout the text. The prefactor in the exponent shows that this map takes ingoing solutions that are analytic in $r$ to outgoing solutions which exhibit branch cuts.

We end this section with a summary of how the quantities that we have introduced behave as they traverse the horizon cap. As noted before, the mock tortoise coordinate $\z$ has a unit jump when it encircles $r = r_{+}$. Noting that the gauge field $\mathcal{A}_{v}$ has the value $-\mu $ at the horizon, we see that the scale factor in $e^{ \b \muq \, \bTh }$ is the fugacity of the boundary theory. To see, recall that there is no $\sqrt{-g}$ in the integrand in the expression for $\bTh$ (See Eq.~\eqref{eq:muqTheta}). So, $\bTh$ suffers a discontinuity of unity around the horizon despite $\mathcal{A}_v$ being analytic on the full RNSK contour. Thus, the fugacity $e^{\b \muq}$ can be seen by evaluating different boundary limits as shown below,
\begin{equation}
    \lim_{\z \to 0} e^{ \b \muq \, \bTh } =1 \ , \qquad  \lim_{\z \to 1} e^{ \b \muq \, \bTh } = e^{\b \muq} \ .
\end{equation}

To check the prescription of RNSK geometry, the authors in \cite{Loganayagam:2020iol} probe this geometry by free Dirac fermions. Later, the same (RNSK) geometry has been used to describe charge diffusion and energy transport in a charged plasma~\cite{He:2021jna, He:2022deg}. However, an important next step is to move beyond free-field dynamics and study how interactions behave in this background. This is precisely the question we turn to next.

\paragraph{Note:} To avoid clutter in the notation, we will occasionally suppress the subscript $q$ on fields indicating their charge, whenever the context makes the charge dependence clear.

\section{Self-interacting complex scalar in RNSK geometry}\label{sec:phiinRNSK}

In this section, we will consider an interacting charged scalar field $\Phi$ against a charged black hole background (i.e., RNSK geometry). As an illustrative example, we will only take a massless complex scalar field interacting via a quartic term,\footnote{This is chosen for simplicity; however, the whole analysis can be easily generalised to the $|\Phi|^n$ interaction.} with the action given by,
\begin{equation}\label{eq:actionPhi4}
    S = \ -\oint d \z \, d^{d}x \sqrt{-g} \left[ |D_{M} \Phi |^2 + \frac{\lambda}{2!\, 2!}|\Phi|^4 \right] \ , 
\end{equation}
with the covariant derivative $D_{M}$, given by,
\begin{equation}
    D_{M} \equiv \partial_{M}-i q \mathcal{A}_{M}  \ ,
\end{equation}
where $q$ and $\lambda$ are the charge and the interaction parameter of the field, respectively. Varying the above action (\ref{eq:actionPhi4}) w.r.t the field, we obtain the following field equation (non-linear Klein-Gordon equation) for $\Phi$:
\begin{equation}
    D_{M}D^{M}\Phi = \frac{\lambda }{2!} \Phi|\Phi|^2  \ .
\end{equation}

Before delving into the full dynamics and their implications, we will begin by examining the free part of the field equation only. Specifically, we will first describe the propagation of the scalar field $\Phi$ in the RNSK geometry, excluding interaction terms. This approach will also allow us to establish the necessary notation for subsequent sections. Later in this section, and the following one, we will use this established notation even with the interactions.

\subsection{Free complex scalar}\label{sec:freescalarRNSK}
Let us start with the linear part of the EOM (free Klein-Gordon equation) and solve it first. Later, we will correct this by adding non-linear terms in the Klein-Gordon (KG) equation and solve it by using the Green's function.

The field equation for a free complex scalar field $\Phi$ is given by,
\begin{equation}\label{eq:freeEOM}
    D_{M}D^{M} \Phi  = 0 \  .
\end{equation}
After the free field equation (\ref{eq:freeEOM}), we now turn to the solutions of this equation under the following double-Dirichlet boundary conditions,
\begin{equation}\label{eq:BCRLforPhi}
    \lim_{\z \to 0 } \Phi(\z,k) = J_{\sL}(k) \ , \qquad \qquad  \lim_{\z \to 1 } \Phi(\z,k) = J_{\sR}(k) \ ,   
\end{equation} 
where $J_{\sR}$ and $J_{\sL}$ are right and left boundary sources, respectively.

As far as our analysis is concerned, note that we are working in boundary momentum space, as 
\begin{equation}
    \Phi(\z,v,\mathbf{x}) =  \int \frac{d^d k}{(2 \pi)^d} \Phi(\z, k) e^{-i k^0 v+i \mathbf{k.x}} \ ,
\end{equation}
where $k$ represents the Fourier transformed tuple $\{k^0, \mathbf{k} \}$ corresponding to the variables $\{ v, \mathbf{x} \}$. In the momentum domain, we find that the free field equation given in Eq.~(\ref{eq:freeEOM}), can be explicitly written as--
\begin{equation}
    \begin{split}
        &\frac{d}{d \z}\left[ r^{d-1} \frac{d \Phi}{d \z}\right] +\frac{\b k^0}{2} \left[ r^{d-1}\frac{d \Phi}{d \z}+\frac{d}{d \z} \left(  r^{d-1}\Phi \right)\right]+r^{d-1} f \left(\frac{\b \mathbf{k}}{2}\right)^2 \Phi \\
        &\hspace{2cm}+\frac{\b q}{2} \left[ r^{d-1} \mathcal{A}_{v} \frac{d \Phi}{d \z}+\frac{d}{d \z} \left(  r^{d-1} \mathcal{A}_{v}  \Phi \right)\right]= 0 \ ,
    \end{split} 
\end{equation}
where we have suppressed the coordinate dependence of the field $\Phi$.

Since the field equation is a second-order ODE, the general solution is determined by two linearly independent solutions. We represent the solution using the ingoing and outgoing solutions, which are related by the \emph{CPT} isometry, given in Eq.~(\ref{eq:intooutmap}). The explicit action of this \emph{CPT} transformation on the field equation is presented in Appendix \ref{app:CPT}.

Thus, the free solution in terms of ingoing $\Gin$ and outgoing $\Gout$ boundary-to-bulk Green's function,\footnote{The approach involving ingoing and outgoing Green's functions was first investigated in \cite{Son:2009vu} and subsequently elaborated upon for the grSK geometry in \cite{Chakrabarty:2019aeu, Jana:2020vyx}.} can be written as,
\begin{equation}\label{Phi0PF}
    \Phi(\z,k)  = \ - \Gin(\z,k) J_{\Fb}(k)+e^{\beta(k^0 -\muq)} \Gout(\z,k) J_{\Pb}(k) \ ,
\end{equation}
with the following ingoing and outgoing Green's function relationship
\begin{equation}\label{eq:GintoGout}
    \Gout_{\rm q}(\z,k) = e^{-\beta k^0 \z} e^{\b \muq \bTh}  \Gin_{\rm -q}(\z,-k) \ , 
\end{equation}
where we have written the boundary sources in \emph{Future-Past} (FP) basis, defined as
\begin{equation}\label{FPbasis}
    J_{\Fb} = -\left[1+\nbe_k \right] J_{\sR} +\nbe_k \,  J_{\sL} \ , \qquad J_{\Pb} = -\nbe_k\left[J_{\sR}-J_{\sL}\right] \ ,
\end{equation}
where $\nbe_k$ is the Bose-Einstein factor at chemical potential, given by
\begin{equation}
    \nbe_k = \frac{1}{e^{\beta(k^0-\muq)}-1} \ .
\end{equation}
From the relation in Eq.~(\ref{eq:GintoGout}), it is clear that the ingoing solution is sufficient to construct the general solution. To solve for the ingoing Green's function, two boundary conditions must be supplied: 
\begin{enumerate}

\item It must be regular on the horizon, explicitly written as:
\begin{equation}
    \frac{d \Gin}{d \z}\bigg|_{r_+} = 0 \ .
\end{equation}

\item It is normalised to unity at the boundary.
\begin{equation}
     \lim_{r \to \infty} \Gin =1 \ .
\end{equation}
\end{enumerate}

\noindent
When written in the \emph{Right-Left} (RL) basis of the sources, the solution takes the Son-Teaney form \cite{Son:2009vu}, as
\begin{equation}\label{Phi0main}
\begin{split}
\Phi(\zeta, k) &\equiv \   g_{\sR}(\zeta,k)\,J_{\sR}(k)-g_{\sL}(\zeta,k)\,J_{\sL}(k) \ ,
\end{split}
\end{equation}
where $g_{\sR,\sL}$ denote the boundary-to-bulk Green's functions from the right and left AdS boundaries, respectively. These functions play a key role in constructing the bulk-to-bulk Green’s functions, as discussed in Appendix~\ref{app:bbGRNSK}. They can be expressed in terms of the ingoing and the outgoing Green's functions, taking the following form,
\begin{equation}\label{gRgLmain}
\begin{split}
&g_{\sR}(\zeta,k)  \equiv  \left(1+\nbe_{k}\right)\left[\Gin(\zeta,k)-\Gout(\zeta,k)  \right]\ ,  \\
&g_{\sL}(\zeta,k) \equiv \nbe_{k}\left[\Gin(\zeta,k)-e^{\beta (k^0-\muq) } \Gout(\zeta,k) \right]  \ .
\end{split}
\end{equation}
As it is clear from the name, the boundary conditions for these Green's functions are as follows:
\begin{equation}\label{eq:LBCgRgL}
    \lim_{\z \to 0} g_{\sR}(\zeta,k) = 0 \ , \qquad \qquad \lim_{\z \to 0} g_{\sL}(\zeta,k) = -1  \ ,
\end{equation}
and
\begin{equation}\label{eq:RBCgRgL}
    \lim_{\z \to 1} g_{\sR}(\zeta,k) = 1 \ , \qquad \qquad \lim_{\z \to 1} g_{\sl}(\zeta,k) = 0  \ .
\end{equation}
Combining the solution with these boundary conditions, we see that in the RL basis, the bulk field $\Phi$ satisfies the correct boundary conditions given in Eq.~\eqref{eq:BCRLforPhi}.

To conclude the discussion on the free Klein-Gordon equation, we also provide the general solution for the free conjugate field $\bar{\Phi}$. Since the analysis for the conjugate field closely mirrors that of the field itself, we will not repeat the steps but directly present the results and move to the interactions.

Before proceeding further, please note that a `bar' on top of a field, Green's function, or boundary source indicates that it is the conjugate version of the corresponding object. The free solution for the complex-conjugate scalar field is given by,
\begin{equation}\label{Phib0PF}
    \Phib(\z,k)  = \ - \Ginb(\z,k) \Jb_{\Fb}(k)+e^{\beta(k^0 +\muq)} \Goutb(\z,k) \Jb_{\Pb}(k) \ ,
\end{equation}
with the \emph{CPT} isometry relates these two Green's functions in the following manner,
\begin{equation}
    \Goutb_{\rm q}(\z,k) = e^{-\beta k^0 \z} e^{-\b \muq \bTh } \Ginb_{\rm -q}(\z,-k) \  ,
\end{equation}
where we have defined the Future-Past conjugate sources as,
\begin{equation}\label{FbPbbasis}
    \Jb_{\Fb} = -\left[1+\nbeb_{k}\right] \Jb_{\sR} +\nbeb_{k} \Jb_{\sL} \ , \qquad \Jb_{\Pb} = -\nbeb_{k}\left[\Jb_{\sR}-\Jb_{\sL}\right] \ ,
\end{equation}
where $\nbeb_k$ is the Bose-Einstein factor at finite chemical potential (with sign of $q$ reversed) given by,
\begin{equation}
    \nbeb_k = \frac{1}{e^{\beta(k^0+\muq)}-1} \ .
\end{equation}
In the right-left basis of the conjugate sources, the solution for the conjugate field is
\begin{equation}\label{Phib0main}
\begin{split}
\Phib(\zeta, k) &\equiv \   \gb_{\sR}(\zeta,k)\,\Jb_{\sR}(k)-\gb_{\sL}(\zeta,k)\,\Jb_{\sL}(k) \ ,
\end{split}
\end{equation}
where $\gb_{\sR,\sL}$ denote the boundary-to-bulk conjugate Green's functions from the right and left AdS boundaries for the conjugate field, which in terms of the ingoing and the outgoing conjugate Green's functions take the form,
\begin{equation}\label{gRbgLbmain}
\begin{split}
&\gb_{\sR}(\zeta,k)  \equiv  \left(1+\nbeb_{k}\right)\left[\Ginb(\zeta,k)-\Goutb(\zeta,k)  \right]\ ,  \\
&\gb_{\sL}(\zeta,k) \equiv \nbeb_{k}\left[\Ginb(\zeta,k)-e^{\beta (k^0+\muq) } \Goutb(\zeta,k) \right]  \ .
\end{split}
\end{equation}
Again, the boundary conditions for these conjugate Green's functions are:
\begin{equation}\label{eq:LBCgRbgLb}
    \lim_{\z \to 0} \gb_{\sR}(\zeta,k) = 0 \ , \qquad  \qquad  \lim_{\z \to 0} \gb_{\sL}(\zeta,k) = -1  \ ,
\end{equation}
and
\begin{equation}\label{eq:RBCgRbgLb}
    \lim_{\z \to 1} \gb_{\sR}(\zeta,k) = 1 \ , \qquad  \qquad \lim_{\z \to 1} \gb_{\sl}(\zeta,k) = 0  \ .
\end{equation}
Thus, it is easy that the following boundary limits hold:
\begin{equation}
    \lim_{\z \to 0 } \Phib(\z,k) = \Jb_{\sL}(k) \ , \qquad \qquad  \lim_{\z \to 1 } \Phib(\z,k) = \Jb_{\sR}(k) \ .  
\end{equation}
Here, we should note that the analysis of the conjugate field (i.e., field equation, solution, etc.) is exactly similar to that of the field itself. The only difference we can spot is the change in the sign of charge, i.e. $q \mapsto -q$.

\subsection{Interacting complex scalar}

Now that we have worked through the free Klein-Gordon equation and its solutions, we can move on to interactions. In this part, we will consider the full quartic theory and use the method of Green's function in the bulk to determine the interactive portion of the solution.

For completeness, we once again write the full Klein-Gordon field equation as, 
\begin{equation}\label{EOM}
    D_{M}D^{M}\Phi = \frac{\lambda }{2!} \Phi|\Phi|^2  \ .
\end{equation}
We will solve this field equation perturbatively in the coupling constant $\lambda$ by expanding the field $\Phi$ as,
\begin{equation}\label{eq:Phipert}
  \Phi = \sum_{i=0}^{\infty} \lambda^{i} \Phi_{(i)} = \Phi_{(0)}+\lambda \Phi_{(1)}+\lambda^{2} \Phi_{(2)}\ldots  \ ,
\end{equation}
under the double-Dirichlet boundary conditions (similar to that of the free theory), given by:
\begin{equation}\label{eq:BCs}
    \lim_{\z \to 0,1} \Phi_{(0)}(\z,k) = J_{\sL,\sR}(k)  \ , \qquad \qquad \lim_{\z \to 0,1} \Phi_{(i)}(\z,k) = 0 \quad \forall \quad i>1 \ ,
\end{equation}
where $J_{\sL}$ and $J_{\sR}$ are the left and the right boundary sources, respectively.

These boundary conditions are \emph{doubled} versions of the standard conditions in AdS/CFT, modified specifically to fit the Schwinger-Keldysh formalism. The leading-order solution $\Phi_{(0)}$ is just the free solution that we have already obtained in Eq.~(\ref{Phi0PF}). Regarding the boundary conditions, $\Phi_{(0)}$ satisfies the GKPW \cite{Gubser:1998bc, Witten:1998qj} boundary conditions, while all higher-order corrections should vanish at both boundaries, consistent with Eq.~(\ref{eq:BCs}).

To find the higher-order solutions, we use the \textit{bulk-to-bulk} Green's function. This function should vanish at both boundaries and is therefore termed \textit{binormalisable}. The binormalisable bulk-to-bulk Green's function\footnote{Here, we adopt the same notation as in \cite{Loganayagam:2024mnj}, but the convention for the bulk-to-bulk Green’s function $\bbG$ used here differs by an overall sign.} $\bbG$ satisfies:
\begin{equation}\label{eq:bbGdef}
    D_{M}D^{M} \bbG(\z|\z',k) = \frac{\delta(\z-\z')}{\sqrt{-g}} \ ,
\end{equation}
where $\delta(\z-\z')$ is the radial delta function on the RNSK contour. The above differential equation should be solved with the following boundary conditions,
\begin{equation}
    \lim_{\z \to 0}\bbG(\z|\z',k)= \lim_{\z \to 1}\bbG(\z|\z',k) =0 \ .
\end{equation}
The explicit calculation for the computation of binormalisable bulk-to-bulk Green's function is given in the Appendix \ref{app:bbGRNSK} and the explicit expression is given in Eq.~(\ref{eq:bbGRLbasis}) as,
\begin{equation}
    \bbG(\z|\z',k)  
     = \frac{e^{\b k^0 \z' -\b \muq \bTh(\z') }}{(1+\nbe_{k}) } \left( \frac{ \Theta(\z>\z') g_{\sL}(\z,k) g_{\sR}(\z',k)+ \Theta(\z<\z') g_{\sR}(\z,k) g_{\sL}(\z',k) }{ \left[ K^{\rm in}_{\rm q}(k) -K^{\rm in}_{\rm -q}(-k)\right] } \right) \ ,
\end{equation}
where $\Theta$ is the radial Heavyside step function on the RNSK contour.

Note that $K^{\rm in}$ is the retarded two-point boundary correlator in the above equation. Additionally, we point out that later (in section~$\S$\ref{sec:extRNSK}) we will also use the \emph{retarded} $\bbGR$ and the \emph{advanced} $\bbGA$ bulk-to-bulk Green's functions. The expressions for two Green's functions are also present in Appendix~\ref{app:bbGRNSK}, along with their relationship with each other \eqref{eq:bbGARreci} and with binormalisable Green's function \eqref{eq:bbGARbbG}. We reiterate that all Green's functions introduced here are defined with a convention that differs by an overall sign from that used in \cite{Loganayagam:2024mnj}.

With the help of the binormalisable bulk-to-bulk Green's function, the remaining higher-order terms can be written as: 
\begin{equation}
    \Phi_{(i)}(\z,k) = \oint_{\z'} \bbG(\z|\z',k) \mathbb{J}_{(i)}(\z',k) \ , \qquad \text{with} \quad  \oint_{\z} \equiv \oint d\z \sqrt{-g} \ ,
\end{equation}
where $\mathbb{J}_{(i)}$ are bulk sources for $i$-th order term in the solution. The first two terms of these bulk sources are as follows:
\begin{equation}\label{bulksources}
\small
    \begin{split}
        \mathbb{J}_{(1)}(\z,k) &=\int \prod_{i=1}^{3}\left(\frac{d^d p_i}{(2\pi)^d}\right)  (2 \pi)^d \delta^d \left(\sum_{i}^{3}p_i  -k\right)  \frac{1}{2!} \Phi_{(0)}(\z,p_1) \bar{\Phi}_{(0)}(\z,p_2)\Phi_{(0)}(\z,p_3) \ , \\
        \mathbb{J}_{(2)}(\z,k) &= \int \prod_{i=1}^{3}\left(\frac{d^d p_i}{(2\pi)^d}\right)  (2 \pi)^d \delta^d \left(\sum_{i}^{3}p_i  -k\right)  \\
        &  \times\left[ \Phi_{(1)}(\z,p_1) \bar{\Phi}_{(0)}(\z,p_2)\Phi_{(0)}(\z,p_3)  +\frac{1}{2!} \Phi_{(0)}(\z,p_1) \bar{\Phi}_{(1)}(\z,p_2)\Phi_{(0)}(\z,p_3)\right] \ .
    \end{split}
\end{equation}
Now that we have expressed the solution in formal form, our next task is to construct the on-shell action for the $|\Phi|^4$ theory in the RNSK background. Once the on-shell action is obtained, boundary correlators can be computed by taking functional derivatives with respect to the boundary sources.

We now turn towards the computation of the on-shell action for the $|\Phi|^4$ theory in the RNSK background.

\section{On-shell action \& Exterior diagrammatics}\label{sec:extdiagram}

Previously, in section~$\S$\ref{sec:phiinRNSK}, we explained the method for solving the bulk field equations to arbitrary orders in the coupling constant $\lambda$. By substituting these solutions into the action, we can compute the on-shell action\footnote{Since our analysis is limited to tree-level diagrams, we use on-shell action and tree-level influence phase interchangeably.} to any desired order in $\lambda$. Our approach, also used in \cite{Martin:2024mdm}, circumvents the need to evaluate the boundary terms in the on-shell action explicitly. It leverages the fact that the boundary value of the higher-order solution vanishes, thereby simplifying the computation.

The bare action $S_{\rm bare}$ is given by
\begin{equation}
    S_{\rm bare} = \ -\oint_{\mathcal{M}} \left( |D_{M} \Phi |^2 + \frac{\lambda}{2!\, 2!} |\Phi|^{4} \right) \  , \qquad \text{where} \quad \oint_{\mathcal{M}} \equiv \oint d \z \, d^d x \sqrt{-g} \ ,
\end{equation}
where $\mathcal{M}$ denotes the RNSK background. Inserting the perturbative solution into the bare action yields the \emph{on-shell action} $S_{\rm os}$ in the following form:
\begin{equation}\label{eq:Sos}
    S_{\rm os} = \ -\int_{\partial\mathcal{M}}  \Phi_{(0)}  \overline{\left( D_{A} \Phi_{(0)} \right)}  + \lambda \oint_{\mathcal{M}}  \Big[ \frac{ 1 }{2!}   \left(\Phib -\Phib_{(0)}\right)\Phi  |\Phi|^{2} - \frac{ 1 }{2! 2!}   |\Phi|^{4} \Big]  \ ,   
\end{equation}
where $\partial \mathcal{M}$ denotes the boundary of $\mathcal{M}$. Note that in deriving Eq.~(\ref{eq:Sos}), we have used the field equations and imposed the appropriate boundary conditions to significantly simplify the expression for the on-shell action. 

The on-shell action $S_{\rm os}$, expanded in powers of $\lambda$ , is expressed as:
\begin{equation}
    S_{\rm os} = S_{(2)} + S_{(4)} + S_{(6)}+\ldots \ ,
\end{equation}
where the subscript indicates the number of boundary sources in each term. For example, the quadratic term $S_{(2)}$, the quartic term $S_{(4)}$ and the term with six sources $S_{(6)}$ are given by:
\begin{equation}\label{eq:S2S4S6}
    \begin{split}
        S_{(2)} &= - \int_{\partial\mathcal{M}}  \Phi_{(0)}  \overline{\left( D_{A} \Phi_{(0)} \right)} \ , \\
        S_{(4)} &=  -\frac{\lambda}{2!\, 2!} \oint_{\mathcal{M}}  \left(\Phib_{(0)} \Phi_{(0)}\right)^{2} \ , \\
        S_{(6)} &= -\frac{\lambda^2}{2} \oint_{\mathcal{M}}    \    \Phib_{(0)} \Phi_{(1)}  \Phib_{(0)} \Phi_{(0)}  \ .
    \end{split}
\end{equation}
The terms above correspond to the free, contact, and single-exchange contributions at the tree level in Feynman diagrams. Within the framework of open effective field theory \cite{Jana:2020vyx, Loganayagam:2022zmq}, $S_{(n)}$ represents the $n$-point influence phase of the boundary theory.

We find it useful to express the on-shell action in the \textit{Past-Future} (PF) basis, as defined in Eqs.~(\ref{FPbasis}) and (\ref{FbPbbasis}). In this basis, the Schwinger-Keldysh collapse (unitarity) and the Kubo-Martin-Schwinger conditions (thermality) manifest as the vanishing of coefficients for terms involving all Future and all Past sources, respectively. Furthermore, this basis exclusively incorporates retarded/advanced propagators in the boundary correlators, ensuring the description of causal scattering processes.

The explicit computation of the on-shell action requires evaluating monodromy integrals on the RNSK contour. This involves a contour integral over the complexified radial coordinate, as illustrated in figure~\ref{fig:fixedv}. The integral simplifies into a single exterior radial integral, with discontinuities arising at the horizon cap. The origin and evaluation of these discontinuities are detailed in Appendix~\ref{app:Disc}. Using the discontinuity formulas listed in the appendix (see Eqs.~\eqref{RNSKtoExt1} and \eqref{RNSKtoExt2}), the on-shell action can be expressed in terms of a single exterior radial integral, as shown in Appendix~\ref{app:onshell}.

To declutter the expressions, we write the terms in the on-shell action in the following way,
\begin{equation}\label{eq:SosIdef}
    \begin{split}
        S_{(2n)} &= \int_{k_{1,2,...,2n}} \sum_{r,s=0}^{n} \mathcal{I}^{(2n)}_{r,s}(k_1,..,k_{2n}) \prod_{i=1}^{r} \Jb_{\Fb}(k_i) \prod_{j=1}^{s} J_{\Fb}(k_j) \prod_{l=r+1}^{n} \Jb_{\Pb}(k_l) \prod_{m=s+1}^{n} J_{\Pb}(k_m) \ ,
    \end{split}
\end{equation}
where we have used the following notation 
\begin{equation}
    \int_{k_{1,2,...,m}} \equiv  \int \frac{d^d k_{1}}{(2 \pi)^d} \int \frac{d^d k_{2}}{(2 \pi)^d}   \ldots \int \frac{d^d k_{m}}{(2 \pi)^d}  \ \delta^{(d)} \left( k_{1}+k_2+\ldots +k_m \right) \ .
\end{equation}

From this point onward, we will only present the expressions for $\mathcal{I}_{r,s}$ derived from RNSK geometry. The physical interpretation of $\mathcal{I}^{(2n)}_{r,s}$ corresponds to a scattering process, where $r$ and $s$ ingoing modes with charges $-$ and $+$, respectively, scatter into $(n- r)$ and $(n- s)$ outgoing modes with charges $+$ and $-$, respectively. Although the results for $\mathcal{I}^{(2n)}_{r,s}$ themselves are intricate, there is a particularly insightful way (i.e., Feynman diagrammatic way given in section~$\S$\ref{sec:extRNSK}) to summarise them that highlights their physical significance. We find that our findings can be concisely represented by a diagrammatic expansion governed by the Feynman rules.

As discussed earlier, the explicit computation of the on-shell action in the bulk naturally gives rise to an \textit{exterior field theory}. This theory exists outside the outermost horizon of the RN-AdS black brane. Next, we will demonstrate how this exterior field theory can be utilised to compute boundary correlation functions. The key idea is to define Feynman diagrammatic rules for the exterior theory and then apply these rules to compute the on-shell action (or boundary generating functional).

Below, we outline the diagrammatic rules for computing the on-shell action (tree-level influence phase).

\subsection{Feynman Rules for Exterior diagrammatics}\label{sec:extRNSK}
The conventions for momentum $k$ flow are as follows: In boundary-to-bulk propagators, momentum flows from the boundary to the bulk (i.e, from up to down) while in bulk-to-bulk propagators, it flows from left to right. Here, it is important to highlight that we will only present the Feynman rules/diagrammatics in the PF basis. However, one could also do the same in other basis, but the rules and diagrams might look different in other basis.

The \textit{boundary-to-bulk propagators} (along with their corresponding source) for complex scalar are:
\begin{equation}
    \begin{tikzpicture}[scale=1.5]
        \draw[blue] (-0.5,0)--(0.5,0);  
        \Sdiode{0}{0}{0}{-1.7};
        \node at (0,-1.7) {$\bullet$};
        \node at (0.2,-1.7) {$r$};
        \node at (0.3,-0.5) {$k$};
        \node[red] at (-0.3,-0.7) {$\uparrow$};
        \node at (1.65, -0.75) {$\equiv \ \frac{\Gin(r,k)}{1+\nq_{k}} \ J_{\Fb}(k)\ ,$};
    \end{tikzpicture}
    \hspace{2cm}
    \begin{tikzpicture}[scale =1.5]
        \draw[blue] (-0.5,0)--(0.5,0);  
        \Sdiode{0}{-1.7}{0}{0};
        \node at (0,-1.7) {$\bullet$};
        \node at (0.2,-1.7) {$r$};
        \node at (0.3,-0.5) {$k$};
        \node[red] at (-0.3,-0.7) {$\uparrow$};
        \node at (2.5, -0.75) {$\equiv  \ \Gout(r,k ) \ J_{\Pb}(k) \ ,$};
    \end{tikzpicture}
\end{equation}
where $r$ denotes the bulk point, and $k$ denotes the momentum carried along the propagator. The red-colored arrow indicates the direction of charge flow. Additionally, note that the orientation of the `diode' (or triangle) symbol distinguishes whether the propagator is ingoing or outgoing. 

The corresponding \textit{boundary-to-bulk propagators} for the conjugate field are:
\begin{equation}
    \begin{tikzpicture}[scale=1.5]
        \draw[blue] (-0.5,0)--(0.5,0);  
        \Sdiode{0}{0}{0}{-1.7};
        \node at (0,-1.7) {$\bullet$};
        \node at (0.2,-1.7) {$r$};
        \node at (0.3,-0.5) {$k$};
        \node[red] at (-0.3,-0.7) {$\downarrow$};
        \node at (1.65, -0.75) {$\equiv \ \frac{\Ginb(r,k)}{1+\nqb_{k}} \ \Jb_{\Fb}(k)\ ,$};
    \end{tikzpicture}
    \hspace{2cm}
    \begin{tikzpicture}[scale =1.5]
        \draw[blue] (-0.5,0)--(0.5,0);  
        \Sdiode{0}{-1.7}{0}{0};
        \node at (0,-1.7) {$\bullet$};
        \node at (0.2,-1.7) {$r$};
        \node at (0.3,-0.5) {$k$};
        \node[red] at (-0.3,-0.7) {$\downarrow$};
        \node at (2.5, -0.75) {$\equiv \ \Goutb(r,k ) \ \Jb_{\Pb}(k) \ ,$};
    \end{tikzpicture}
\end{equation}
and the retarded and advanced \textit{bulk-to-bulk propagators} are:
\begin{equation}
\begin{split}
    &\begin{tikzpicture}[scale=1.7]
        \Sdiode{-1}{0}{1}{0};
        \node[red] at (-0.3,-0.2) {$\rightarrow$};
        \node at (-0.3,0.2) {$k$};
        \node at (-1,-0.2) {$r_1$};
        \node at (1,-0.2) {$r_2$};
        \node at (-1,0) {$\bullet$};
        \node at (1,0) {$\bullet$};
        \node at (2.5,0) { \ $ \equiv \ i \, \bbGR (r_2|r_1,k) \ ,$};
    \end{tikzpicture}
     \ \\
    &\begin{tikzpicture}[scale=1.7]
        \Sdiode{1}{0}{-1}{0};
        \node[red] at (-0.3,-0.2) {$\rightarrow$};
        \node at (-0.3,0.2) {$k$};
        \node at (-1,-0.2) {$r_1$};
        \node at (1,-0.2) {$r_2$};
        \node at (-1,0) {$\bullet$};
        \node at (1,0) {$\bullet$};
        \node at (2.5,0) { \ $ \equiv \ i \, \bbGA (r_2|r_1,k) \ .$};
    \end{tikzpicture}
\end{split}    
\end{equation}
where $r_1$ and $r_2$ are two bulk points connected by these propagators. The explicit expressions for retarded $\bbGR$ and advanced $\bbGA$ bulk-to-bulk Green's function are given in Eqs.~\eqref{eq:bbGRq} and \eqref{eq:bbGAq}, respectively. Note that the bulk-to-bulk propagators in our case are given by $i \bbG$, rather than $-i \bbG$ as used in \cite{Loganayagam:2024mnj}. This discrepancy arises because our convention for the Green's function differs from that of \cite{Loganayagam:2024mnj}: specifically, we define the Green's function as the inverse of the Laplacian operator, rather than the negative of it (See Eq.~\eqref{eq:bbGdef}).

It is important to note that all these propagators are expressed in a basis that explicitly reveals their causal properties. As a result, the causal structure of Witten diagrams\footnote{Witten diagrams are the AdS/CFT analogues of Feynman diagrams, with the key distinction that their external legs are anchored on the AdS boundary.} will naturally emerge during their construction, thereby simplifying the understanding of scattering processes against black holes.

The final ingredient needed to construct the Witten diagrams is the set of bulk vertices. Before specifying the vertices in this exterior field theory, it is useful first to define,
\begin{equation}
    n_{k, \alpha} \equiv \frac{1}{ e^{\b (k^0 -\alpha \, \muq )} -1} \ , \qquad \text{and} \quad k_{12...m} \equiv k_1 +k_2 +... k_m \  , 
\end{equation}
where we note $n_{k,1} = \nq_{k}$ and $n_{k,-1} = \nqb_{k}$, and we will use the notation in these two relations interchangeably later. 

Before we write down the bulk vertices, we would like the reader to note that we will exclusively use the $n_{k, \alpha}$ notation in describing vertices. Now, let us come to the \emph{vertices} in this exterior field theory, given as follows:
\begin{equation}
    \begin{tikzpicture}[scale=1.2]
        \node at (0,0) {$\bullet$};
        \Ssemicap{-1}{1}{0}{0}
        \node at (-0.7,0.4) {$k_1$};
        \draw[red, arrows={->[scale=1.2,red]}] (-0.5,0.7) -- (-0.2,0.4);
        \Sdiodearrow{0}{0}{1}{1}
        \node at (0.8,0.4) {$k_2$};
        \draw[red, arrows={->[scale=1.2,red]}] (0.2,0.4) -- (0.5,0.7);
        \Sdiodearrow{0}{0}{1}{-1}
        \node at (0.8,-0.3) {$k_4$};
        \draw[red, arrows={->[scale=1.2,red]}] (0.3,-0.5) -- (0.6,-0.8);
        \Sdiodearrow{0}{0}{-1}{-1}
        \node at (-0.7,-0.3) {$k_3$};
        \draw[red, arrows={->[scale=1.2,red]}] (-0.5,-0.7) -- (-0.2,-0.4);
        \node at (3,0) {$=  i \lambda  \frac{n_{-k_1, -1}  }{ n_{k_{234},1}  } \ ,$};
    \end{tikzpicture}
    \qquad \qquad
    \begin{tikzpicture}[scale=1.2]
        \node at (0,0) {$\bullet$};
        \Sdiodearrow{0}{0}{-1}{1}
        \node at (-0.7,0.4) {$k_1$};
        \draw[red, arrows={->[scale=1.2,red]}] (-0.5,0.7) -- (-0.2,0.4);
        \Ssemicap{1}{1}{0}{0}
        \node at (0.8,0.4) {$k_2$};
        \draw[red, arrows={->[scale=1.2,red]}] (0.2,0.4) -- (0.5,0.7);
        \Sdiodearrow{0}{0}{1}{-1}
        \node at (0.8,-0.3) {$k_4$};
        \draw[red, arrows={->[scale=1.2,red]}] (0.3,-0.5) -- (0.6,-0.8);
        \Sdiodearrow{0}{0}{-1}{-1}
        \node at (-0.7,-0.3) {$k_3$};
        \draw[red, arrows={->[scale=1.2,red]}] (-0.5,-0.7) -- (-0.2,-0.4);
        \node at (3,0) {$=   i \lambda  \frac{n_{-k_2,1}  }{ n_{k_{134},-1}  }  \ ,$};
    \end{tikzpicture}
\end{equation}
and
\begin{equation}
    \begin{tikzpicture}[scale=1.2]
        \node at (0,0) {$\bullet$};
        \Ssemicap{-1}{1}{0}{0}
        \node at (-0.7,0.4) {$k_1$};
        \draw[red, arrows={->[scale=1.2,red]}] (-0.5,0.7) -- (-0.2,0.4);
        \Sdiodearrow{0}{0}{1}{1}
        \node at (0.8,0.4) {$k_2$};
        \draw[red, arrows={->[scale=1.2,red]}] (0.2,0.4) -- (0.5,0.7);
        \Sdiodearrow{0}{0}{1}{-1}
        \node at (0.8,-0.3) {$k_4$};
        \draw[red, arrows={->[scale=1.2,red]}] (0.3,-0.5) -- (0.6,-0.8);
        \Ssemicap{-1}{-1}{0}{0}
        \node at (-0.7,-0.3) {$k_3$};
        \draw[red, arrows={->[scale=1.2,red]}] (-0.5,-0.7) -- (-0.2,-0.4);
        \node at (3,0) {$=   i \lambda  \frac{n_{-k_1,-1} n_{-k_3,-1} }{ n_{k_{24},2}  } \ ,$};
    \end{tikzpicture}
    \qquad \qquad
    \begin{tikzpicture}[scale=1.2]
        \node at (0,0) {$\bullet$};
        \Sdiodearrow{0}{0}{-1}{1}
        \node at (-0.7,0.4) {$k_1$};
        \draw[red, arrows={->[scale=1.2,red]}] (-0.5,0.7) -- (-0.2,0.4);
        \Ssemicap{1}{1}{0}{0}
        \node at (0.8,0.4) {$k_2$};
        \draw[red, arrows={->[scale=1.2,red]}] (0.2,0.4) -- (0.5,0.7);
        \Ssemicap{1}{-1}{0}{0}
        \node at (0.8,-0.3) {$k_4$};
        \draw[red, arrows={->[scale=1.2,red]}] (0.3,-0.5) -- (0.6,-0.8);
        \Sdiodearrow{0}{0}{-1}{-1}
        \node at (-0.7,-0.3) {$k_3$};
        \draw[red, arrows={->[scale=1.2,red]}] (-0.5,-0.7) -- (-0.2,-0.4);
        \node at (3,0) {$=  i \lambda \frac{n_{-k_2, 1} n_{-k_4, 1} }{ n_{k_{13}, -2} }  \ ,$};
    \end{tikzpicture}
\end{equation}
\begin{equation}
    \begin{tikzpicture}[scale=1.2]
        \node at (0,0) {$\bullet$};
        \Ssemicap{-1}{1}{0}{0}
        \node at (-0.7,0.4) {$k_1$};
        \draw[red, arrows={->[scale=1.2,red]}] (-0.5,0.7) -- (-0.2,0.4);
        \Ssemicap{1}{1}{0}{0}
        \node at (0.8,0.4) {$k_2$};
        \draw[red, arrows={->[scale=1.2,red]}] (0.2,0.4) -- (0.5,0.7);
        \Sdiodearrow{0}{0}{1}{-1}
        \node at (0.8,-0.3) {$k_4$};
        \draw[red, arrows={->[scale=1.2,red]}] (0.3,-0.5) -- (0.6,-0.8);
        \Sdiodearrow{0}{0}{-1}{-1}
        \node at (-0.7,-0.3) {$k_3$};
        \draw[red, arrows={->[scale=1.2,red]}] (-0.5,-0.7) -- (-0.2,-0.4);
        \node at (3,0) {$=   i \lambda  \frac{n_{-k_1,-1} n_{-k_2,1} }{ n_{k_{34},0}  } \ ,$};
    \end{tikzpicture}
    \qquad
\end{equation}
and 
\begin{equation}
    \begin{tikzpicture}[scale=1.2]
        \node at (0,0) {$\bullet$};
        \Ssemicap{-1}{1}{0}{0}
        \node at (-0.7,0.4) {$k_1$};
        \draw[red, arrows={->[scale=1.2,red]}] (-0.5,0.7) -- (-0.2,0.4);
        \Ssemicap{1}{1}{0}{0}
        \node at (0.8,0.4) {$k_2$};
        \draw[red, arrows={->[scale=1.2,red]}] (0.2,0.4) -- (0.5,0.7);
        \Ssemicap{1}{-1}{0}{0}
        \node at (0.8,-0.3) {$k_4$};
        \draw[red, arrows={->[scale=1.2,red]}] (0.3,-0.5) -- (0.6,-0.8);
        \Sdiodearrow{0}{0}{-1}{-1}
        \node at (-0.7,-0.3) {$k_3$};
        \draw[red, arrows={->[scale=1.2,red]}] (-0.5,-0.7) -- (-0.2,-0.4);
        \node at (3,0) {$=  i \lambda  \frac{n_{-k_1,-1} n_{-k_2,1} n_{-k_4,1} }{n_{k_{3},-1}}\ ,$};
    \end{tikzpicture}
    \qquad \qquad
    \begin{tikzpicture}[scale=1.2]
        \node at (0,0) {$\bullet$};
        \Ssemicap{-1}{1}{0}{0}
        \node at (-0.7,0.4) {$k_1$};
        \draw[red, arrows={->[scale=1.2,red]}] (-0.5,0.7) -- (-0.2,0.4);
        \Ssemicap{1}{1}{0}{0}
        \node at (0.8,0.4) {$k_2$};
        \draw[red, arrows={->[scale=1.2,red]}] (0.2,0.4) -- (0.5,0.7);
        \Sdiodearrow{0}{0}{1}{-1}
        \node at (0.8,-0.3) {$k_4$};
        \draw[red, arrows={->[scale=1.2,red]}] (0.3,-0.5) -- (0.6,-0.8);
        \Ssemicap{-1}{-1}{0}{0}
        \node at (-0.7,-0.3) {$k_3$};
        \draw[red, arrows={->[scale=1.2,red]}] (-0.5,-0.7) -- (-0.2,-0.4);
        \node at (3,0) {$=  i \lambda  \frac{n_{-k_1,-1} n_{-k_2,1} n_{-k_3,-1} }{n_{k_{4},1}}  \ ,$};
    \end{tikzpicture}
\end{equation}
where one can easily see the charge conservation at the bulk vertex. For each of the vertices above, we have chosen the convention that all momenta flow into the vertex. Only these vertices contribute to non-zero terms, while the rest vanish. More precisely, vertices with the same legs result in no contribution, as represented below:
\begin{equation}
    \begin{tikzpicture}[scale=1.2]
        \node at (0,0) {$\bullet$};
        \Sdiodearrow{0}{0}{-1}{1}
        \node at (-0.7,0.4) {$k_1$};
        \draw[red, arrows={->[scale=1.2,red]}] (-0.5,0.7) -- (-0.2,0.4);
        \Sdiodearrow{0}{0}{1}{1}
        \node at (0.8,0.4) {$k_2$};
        \draw[red, arrows={->[scale=1.2,red]}] (0.2,0.4) -- (0.5,0.7);
        \Sdiodearrow{0}{0}{1}{-1}
        \node at (0.8,-0.3) {$k_4$};
        \draw[red, arrows={->[scale=1.2,red]}] (0.3,-0.5) -- (0.6,-0.8);
        \Sdiodearrow{0}{0}{-1}{-1}
        \node at (-0.7,-0.3) {$k_3$};
        \draw[red, arrows={->[scale=1.2,red]}] (-0.5,-0.7) -- (-0.2,-0.4);
        \node at (3,0) {$\ = \quad 0 \quad = \ $};
    \end{tikzpicture}
    \qquad
    \begin{tikzpicture}[scale=1.2]
        \node at (0,0) {$\bullet$};
        \Ssemicap{-1}{1}{0}{0}
        \node at (-0.7,0.4) {$k_1$};
        \draw[red, arrows={->[scale=1.2,red]}] (-0.5,0.7) -- (-0.2,0.4);
        \Ssemicap{1}{1}{0}{0}
        \node at (0.8,0.4) {$k_2$};
        \draw[red, arrows={->[scale=1.2,red]}] (0.2,0.4) -- (0.5,0.7);
        \Ssemicap{1}{-1}{0}{0}
        \node at (0.8,-0.3) {$k_4$};
        \draw[red, arrows={->[scale=1.2,red]}] (0.3,-0.5) -- (0.6,-0.8);
        \Ssemicap{-1}{-1}{0}{0}
        \node at (-0.7,-0.3) {$k_3$};
        \draw[red, arrows={->[scale=1.2,red]}] (-0.5,-0.7) -- (-0.2,-0.4);
    \end{tikzpicture}
\end{equation}
The fact that vertices with only `bar' legs or `triangle' legs vanish follows from the collapse rule and the KMS condition, respectively. Notably, these vertices closely resemble those found in thermal field theory \cite{Gelis:1997zv, Carrington:2006xj, Gelis:2019yfm}. Here, one can interpret the legs with triangles as a single effective leg carrying the combined momentum and charge, which explains why some scattering processes seem to involve particles with zero or double the charge. This feature is quite general and holds for fermions as well, with the additional observation that two fermions with opposite quantum numbers behave like a boson \cite{Martin:2024mdm}.  

Along with the above Feynman rules for propagators and vertices, to obtain the \textit{boundary Schwinger-Keldysh generating functional} (on-shell action), we must further supply with following rules:
\begin{enumerate}
    \item  Multiply every diagram by $-i$.
    
    \item The vertices are integrated over the exterior of the black brane, with the following  radial exterior integral
\begin{equation}\label{eq:extint}
    \int_{\rm ext} = \int_{r_+}^{r_c}dr \, r^{d-1}  \ ,
\end{equation}
where $r_c$ is the radial cut-off required for holographic renormalisation \cite{Skenderis:2008dg}.
    \item  Impose momentum conservation at each vertex and integrate over all momenta.

    \item Divide by the symmetry factor of the diagram.
\end{enumerate}

\noindent
Using these Feynman rules for Witten diagrams outlined above, we can calculate the boundary SK generating functional to any desired order in $\lambda$. In this note, we only present the result up to quadratic order in $\lambda$, incorporating contributions from contact and single-exchange diagrams.

\subsection{Contact \& Exchange terms}

We begin with the contact diagrams contributing to the four-point influence phase $S_{(4)}$. The corresponding four-point contact term in the on-shell action is given by:
\begin{equation}
    S_{(4)} =  -\frac{\lambda}{2!\, 2!} \int_{k_{1,2,3,4} } \oint_{\z}  \   \Phib_{(0)}(\z,k_1)  \Phi_{(0)}(\z,k_2) \Phib_{(0)}(\z,k_3)  \Phi_{(0)}(\z,k_4)
\end{equation} 
where we have taken $S_{(4)}$ from the Eq.~(\ref{eq:S2S4S6}) and written in the momentum space.

Using the solutions of the Klein-Gordon equation and performing the RNSK radial contour integral, as explained above in this section. We obtain the on-shell action at linear order in coupling constant $S_{(4)}$, and read off coefficients $\mathcal{I}^{(4)}_{r,s}(k_1,k_2,k_3,k_4)$ as explained in the Appendix \ref{app:onshell}.\footnote{Recall that symbol $\mathcal{I}^{(4)}_{r,s}$ labels the coefficients of quartic influence phase, defined in Eq.~\eqref{eq:SosIdef}.} All the diagrams contributing to the $S_{(4)}$ are:

\begin{figure}[H]
    \centering
    \begin{subfigure}{0.4\textwidth}
    \centering
        \fourptPFcmplxphifour{1}{1}{1}{-1}
        \caption{$\mathcal{I}^{(4)}_{2,1}$}
    \end{subfigure} \qquad
    \begin{subfigure}{0.4\textwidth}
        \centering
        \fourptPFcmplxphifour{1}{1}{-1}{1}
        \caption{$\mathcal{I}^{(4)}_{1,2}$}
    \end{subfigure}\\ \vspace{0.5cm}
    \begin{subfigure}{0.4\textwidth}
        \centering
        \fourptPFcmplxphifour{1}{1}{-1}{-1}
        \caption{$\mathcal{I}^{(4)}_{1,1}$}
    \end{subfigure} \qquad
    \begin{subfigure}{0.4\textwidth}
        \centering
        \fourptPFcmplxphifour{1}{-1}{1}{-1}
        \caption{$\mathcal{I}^{(4)}_{2,0}$}
    \end{subfigure} \\ \vspace{0.5cm}
    \begin{subfigure}{0.4\textwidth}
        \centering
        \fourptPFcmplxphifour{-1}{1}{-1}{1}
        \caption{$\mathcal{I}^{(4)}_{0,2}$}
    \end{subfigure} \qquad
    \begin{subfigure}{0.4\textwidth}
        \centering
        \fourptPFcmplxphifour{1}{-1}{-1}{-1}
        \caption{$\mathcal{I}^{(4)}_{1,0}$}
    \end{subfigure}  \\ \vspace{0.5cm}
    \begin{subfigure}{0.4\textwidth}
        \centering
        \fourptPFcmplxphifour{-1}{1}{-1}{-1}
        \caption{$\mathcal{I}^{(4)}_{0,1}$}
    \end{subfigure}
    \caption{Witten diagrams contributing to the four-point influence phase $S_{(4)}$. Each sub-figure corresponds to a distinct interaction vertex and its associated term in the influence phase.}
    \label{fig:fourpointcontact}
\end{figure}
\noindent
Note that these are the only non-zero contributions to the quartic influence phase $S_{(4)}$ at linear order in the coupling $\lambda$. The terms corresponding to these diagrams (recall the definition in Eq.~\eqref{eq:SosIdef}), listed in the same order, are given as follows:

\begin{equation}\label{eq:I4}
    \begin{split}
        \mathcal{I}^{(4)}_{2,1} &= - \frac{\lambda}{2} \frac{1}{n_{k_4}} \int_{\rm ext} \Ginb(r,k_1) \Gin(r,k_2)   \Ginb(r,k_3) \Gout(r,k_4) \ , \\
        \mathcal{I}^{(4)}_{1,2} &= - \frac{\lambda}{2} \frac{1}{\nqb_{k_3}} \int_{\rm ext} \Ginb(r,k_1) \Gin(r,k_2) \Goutb(r,k_3)  \Gin(r,k_4) \ ,  \\
        \mathcal{I}^{(4)}_{1,1} &=  \lambda \frac{1}{n_{k_{34},0}} \int_{\rm ext}  \Ginb(r,k_1) \Gin(r,k_2)   \Goutb(r,k_3) \Gout(r,k_4) \ ,  \\
        \mathcal{I}^{(4)}_{2,0} &= \lambda \frac{1}{n_{k_{24},2}} \int_{\rm ext}  \Ginb(r,k_1)   \Gout(r,k_2)   \Ginb(r,k_3)  \Gout(r,k_4) \ , \\
        \mathcal{I}^{(4)}_{0,2} &=  \lambda  \frac{1}{n_{k_{13},-2}} \int_{\rm ext}  \Goutb(r,k_1) \Gin(r,k_2)   \Goutb(r,k_3) \Gin(r,k_4) \ , \\
        \mathcal{I}^{(4)}_{1,0} &= - \frac{\lambda}{2} \frac{1}{n_{k_{234}}} \int_{\rm ext}  \Ginb(r,k_1) \Gout(r,k_2)   \Goutb(r,k_3) \Gout(r,k_4)  \ , \\
        \mathcal{I}^{(4)}_{0,1} &= - \frac{\lambda}{2} \frac{1}{\nqb_{k_{134}}} \int_{\rm ext}  \Goutb(r,k_1) \Gin(r,k_2)    \Goutb(r,k_3) \Gout(r,k_4) \ , \\
    \end{split}
\end{equation}
where we have used $n_{k,1} = \nq_{k}$ and $n_{k,-1} = \nqb_{k}$ and suppressed the momentum dependence in $\mathcal{I}^{(4)}_{r,s}$ to shorten the expressions. Here, one could notice that the diagrams with either all the Future sources or all the Past sources vanish,
\begin{figure}[H]
\begin{center}
\begin{tikzpicture}

\begin{scope}[shift={(0,0)},scale=1.4]
\coordinate (z) at (0,-0.6);
    \draw[blue] (-1.75,1)--(1.75,1);
    \Sdiode{-1.5}{1}{0}{-0.6};
    \Sdiode{-0.5}{1}{0}{-0.6};
    \Sdiode{0.5}{1}{0}{-0.6};
    \Sdiode{1.5}{1}{0}{-0.6};
    \node at (z) {$\bullet$};
    \draw[red, arrows={->[scale=1.2,red]}] (-1.55,0.8) -- (-1.2,0.4);
    \draw[red, arrows={->[scale=1.2,red]}] (-0.50,0.4) -- (-0.6,0.8);
    \draw[red, arrows={->[scale=1.2,red]}] (0.55,0.8) -- (0.4,0.4);
    \draw[red, arrows={->[scale=1.2,red]}] (1.2,0.4) -- (1.55,0.8);

\end{scope}

\begin{scope}[shift={(4,0)},scale=1.2]

\node at (0,0) {\large $= \ 0= \ $};

\end{scope}

\begin{scope}[shift={(8,0)},scale=1.4]

\coordinate (z) at (0,-0.6);
    \draw[blue] (-1.75,1)--(1.75,1);
    \Sdiode{0}{-0.6}{-1.5}{1};
    \Sdiode{0}{-0.6}{-0.5}{1};
    \Sdiode{0}{-0.6}{0.5}{1};
    \Sdiode{0}{-0.6}{1.5}{1};
    \node at (z) {$\bullet$};
    \draw[red, arrows={->[scale=1.2,red]}] (-1.55,0.8) -- (-1.2,0.4);
    \draw[red, arrows={->[scale=1.2,red]}] (-0.50,0.4) -- (-0.6,0.8);
    \draw[red, arrows={->[scale=1.2,red]}] (0.55,0.8) -- (0.4,0.4);
    \draw[red, arrows={->[scale=1.2,red]}] (1.2,0.4) -- (1.55,0.8);

\node at (2,0) {\large $ \ . $};
\end{scope}

\end{tikzpicture}
\end{center}
\label{fig:collpseKMS}
    \caption{Two Witten diagrams that yield no contribution to the four-point influence phase $S_{(4)}$, due to the SK collapse and the KMS condition, respectively.}
\end{figure}

\noindent
This is nothing but the SK collapse condition and the KMS condition at four-point functions. There are two equivalent ways to see why these diagrams vanish:
\begin{itemize}
    \item \textbf{On the RNSK contour:} The underlying reason why the SK collapse rules and the KMS relations hold lies in the analyticity properties of the ingoing propagator. Specifically, the integrand is analytic, and its integral over the RNSK radial contour vanishes. To illustrate this, consider an analytic function $F(r)$ integrated over the radial RNSK contour:
    \begin{equation}
        \oint d r \  F(r) \ ,
    \end{equation}
\noindent
where $r$ runs over the full RNSK radial contour.

    Breaking the full contour integral into its different parts, we see
\begin{equation}
\begin{split}
    \oint d r \  F(r) 
    &= \lim_{\epsilon \to 0}\left[ \int_{\infty+ i \epsilon}^{r_+ + i \epsilon} F(r) +   \cancel{\int^{r_+ - i \epsilon}_{r_+ + i \epsilon} F(r)}+   \int^{\infty- i \epsilon}_{r_+ - i \epsilon} F(r) \right] \\
    &=  \lim_{\epsilon \to 0}\left[ \int_{\infty+ i \epsilon}^{r_+ + i \epsilon} F(r)-   \int_{\infty- i \epsilon}^{r_+ - i \epsilon} F(r) \right] \\
    &=  \lim_{\epsilon \to 0}\left[ \int_{\infty+ i \epsilon}^{r_+ + i \epsilon} F(r)-   \int_{\infty+ i \epsilon}^{r_+ + i \epsilon} F(r) \right] =0 \ ,
\end{split}    
\end{equation}
where we have used the analyticity of $F(r)$ in the above steps.    

    \item \textbf{From exterior field theory:} The reason for the SK collapse and KMS condition lies behind the disappearance of corresponding effective vertices. These vertices vanish because they contain the inverse of the Bose-Einstein factor evaluated at $k^0=0$, i.e. $\frac{1}{n_{k,0}}\Big|_{k^0=0} = 0$.
\end{itemize}
\noindent
Next, we come to the exchange diagrams contributing to the six-point influence phase $S_{(6)}$. Here, we will restrict ourselves to single exchange diagrams only that correspond to a single bulk-to-bulk propagator in an algebraic expression. The corresponding term in the on-shell action is given by:
\begin{equation}
    \begin{split}
        S_{(6)} &=  -\frac{\lambda^2}{4} \int_{ k_{1,2,3,4,5,6}} \oint_{\z,\z'} \Phib_{(0)}(\z,k_1)\Phi_{(0)}(\z,k_2) \Phib_{(0)}(\z,k_3) \\
        &\hspace{1.3cm}\times  \bbG(\z|\z',k_{456} ) \Phi_{(0)}(\z',k_4) \Phib_{(0)}(\z',k_5) \Phi_{(0)}(\z',k_6) \ .
    \end{split}
\end{equation}
\noindent
Again, performing the RNSK radial integral to obtain the exterior radial integral using Eqs.~\eqref{RNSKtoExt1} and \eqref{RNSKtoExt2} and the details are given in the Appendix \ref{app:onshell} (for instance, see Eqs. after \eqref{eq:appDS6one}). But here, we directly draw the diagrams and then write the corresponding expressions using the Feynman rules given above. For simplicity, we only illustrate two diagrams that contribute to $S_{(6)}$, while noting that other diagrams can be similarly derived using the same Feynman rules. Below are these diagrams:
\begin{figure}[H]
    \begin{subfigure}{0.33\textwidth}
        \sixptPFcmplxphifour{1}{1}{1}{1}{1}{1}{-1}  
        \caption{$\mathcal{I}^{(6)}_{3,2}$} 
    \end{subfigure} \hspace{3cm}
    \begin{subfigure}{0.33\textwidth}
        \sixptPFcmplxphifour{1}{1}{1}{1}{1}{-1}{1}  
        \caption{$\mathcal{I}^{(6)}_{2,3}$}
    \end{subfigure}
    \caption{Two particular Witten diagrams that contribute to six-point influence phase $S_{(6)}$.}
    \label{fig:sixpointexchange}
\end{figure}
\noindent
and the terms corresponding to these diagrams are:
\begin{equation}\label{eq:I6}
    \begin{split}
         \mathcal{I}^{(6)}_{3,2} &=  \frac{\lambda^2}{2} \frac{1}{n_{k_6}}\int_{\rm ext} \int_{\rm ext} \bbGR(r|r',k_{123}) \\
        & \hspace{2cm}\times\left[ \Ginb(r,k_1) \Gin(r,k_2) \Ginb(r,k_3) \Gin(r',k_4) \Ginb(r',k_5) \Gout(r',k_6)  \right] \ ,\\
         \mathcal{I}^{(6)}_{2,3} &= \frac{\lambda^2}{2} \frac{1}{\nqb_{k_5}}\int_{\rm ext} \int_{\rm ext}\bbGR(r|r',k_{123}) \\
         &  \hspace{2cm}\times \left[ \Ginb(r,k_1) \Gin(r,k_2) \Ginb(r,k_3) \Gin(r',k_4) \Goutb(r',k_5) \Gin(r',k_6)  \right] \ ,
    \end{split}
\end{equation}
where the two `ext' integrals denote integration over the two bulk points in the exterior region.

Similarly, higher-point exchange diagrams contributing to the higher-point influence phase can be built using the Feynman rules described above. This concludes our analysis of exterior diagrammatics, and we now shift focus to constructing solutions explicitly within the gradient expansion.

\section{The Gradient Expansion}\label{sec:gradexp}

In this section, we solve the free complex Klein-Gordon equation by expressing the field in the gradient expansion. To derive the \textit{fluctuation-dissipation relations} (FDRs), we only need to assume that the field $\Phi$ is independent of the boundary space coordinates $\mathbf{x}$. This is equivalent to setting its Fourier counterpart $\mathbf{k}=0$ in momentum space. Under this assumption, the free complex Klein-Gordon equation (\ref{eq:freeEOM}) reduces to the explicit form given below:
\begin{equation}\label{eq:expKGeq}
    \begin{split}
        \frac{d}{d \z}\left[ r^{d-1} \frac{d \Phi}{d \z}\right] &+\frac{\b k^0}{2} \left[ r^{d-1}\frac{d \Phi}{d \z}+\frac{d}{d \z} \left(  r^{d-1}\Phi \right)\right] \\
        &-\frac{\b \mu_{q}}{2} \left[ r \, r_{+}^{d-2} \frac{d \Phi}{d \z}+\frac{d}{d \z} \left(  r \,  r_{+}^{d-2} \Phi \right)\right]= 0 \ ,
    \end{split} 
\end{equation}
where we have used the explicit expression for the gauge field $\mathcal{A}_v$ with $r_+$ as outer horizon size, given in Eq.~(\ref{eq:gA}).

\medskip

\textbf{Note:} We have omitted the subscript $(0)$, which indicates that we are working at zeroth order in the coupling $\lambda$, as this will be clear from the context. We will continue this convention throughout the remainder of the section.

We begin by writing the most general solution (at $\mathbf{k} = 0$) as a gradient expansion in frequency $k^0$. Instead of expanding around zero frequency (i.e, $k^0 \sim 0$), we find it technically simpler to solve the field equations by expanding around the chemical potential, i.e., for frequencies $k^0 \sim \muq$, up to second order, as
\begin{equation}\label{eq:Expfreqnearmu}
    \Phi = \sum_{i=0}^{2}\left(\frac{\b (k^0-\mu_{q}) }{2}\right)^i  \phi_{i}(\z) \ . 
\end{equation}
This type of expansion is natural when studying quasiparticle dynamics near the Fermi surface, where relevant excitations occur close to the chemical potential. Although our system does not contain fermions, we adopt this expansion purely for technical convenience.

Later, we will consider the small chemical potential limit ($\muq \to 0$). In that limit, the expansion around chemical potential (i.e, $k^0 \sim \muq$) effectively becomes an expansion around $k^0 \sim 0$. However, it is important to emphasize that these are distinct regimes: expanding in small $(k^0-\muq)$ is not the same as expanding around $k^0 = 0$ at fixed $\muq$. In particular, for arbitrary or large values of $\muq$, the behaviour of the system near $k^0 = \muq$ might differ from that near $k^0 = 0$. A more detailed analysis in the regime of arbitrary chemical potential is left for future work.

Upon using this expansion \eqref{eq:Expfreqnearmu} in the field equations (\ref{eq:expKGeq}), we can obtain the field equations at different orders. Subsequently, the field equations at zeroth order in the $\b (k^0- \muq)$ expansion, is
\begin{equation}
    \begin{split}
         & \frac{d}{d \z}\left[ r^{d-1} \frac{d \phi_{0} }{d \z}+ \b \mu_{q} \left(r^{d-1}-r \, r_{+}^{d-2}  \right) \phi_{0} \right]  = \frac{\b \mu_{q}}{2} \left(r^{d-1}-r \, r_{+}^{d-2}  \right)' \phi_{0} \ , 
    \end{split} 
\end{equation}
and at the first order in the $\b (k^0- \muq)$ expansion, is
\begin{equation}
    \begin{split}
          \frac{d}{d \z}\Bigg[ r^{d-1} \frac{d \phi_{1} }{d \z}+ \b \mu_{q} \left(r^{d-1}-r  \, r_{+}^{d-2}  \right) \phi_{1} &+2 \, r^{d-1} \phi_{0}  \Bigg]  \\
          &= \left[r^{d-1}\right]' \phi_{0} + \frac{\b \mu_{q}}{2} \left[r^{d-1}-r \, r_{+}^{d-2}  \right]' \phi_{1} \ ,
    \end{split} 
\end{equation}
and at the second order in the $\b (k^0- \muq)$ expansion, is
\begin{equation}
    \begin{split}
          \frac{d}{d \z}\Bigg[ r^{d-1} \frac{d \phi_{2} }{d \z}+ \b \mu_{q} \left(r^{d-1}-r  \, r_{+}^{d-2}  \right) \phi_{2} &+2 \, r^{d-1} \phi_{1}  \Bigg]  \\
          &= \left[r^{d-1}\right]' \phi_{1} + \frac{\b \mu_{q}}{2} \left[r^{d-1}-r \, r_{+}^{d-2}  \right]' \phi_{2} \ ,
    \end{split} 
\end{equation}
where $'$ denotes the derivative w.r.t $\z$ variable in all above-mentioned equations.

Next, to simplify further, we perform an expansion in small chemical potential (or equivalently, small charge) by taking $\muq \sim 0$, up to first order, as:
\begin{equation}\label{eq:Expsmallcharge}
\phi_{i} = \sum_{j=0}^{1}\left(\b \mu_{q} \right)^j \phi_{i,j}(\z) \ .
\end{equation}
Taken together, Eqs.~\eqref{eq:Expfreqnearmu} and \eqref{eq:Expsmallcharge} define what we will refer to as the \emph{near-$\muq$ gradient expansion in the small charge regime}. Note that these are nested expansions: we first expand in frequency near $\muq$, and then expand each coefficient $\phi_i$ in powers of $\muq$.

As discussed in the previous sections (e.g., see section~$\S$\ref{sec:freescalarRNSK}), it suffices to evaluate the ingoing solution, since the most general solution can be constructed using the \emph{CPT} isometry. To determine the ingoing boundary-to-bulk Green's function $\Gin$, we will solve the field equations (\ref{eq:expKGeq}) under the following ingoing boundary conditions:
\begin{equation}
    \Gin \big|_{r \to \infty} =1 \ , \qquad \qquad \frac{d \Gin}{d \z}\bigg|_{r_+} = 0 \ .
\end{equation}
Under the restriction that the ingoing Green's function $\Gin$ is independent of the boundary space coordinates\footnote{Since our goal is to derive FDRs, the boundary spatial directions $\mathbf{x}$ do not play a role.} $\mathbf{x}$, the $\Gin$ can be solved in the expansion near the chemical potential along with the small charge expansion. Here, we will only present the answer for $\Gin$ without going into the details of obtaining it.

Thus, the ingoing boundary-to-bulk Green's function in the near-$\muq$ gradient expansion in the small charge regime is,
\begin{equation}\label{eq:Ginexp}
    \Gin = \sum_{i=0}^{2}\sum_{j=0}^{1} \Gin_{i,j}(\z)  \left(\frac{\b (k^0-\mu_{q}) }{2}\right)^i \left(\b \mu_{q} \right)^j  \ , 
\end{equation}
with the coefficients $\Gin_{i,j}$ explicitly given below
\begin{equation}\label{eq:Ginpart1}
    \begin{split}
     &\Gin_{0,0}=1 \ ,   \hspace{2.5cm} \Gin_{0,1} =  \frac{1}{2} \mathbb{F}_{2}(r) \  , \hspace{2.5cm} \Gin_{1,0} =  \mathbb{F}_{1}(r)  \ , \\
     &\Gin_{2,0} = \int_{\z_c}^{\z} \frac{d \bar{\z}}{\bar{r}^{d-1}}\int_{\z_h}^{\bar{\z}}d \hat{\z} \left[\hat{r}^{d-1}\right]' \mathbb{F}_{1}(\hat{r})-2 \, \int_{\z_c}^{\z}d \bar{\z} \   \left( \mathbb{F}_{1}(r)-\frac{r_{+}^{d-1} }{r^{d-1} }\mathbb{F}_{1}(r_+) \right)  \ ,
    \end{split}
\end{equation}
where we have defined
\begin{equation}\label{eq:Fndef}
    \mathbb{F}_{n}(r) \equiv  \int_{\z_c}^{\z} d \bar{\z} \left[ \left( \frac{r_{+} }{\bar{r}} \right)^{d-n}-1 \right] \ , \qquad  \quad \z_h \equiv \z(r_+)  \ ,
     \qquad     \z_c \equiv \z(r_c) \  .
\end{equation}
The $\mathbb{F}_{n}$ is analytic on the radial RNSK contour as long as $n<d$. The other remaining terms are written below:
\begin{equation}\label{eq:Ginpart2}
\small
    \begin{split}
        &    \Gin_{1,1} + \int_{\z_c}^{\z} d\bar{\z} \left(1-\frac{\bar{r}  \, r_{+}^{d-2}}{\bar{r}^{d-1}} \right)  \mathbb{F}_{1}(\bar{r}) +   \int_{\z_c}^{\z} d\bar{\z}  \left( \mathbb{F}_{2}(r)-\frac{r_{+}^{d-1} }{r^{d-1} }\mathbb{F}_{2}(r_+) \right)  = \\
        & \hspace{1cm} + \frac{1}{2} \int_{\z_c}^{\z} \frac{ d\bar{\z}}{\bar{r}^{d-1}} \int_{\z_h}^{\bar{\z}} d \hat{\z} \left[\hat{r}^{d-1}\right]'  \mathbb{F}_{2}(\hat{r}) + \frac{1}{2}  \int_{\z_c}^{\z}  \frac{ d\bar{\z}}{\bar{r}^{d-1}}  \int_{\z_h}^{\bar{\z} } d\hat{\z}\left[\hat{r}^{d-1}- \hat{r} \, r_{+}^{d-2} \right]'  \mathbb{F}_{1}(\hat{r})  \  ,
    \end{split}
\end{equation}
and
\begin{equation}\label{eq:Ginpart3}
\small
    \begin{split}
        &    \Gin_{2,1}+ \int_{\z_c}^{\z} d\bar{\z}\left[1- \left( \frac{r_{+}}{\bar{r}}\right)^{d-2} \right]  \Gin_{2,0}(\bar{r}) + 2 \int_{\z_c}^{\z} d\bar{\z}  \left( \Gin_{1,1}(\bar{r})-\frac{r_{+}^{d-1} }{\bar{r}^{d-1} }\Gin_{1,1}(r_+) \right)    \\
        &\hspace{0.2cm} = \int_{\z_c}^{\z} d\bar{\z}\frac{1}{\bar{r}^{d-1}} \int_{\z_h}^{\bar{\z}} d\hat{\z} \left[\hat{r}^{d-1}\right]'  \Gin_{1,1}(\hat{r}) +\frac{1}{2} \int_{\z_h}^{\bar{\z}} d\bar{\z} \frac{1}{\bar{r}^{d-1}} \int_{\bar{\z}_h}^{\z} d\hat{\z}\left[\hat{r}^{d-1}- \hat{r} \, r_{+}^{d-2} \right]'  \Gin_{2,0}(\hat{r})  \  .
    \end{split}
\end{equation}
It can be readily checked that the ingoing Green's function given in Eq.~(\ref{eq:Ginexp}), along with the coefficients mentioned above, solves the Klein-Gordon equation with ingoing boundary conditions. Similarly, the ingoing boundary-to-bulk conjugate Green's function can be obtained by replacing $q \to -q$. 

We will now express the field in the near--$\muq$ gradient expansion in the small charge regime. The approach involves first constructing the ingoing solution and then applying the \emph{CPT} isometry \eqref{eq:GintoGout} to obtain the full solution. We find it helpful to represent the solution (\ref{Phi0main}) in the following form,
\begin{equation}
    \Phi = \Gin_{\rm q}(k) \left[ \Ja(k) + \left( n_{k}+\frac{1}{2} \right) \Jd(k) \right] -e^{\b \mu_{q} [\mathbb{\Theta}-1]} e^{-\b k^0 (\z-1)} \Gin_{\rm -q}(-k)  n_{k} \  \Jd(k) \ , 
\end{equation}
where we suppressed the radial dependence from the ingoing Green's function and also defined the sources in \textit{average-difference} basis as,
\begin{equation}
    \Ja \equiv \frac{1}{2} \left( J_{\sR} + J_{\sL} \right)  \ , \hspace{2cm} \Jd \equiv  J_{\sR} - J_{\sL} \ .
\end{equation}

In what follows, we retain only the leading-order terms in the nested expansion \eqref{eq:Ginexp}, keeping contributions up to linear order in both $k^0$ and $\muq$, while discarding higher-order and mixed terms such as $\mathcal{O}((k^0)^2)$, $\mathcal{O}(\muq^2)$, and $\mathcal{O}(k^0 \muq)$. This truncation isolates the dominant behaviour in the small frequency and small charge regime, and is sufficient for our purposes. The resulting solution, keeping only the terms specified above, is given by: 
\begin{equation}\label{Phigrdexpmom}
\small
\begin{split}
    \Phi &= \left[\Ja +\left( \z-\frac{1}{2}+ \mathbb{F}_{1} \right) \Jd \right]+ \frac{\b k^0}{2} \left[ \mathbb{F}_{1}  \Ja -\left( \z (\z-1)+ \left[ \z -\frac{1}{2} \right] \mathbb{F}_{1} \right)\Jd \right]  \\
    &+\frac{\b \mu_{q}}{2}\left[  \left( \mathbb{F}_{2} -\mathbb{F}_{1} \right) \Ja +\left( \z (\z-1)+ \left[ \z -\frac{1}{2} \right] \mathbb{F}_{2} +  \left( \z-\frac{1}{2}+ \mathbb{F}_{2} \right) \mathbb{F}_{1} \right)\Jd \right]  \ .
\end{split}
\end{equation}
To conclude the discussion on gradient expansion, we provide the solution in position space as well. This form will be useful when we evaluate the non-linear FDRs in the next section.

In position-space coordinates, the solution can be written as
\begin{equation}\label{Phigrdexppos}
\small
\begin{split}
    \Phi &= \left[\Ja + \left( \z-\frac{1}{2}+ \mathbb{F}_{1} \right) \Jd \right]+ \frac{ i \b}{2} \left[ \mathbb{F}_{1}  \dt \Ja -\left( \z (\z-1)+ \left[ \z -\frac{1}{2} \right] \mathbb{F}_{1} \right) \dt \Jd \right]  \\
    &+\frac{\b \mu_{q}}{2}\left[  \left( \mathbb{F}_{2} -\mathbb{F}_{1} \right) \Ja +\left( \z (\z-1)+ \left[ \z -\frac{1}{2} \right] \mathbb{F}_{2} +  \left( \z-\frac{1}{2}+ \mathbb{F}_{2} \right) \mathbb{F}_{1} \right)\Jd \right] \ ,
\end{split}
\end{equation}
where we have replaced $k^0 \mapsto i \partial_{t}$ to obtain the above result.

\subsection{On-shell action in gradient expansion}

Using the solutions given in Eqs.~(\ref{Phi0PF}) and (\ref{Phib0PF}), we proceed to compute the on-shell action (or influence phase), starting with the quadratic (free) contribution $S_{(2)}$, given as
\begin{equation}
    S_{(2)} = -\int d^d x  \sqrt{-g}   \  \Phi_{(0)}  \left(\overline{ \mathbb{D} \Phi_{(0)} } \right) \ \bigg|^{\z=1}_{\z=0} \ .
\end{equation}
Upon transitioning to momentum space, the on-shell action at quadratic order in the \textit{average-difference} basis is found to be:
\begin{equation}\label{eq:S2osaction}
    S_{(2)} = \int \frac{d^d k}{(2 \pi)^d }  \left[ \widehat{\mathcal{J}}_{ad} \  
     \Jbd(-k) \Ja (k)  + \widehat{\mathcal{J}}_{da} \ 
    \Jba(-k) \Jd(k)  + \widehat{\mathcal{J}}_{dd} \  \Jbd(-k) \Jd(k) \right] \ ,
\end{equation}
with $\widehat{\mathcal{J}}_{ad}$, $\widehat{\mathcal{J}}_{da}$ and $\widehat{\mathcal{J}}_{dd}$ are the boundary Green's functions given below,\footnote{Here, the hat notation $\widehat{\mathcal{J}}$ denotes that $\mathcal{J}$ has been integrated over the radial direction and thus no longer depends on the radial coordinate, i.e. $\widehat{\mathcal{J}}\equiv \oint d\z \sqrt{-g} \  \mathcal{J}$.  }
\begin{equation}\label{eq:Kretdef}
    \begin{split}
        \widehat{\mathcal{J}}_{ad} =  K_{\rm ret}(k) \ , \qquad 
        \widehat{\mathcal{J}}_{da}  = K_{\rm adv}(k) \ ,\qquad \widehat{\mathcal{J}}_{dd} =K_{\rm kel} \ ,
    \end{split}
\end{equation}
where $ K_{\rm ret}$, $ K_{\rm adv}$ and $ K_{\rm kel}$ denote the two-point retarded, advanced and Keldysh boundary correlators respectively. The retarded and advanced components only capture spectral information and do not depend on the occupation number $n_{k}$, as opposed to the Keldysh component that explicitly involves 
$n_{k}$. The absence of $\widehat{\mathcal{J}}_{aa}$ follows from the Schwinger-Keldysh collapse condition. It is worth noting that the on-shell action expressed in the average-difference basis naturally produces boundary real-time correlators. The final term, $\widehat{\mathcal{J}}_{dd}$, representing the effects of fluctuations, as expected from the two-point KMS condition, is given by,
\begin{equation}\label{eq:2KMS}
        \widehat{\mathcal{J}}_{dd} = \frac{1}{2} \coth \left(\frac{\b(k^0-\muq)}{2}\right)\left[ \widehat{\mathcal{J}}_{ad} - \widehat{\mathcal{J}}_{da} \right]  \ . 
\end{equation}

With the quadratic order on-shell action established and its coefficients identified, we now focus our attention on the quartic on-shell action $S_{(4)}$. It receives contributions from the contact diagrams illustrated in figure~\ref{fig:fourpointcontact}. Using the contact term from the second line of Eq.~(\ref{eq:S2S4S6}) and considering terms up to linear order in both $\dt$ and $\muq$ while discarding higher-order and mixed terms such as $\mathcal{O}(\dt^2)$, $\mathcal{O}(\muq^2)$, and $\mathcal{O}( \muq \dt)$, we obtain the following expression,
\begin{equation}\label{eq:S4linear}
\begin{split}
    S_{(4)} &= -\lambda\int d^d x\oint d\z \sqrt{-g}\frac{|\Phi_{(0)}|^4}{4} \\
    &= -\lambda \int d^d x \int_{\rm ext}  \mathcal{L}_{0} -\lambda \int d^d x \int_{\rm ext} \mathcal{L}_{1}+ \mathcal{O}((\beta \dt)^2)+ \mathcal{O}((\beta \muq)^2)+ \mathcal{O}( \beta^2 \muq \dt) \ ,
\end{split}    
\end{equation}
where the terms $\mathcal{L}_{0}$ and $\mathcal{L}_{1}$ represent the non-derivative and first-derivative contributions to $S_{(4)}$, respectively. Note that the quartic on-shell action has been expressed in position space using Eq.~(\ref{Phigrdexppos}).

The explicit form of the non-derivative contribution  $\mathcal{L}_{0}$ is given by,
\begin{equation}
    \begin{split}
        \mathcal{L}_{0} &=  \sum_{r,s=0}^{2} \mathcal{G}_{r,s} \  \left[\Jba\right]^{r} \left[\Ja\right]^{s} 
        \left[\Jbd\right]^{2-r}  \left[\Jd\right]^{2-s}     \ ,
    \end{split}
\end{equation}
and we express the first-derivative contribution to the quartic on-shell action $\mathcal{L}_{1}$ as,
\begin{equation}
    \begin{split}
        \mathcal{L}_{1} &=  \sum_{r=0}^{1} \sum_{s=0}^{2}  \Big\{ \mathcal{G}_{\dot{r},s}  \left(\dt \Jba \right) 
        + \mathcal{H}_{\dot{r},s}  \left(\dt \Jbd \right)  \Big\} \left[\Jba\right]^{r} \left[\Ja\right]^{s} 
        \left[\Jbd\right]^{1-r}  \left[\Jd\right]^{2-s}     \\
        &+  \sum_{r=0}^{2} \sum_{s=0}^{1}            \Big\{  \mathcal{G}_{r,\dot{s}}  \left(\dt \Ja \right)  + \mathcal{H}_{r,\dot{s}}  \left(\dt \Jd \right)  \Big\}  \left[\Jba\right]^{r} \left[\Ja\right]^{s} \left[\Jbd\right]^{2-r}  \left[\Jd\right]^{1-s}  
    \end{split} \ ,
\end{equation}
where the coefficients for all terms are provided in Appendix \ref{app:explicitL0L1}, and relationships among them due to charge conjugation as outlined below:
\begin{equation}\label{eq:relationship}
    \mG_{r,s} = \mG_{s,r}\Big|_{q \to -q}  \ , \qquad \mG_{r,\dot{s}} = \mG_{\dot{s},r} \ ,  \qquad \mathcal{H}_{r,\dot{s}} = \mathcal{H}_{\dot{s},r} \ .
\end{equation}
Note that since we are working only up to linear order in both $\dt$ and $\muq$ while discarding higher-order and mixed terms such as $\mathcal{O}((\dt)^2)$, $\mathcal{O}(\muq^2)$, and $\mathcal{O}( \muq \dt)$, the term $\mathcal{L}_1$ cannot contain any $\mu$-dependent contributions. However, if we go beyond this order, $\mathcal{L}_1$ would acquire $\mu$-dependent terms, and the relationship given in Eq.~(\ref{eq:relationship}) would be modified accordingly.

We have outlined the on-shell action (influence phase) to the leading order in the gradient expansion, highlighting the identification of several coefficients. The subsequent step is to examine how these coefficients of the quartic influence phase are related, in a manner analogous to the quadratic case discussed in Eq.~(\ref{eq:2KMS}), albeit in the high-temperature limit. The next section will clarify how these relationships pave the way for the emergence of holographic non-linear FDRs at finite density under small charge expansion.

\section{Holographic FDTs at finite density  }\label{sec:holFDR}

In the previous sections, we examined interactions in the RNSK geometry and how it gives rise to exterior field theory. In this section, we now use those insights to derive fluctuation-dissipation theorems for holographic systems at finite density. Our central question is:

\begin{center}

\emph{Do FDTs exist in holographic theories at finite temperature and finite density?}

\end{center}

To address this question, we begin with a brief review of the holographic FDTs in a zero-density holographic system, following the work \cite{Jana:2020vyx}. 

\subsection{Review of Holographic FDTs at zero-density}\label{sec:realatmuzeror}

Let us begin by considering a massless self-interacting real scalar field $\phi$ in the zero-density case (i.e., in the grSK geometry instead of RNSK geometry), with the action given by:
\begin{equation}
    S = - \oint_{\rm grSK} \left( \frac{1}{2}(\partial_{M} \phi)^2 + \frac{g}{4!} \phi^{4} \right) \ ,
\end{equation}
where $g$ is the coupling constant. Note that while \cite{Jana:2020vyx} considers general $\phi^n$ interactions, we restrict ourselves to the quartic case for simplicity.\footnote{This simplification is for convenience; the analysis can be extended to general $\phi^n$ interactions.} In what follows, we adopt the same notation and conventions as used for the complex scalar field system introduced in sections~$\S$\ref{sec:phiinRNSK} and $\S$\ref{sec:extdiagram}, and will not elaborate on them unless ambiguity arises.

We now present the solution for this system. The zeroth-order (in coupling $g$) solution for the real scalar field $\phi$, to linear order in the gradient expansion (with $\textbf{k}=0$), is:
\begin{equation}
\begin{split}
    \phi_{(0)} &= \left[\Ja + \left( \z-\frac{1}{2}+ \mathbb{F}_{1} \right) \Jd \right]+ \frac{ i \b}{2} \left[ \mathbb{F}_{1}  \dt \Ja -\left( \z (\z-1)+ \left[ \z -\frac{1}{2} \right] \mathbb{F}_{1} \right) \dt \Jd \right]  \ ,
\end{split}
\end{equation}
where we recall the $\mathbb{F}_1$ is defined in Eq.~\eqref{eq:Fndef}, and the corresponding expression of quartic on-shell action is,
\begin{equation}
    S^{\phi}_{(4)} = -\frac{g}{4!}\oint_{\rm grSK} d\z d^dx \sqrt{-g} \ \phi_{(0)}^4  \ ,
\end{equation}
where the superscript indicates that we are working with the real scalar field $\phi$. Upon evaluating the radial contour integrals in the grSK geometry as discussed in \cite{Jana:2020vyx} (See similar discussion in Appendix \ref{app:Disc}), the quartic on-shell action becomes up to linear order in derivatives,
\begin{equation}\label{eq:Sosrealphi4raw}
\begin{split}
    S^{\phi}_{(4)} &= -g\int d^d x \Bigg\{ \sum_{k=1}^{4} \Theta_{k} 
    \frac{\Ja^{4-k}}{(4-k)!} \frac{(i\Jd)^k}{k!} +\sum_{k=1}^{3}\bar{\Theta}^{(a)}_{k} \ (\dt \Ja )\frac{\Ja^{3-k}}{(3-k)!} \frac{(i\Jd)^{k}}{k!} \\
    &\hspace{3cm} + \sum_{k=1}^{3}\bar{\Theta}^{(d)}_{k} \frac{\Ja^{4-k}}{(4-k)!} \frac{(i\Jd)^{k-1}}{(k-1)!}\dt (i\Jd)\Bigg\} +\mathcal{O}\left( (\beta \dt)^2\right) \ ,
\end{split}    
\end{equation}
where the zeroth-order coefficients in the gradient-expansion are:
\begin{equation}\label{eq:Thetadef}
    \Theta_{k} = \frac{1}{i^{k}}\oint_{\rm grSK} d\z \ \sqrt{-g} \left(\z-\frac{1}{2}+\mathbb{F}_1\right)^{k} \ ,
\end{equation}
where we recall that $\mathbb{F}_1$ is defined in Eq.~\eqref{eq:Fndef}, keeping in mind the difference that we are now working with grSK geometry instead of RNSK geometry.

The first-order coefficients in the gradient expansion are given by:
\begin{equation}\label{eq:Thetabaddef}
\begin{split}
    \bar{\Theta}^{(a)}_{k} &= \frac{1}{i^{k-1}}\frac{\beta}{2} \oint_{\rm grSK} d\z \ \sqrt{-g} \ \mathbb{F}_1  \left(\z-\frac{1}{2}+\mathbb{F}_1\right)^{k} \ , \\
    \bar{\Theta}^{(d)}_{k} &= -\frac{1}{i^{k-1}}\frac{\beta}{2} \oint_{\rm grSK} d\z \ \sqrt{-g} \ \left[ \z(\z-1)+\left(\z-\frac{1}{2}\right) \mathbb{F}_1 \right]  \left(\z-\frac{1}{2}+\mathbb{F}_1\right)^{k-1} \ , 
\end{split}    
\end{equation}
where the superscripts $(a)$ and $(d)$ indicate that the time derivative $\partial_t$ acts on the average source $\Ja$ and the difference source $\Jd$, respectively.

By performing integration by parts in Eq.~\eqref{eq:Sosrealphi4raw} and ignoring total derivative terms, we can shift the derivatives from $\Ja$ onto $\Jd$, yielding
\begin{equation}
    S^{\phi}_{(4)} = -g\int d^d x \Bigg\{ \sum_{k=1}^{4} \Theta_{k} 
    \frac{\Ja^{4-k}}{(4-k)!} \frac{(i\Jd)^k}{k!}  + \sum_{k=1}^{3} \left( \bar{\Theta}^{(d)}_{k}- \bar{\Theta}^{(a)}_{k}\right) \frac{\Ja^{4-k}}{(4-k)!} \frac{(i\Jd)^{k-1}}{(k-1)!}\dt (i\Jd) \Bigg\} \ .
\end{equation}
Now, we define $\bar{\Theta}_{k} \equiv \left( \bar{\Theta}^{(a)}_{k}- \bar{\Theta}^{(d)}_{k}\right)$ to simplify the above expression,
\begin{equation}
    S^{\phi}_{(4)} = -g\int d^d x \Bigg\{ \sum_{k=1}^{4} \Theta_{k} 
    \frac{\Ja^{4-k}}{(4-k)!} \frac{(i\Jd)^k}{k!} -\sum_{k=1}^{3}\bar{\Theta}_{k} \frac{\Ja^{4-k}}{(4-k)!} \frac{(i\Jd)^{k-1}}{(k-1)!}\dt (i\Jd)  \Bigg\} \ .
\end{equation}
Then, it is not hard to check that the coefficients are related by the following equation,
\begin{equation}\label{eq:FDRTheta}
    \frac{2}{\b}\bar{\Theta}_{k} + \Theta_{k+1} + \frac{1}{4} \Theta_{k-1}=0 \ ,
\end{equation}
where we note that these relations hold even before performing the radial grSK integral, i.e., they are valid at the level of the integrands. In fact, they reproduce the same FDRs derived in \cite{Jana:2020vyx} for a system at zero density.\footnote{Throughout this work, we use the terms FDT and FDR interchangeably and will continue to do so consistently.}

That said, it is important to emphasize that the coefficients $\{\Theta_{k}, \bar{\Theta}_{k}\}$ appearing here are not identical to those defined in Eq.~\eqref{eq:nlinLangevin}, namely $\{\theta_{k}, \bar{\theta}_{k}\}$, but are proportional to them.\footnote{The proportionality factors can be determined explicitly, but since they do not affect our discussion, we will not elaborate on them here.} The explicit computation of these coefficients requires evaluating a radial grSK integral, as detailed in Appendix (E.1) of \cite{Jana:2020vyx}. Making use of those results, we obtain the following expressions for $\Theta_k$:
\begin{equation}\label{eq:Thetaexplicit}
\begin{aligned}
    \Theta_{1} &= \ \frac{1}{i}\frac{1}{d}\left(r^d_c -r^d_h\right) \ ,\
    &&\Theta_{2} = \ -\frac{1}{i^2}\frac{r^d_h}{\pi i} \log \frac{r_c}{r_h} \ , \\
    \Theta_{3} &= \ \frac{1}{i^3}\left(  \frac{r^d_c}{4d} - \frac{r^d_h}{2d}\right)  \ ,
    &&\Theta_{4} = \ \frac{1}{i^4} \left( - \frac{r^d_h}{2 \pi i} \log\frac{r_c}{r_h} + \frac{1}{2 \pi i} \frac{r_h^d}{d} \frac{6}{\pi^2} \z(3)+\mathcal{O}\left(\frac{1}{r^d_c}\right) \right) \ .
\end{aligned}
\end{equation}
The remaining $\bar{\Theta}_{k}$ terms can be obtained using the non-linear FDRs given in Eq.~\eqref{eq:FDRTheta}.

These non-linear FDRs can also be represented diagrammatically. For example, in the case of $k=2$, Eq.~\eqref{eq:FDRTheta} yields the following schematic representation,
\begin{figure}[H]
    \centering
    
\tikzset{every picture/.style={line width=0.75pt}} 

\begin{tikzpicture}[x=0.75pt,y=0.75pt,yscale=-0.9,xscale=0.9]

\draw [color={rgb, 255:red, 74; green, 144; blue, 226 }  ,draw opacity=1 ][line width=1.5]    (30,60.27) -- (187,60.27) ;
\draw    (43.28,60.27) -- (101.25,150.8) ;
\draw  [dash pattern={on 4.5pt off 4.5pt}]  (153.18,61.23) -- (101.25,150.8) ;
\draw  [dash pattern={on 4.5pt off 4.5pt}]  (119.37,61.23) -- (101.25,150.8) ;
\draw    (79.52,61.23) -- (101.25,150.8) ;
\draw [color={rgb, 255:red, 74; green, 144; blue, 226 }  ,draw opacity=1 ][line width=1.5]    (480,59.27) -- (636,59.27) ;
\draw    (493.2,59.27) -- (550.8,149.8) ;
\draw  [dash pattern={on 4.5pt off 4.5pt}]  (602.4,60.23) -- (550.8,149.8) ;
\draw    (568.8,60.23) -- (550.8,149.8) ;
\draw    (529.2,60.23) -- (550.8,149.8) ;
\draw [color={rgb, 255:red, 74; green, 144; blue, 226 }  ,draw opacity=1 ][line width=1.5]    (246,58.87) -- (414,58.87) ;
\draw    (260.22,58.87) -- (322.25,147.8) ;
\draw  [dash pattern={on 4.5pt off 4.5pt}]  (377.82,59.81) -- (322.25,147.8) ;
\draw  [dash pattern={on 4.5pt off 4.5pt}]  (341.63,59.81) -- (322.25,147.8) ;
\draw  [dash pattern={on 4.5pt off 4.5pt}]  (298.98,59.81) -- (322.25,147.8) ;

\draw (0,70) node [anchor=north west][inner sep=0.75pt]   [align=left] {{ \Large $\frac{2}{\beta}$}};
\draw (190,70) node [anchor=north west][inner sep=0.75pt]   [align=left] {{\Large $=$}};
\draw (220,70) node [anchor=north west][inner sep=0.75pt]   [align=left] {{ \Large$-$}};
\draw (440,70) node [anchor=north west][inner sep=0.75pt]   [align=left] {{ \Large $-\frac{1}{4}$}};

\draw (36.39,38.04) node [anchor=north west][inner sep=0.75pt]   [align=left] {{\footnotesize $\Ja$}};
\draw (143.87,36.23) node [anchor=north west][inner sep=0.75pt]   [align=left] {{\footnotesize $\dot{\Jd}$}};
\draw (111.27,39.23) node [anchor=north west][inner sep=0.75pt]   [align=left] {{\footnotesize $ \Jd$}};
\draw (70.2,38.04) node [anchor=north west][inner sep=0.75pt]   [align=left] {{\footnotesize $\Ja$}};
\draw (486.3,37.04) node [anchor=north west][inner sep=0.75pt]   [align=left] {{\footnotesize Ja}};
\draw (593.1,38.23) node [anchor=north west][inner sep=0.75pt]   [align=left] {{\footnotesize $\Jd$}};
\draw (560.7,38.23) node [anchor=north west][inner sep=0.75pt]   [align=left] {{\footnotesize $\Ja$}};
\draw (519.9,37.04) node [anchor=north west][inner sep=0.75pt]   [align=left] {{\footnotesize $\Ja$}};
\draw (253.36,36.87) node [anchor=north west][inner sep=0.75pt]   [align=left] {{\footnotesize $\Ja$}};
\draw (368.38,38.05) node [anchor=north west][inner sep=0.75pt]   [align=left] {{\footnotesize $\Jd$}};
\draw (333.48,38.05) node [anchor=north west][inner sep=0.75pt]   [align=left] {{\footnotesize $\Jd$}};
\draw (289.55,36.87) node [anchor=north west][inner sep=0.75pt]   [align=left] {{\footnotesize $\Jd$}};

\end{tikzpicture}
    \caption{Schematic representation of a non-linear FDR for the case of $k=2$. Solid and dashed lines represent boundary-to-bulk propagators that terminate on average $\Ja$ and difference $\Jd$ boundary sources, respectively.}
\end{figure}

Here, solid lines denote average sources ($\Ja$), dashed lines denote difference sources ($\Jd$), and dots indicate time derivatives. Similar diagrammatic representations can be constructed for other cases ($k \neq 2$), rendering the structure of the non-linear FDRs manifest in terms of effective vertices.

We now conclude our discussion of FDRs at zero density with a brief motivation to study their counterparts at finite density.

\medskip

To the best of our knowledge, this remains the only available method for deriving FDTs in strongly coupled thermal baths. However, to build greater confidence in these results, it is important to explore other settings where such relations may hold. A natural next step is to extend this analysis to finite-density holographic baths and investigate whether analogous FDRs emerge.

At the outset, it is not obvious that such FDTs should persist in these systems. Finite-density holographic theories introduce qualitatively new features --- such as charged black hole backgrounds, multiple horizons, and chemical potentials --- that might affect the dynamics of fluctuations and dissipation relative to the zero-density case. This leads us to the following central question:

\begin{center}
    \emph{Do FDTs hold in finite-density holographic systems? If so, what form do they take, and how do they differ from the zero-density case?}
\end{center}

\noindent
We now turn to this question in the next section.

\subsection{Holographic FDTs  at small but finite density}

We begin this section by revisiting the setup and results introduced in section~\ref{sec:phiinRNSK}. Recall, we considered a self-interacting complex scalar field propagating in the RNSK geometry, with the action given by,
\begin{equation}
S = \ -\oint_{\rm RNSK} \left( |D_{M} \Phi |^2 + \frac{\lambda}{2! 2!} |\Phi|^{4} \right) \ .
\end{equation}
We employed perturbation theory to study the dynamics of this theory and computed the on-shell bulk action, which corresponds to the Schwinger-Keldysh generating functional for the dual boundary theory.

The on-shell action $S_{\rm os}$ can be expanded in powers of the coupling constant $\lambda$ as:
\begin{equation}
S_{\rm os} = S_{(2)} +  S_{(4)} + S_{(6)} + \ldots \ ,
\end{equation}
where the subscript denotes the number of boundary sources involved in each term. To streamline the notation, we expressed the general $2n$-point contribution to the on-shell action as:
\begin{equation}
\small
\begin{split}
S_{(2n)} &= \int_{k_{1,2,\ldots,2n}} \sum_{r,s=0}^{n} \mathcal{I}^{(2n)}_{r,s}(k_1,\ldots,k_{2n}) \prod_{i=1}^{r} \Jb_{\Fb}(k_i) \prod_{j=1}^{s} J_{\Fb}(k_j) \prod_{l=r+1}^{n} \Jb_{\Pb}(k_l) \prod_{m=s+1}^{n} J_{\Pb}(k_m) \ ,
\end{split}
\end{equation}
where the coefficients $\mathcal{I}^{(2n)}_{r,s}$ are determined from the Feynman rules of the exterior field theory derived from the bulk, as seen in Eqs.~\eqref{eq:I4} and \eqref{eq:I6}.

Our goal now is to extract the relationships between fluctuation and dissipation coefficients --- that is, to identify holographic FDTs for this system. To do this, it is convenient to work in the average-difference basis of the SK formalism. This basis is particularly important because the average field maps to the classical stochastic variable under an inverse Martin-Siggia-Rose (MSR) transformation \cite{Martin:1973zz}, while the difference field corresponds to the noise source.\footnote{For details on this mapping, see section~$\S$3.2 of  \cite{Jana:2020vyx}.}

However, it is worth emphasizing that both the on-shell action (or SK generating functional) and the corresponding stochastic equation contain the same physical information, simply packaged differently. Therefore, instead of performing the MSR transformation, we can equivalently analyse the on-shell action directly and look for relations among various terms --- these relations are the sought-after holographic FDTs.

We begin with the simplest case: the quadratic part of the on-shell action in average-difference basis as given in Eq.~(\ref{eq:S2osaction}),
\begin{equation}
    S_{(2)} = \int \frac{d^d k}{(2 \pi)^d }  \left[ K_{\rm ret} \  
     \Jbd(-k) \Ja (k)  + K_{\rm adv} \ 
    \Jba(-k) \Jd(k)  + K_{\rm kel} \  \Jbd(-k) \Jd(k) \right] \ .
\end{equation}
This part contains two-point correlations and, as seen earlier in Eq.~\eqref{eq:S2osaction}, yields the KMS relation given in Eq.~\eqref{eq:2KMS}. At finite density, the two-point KMS relation takes the following form:
\begin{equation}\label{2ptFDR}
\begin{split}
   \frac{K_{\rm kel}}{2}  & = \frac{1}{4} \coth \left(\frac{\b(k^0-\muq)}{2}\right)\left[ K_{\rm ret} - K_{\rm adv} \right] \ , \\
     & = \frac{i}{2} \coth \left(\frac{\b(k^0-\muq)}{2}\right) \text{Im} \left[ K_{\rm ret} \right] \ .
\end{split}   
\end{equation}
Here, $K_{\rm ret}$ and $K_{\rm adv}$ are the retarded and advanced propagators, and $K_{\rm kel}$ is the Keldysh (or symmetric) propagator, which captures the fluctuations. This relation is the finite-density generalisation of the linear fluctuation-dissipation theorem. To understand this, recall that $\text{Im} \left[ K_{\rm ret} \right]$ corresponds to the spectral function at finite temperature, while $K_{\rm kel}$ represents the anti-commutator (refer to \cite{kamenev_2011, Chaudhuri:2018ymp} for details). Commutators present in response functions naturally measure dissipation and transport, whereas anti-commutators capture fluctuations \cite{Haehl:2017eob}. Consequently, Eq.~(\ref{2ptFDR}) provides a correct description of the linear fluctuation-dissipation relation in finite-density systems.

In the zero-density case ($\mu = 0$), and assuming the field is real, the two-point KMS condition in the high-temperature limit ($\beta \to 0$) takes the form of linear FDT, which can be diagrammatically described as,
\begin{figure}[H]
\centering

\tikzset{every picture/.style={line width=0.75pt}} 

\begin{tikzpicture}[x=0.75pt,y=0.75pt,yscale=-0.7,xscale=0.7]

\draw [color={rgb, 255:red, 74; green, 144; blue, 226 }  ,draw opacity=1 ][line width=1.5]    (31,102.06) -- (265,102.06) ;
\draw  [dash pattern={on 4.5pt off 4.5pt}]  (52.6,102.06) .. controls (72.4,209.53) and (95.8,235.58) .. (137.2,229.07) ;
\draw  [dash pattern={on 4.5pt off 4.5pt}]  (137.2,229.07) .. controls (182.2,225.81) and (198.4,212.78) .. (212.8,105.32) ;
\draw [color={rgb, 255:red, 74; green, 144; blue, 226 }  ,draw opacity=1 ][line width=1.5]    (408,101.55) -- (646,101.55) ;
\draw    (429.97,101.55) .. controls (450.11,209.01) and (473.91,235.07) .. (516.02,228.55) ;
\draw  [dash pattern={on 4.5pt off 4.5pt}]  (516.02,228.55) .. controls (561.78,225.3) and (578.26,212.27) .. (592.91,104.8) ;

\draw (300,150) node [anchor=north west][inner sep=0.75pt]   [align=left] {{$ \propto$}};

\draw (44.2,58) node [anchor=north west][inner sep=0.75pt]   [align=left] {{ $\Jd$}};
\draw (202.6,58) node [anchor=north west][inner sep=0.75pt]   [align=left] {{ $\Jd$}};
\draw (110,180) node [anchor=north west][inner sep=0.75pt]   [align=left] {{  $K_{\rm kel}$}};

\draw (564.66,58) node [anchor=north west][inner sep=0.75pt]   [align=left] {{ $\Jd$}};
\draw (425.22,58) node [anchor=north west][inner sep=0.75pt]   [align=left] {{ $\Ja$}};
\draw (455,180) node [anchor=north west][inner sep=0.75pt]   [align=left] {{  
 \ \ $\text{Im}[K_{\rm ret}]$}};

\end{tikzpicture}
    \caption{Schematic representation of the linear FDT. Solid and dashed lines correspond to average $\Ja$ and difference $\Jd$ sources, respectively.}
\end{figure}

In what follows, we explore how the fluctuation-dissipation paradigm extends to higher-point functions, and whether non-linear FDRs persist in strongly coupled systems at finite density. This question captures the novelty of our work.

Just as we analysed the quadratic on-shell action in the linear FDT case, we now turn to the next non-vanishing term: the quartic term in the on-shell action. Due to the parity structure of the action, only even-order terms contribute, making the quartic term the natural next step.

This quartic contribution to the on-shell action arises from the contact Witten diagrams, depicted in figure~\ref{fig:fourpointcontact}. Each diagram contributes a specific term to the on-shell action, written explicitly using Feynman rules in the Past-Future (PF) basis as shown in Eq.~\eqref{eq:I4}. However, to extract FDRs, it is more appropriate to express the on-shell action in the average-difference basis.

Unlike the quadratic case, the quartic expressions become significantly more involved. Fortunately, since the FDRs only require terms up to first order in derivatives, we can simplify our analysis by keeping only the leading derivative expansion. As shown in Eq.~\eqref{eq:S4linear}, the quartic on-shell action up to linear order in near-$\muq$ gradient and charge expansion while discarding mixed terms, can be written as:
\begin{equation}\label{eq:IFquartic}
\begin{split}
    S_{(4)} 
    &\approx -\lambda \int d^d x \int_{\rm ext}  \mathcal{L}_{0} -\lambda \int d^d x \int_{\rm ext} \mathcal{L}_{1} \ ,
\end{split}    
\end{equation}
where the terms $\mathcal{L}_{0}$ and $\mathcal{L}_{1}$ represent the non-derivative and first-derivative contributions to $S_{(4)}$, respectively. The $\mathcal{L}_{0}$ and $\mathcal{L}_{1}$ are given by,
\begin{equation}\label{eq:L0L1terms}
    \begin{split}
        \mathcal{L}_{0} &=  \sum_{r,s=0}^{2} \mathcal{G}_{r,s} \  \left[\Jba\right]^{r} \left[\Ja\right]^{s} 
        \left[\Jbd\right]^{2-r}  \left[\Jd\right]^{2-s}     \ , \\
        \mathcal{L}_{1} &=  \sum_{r=0}^{1} \sum_{s=0}^{2}  \Big\{ \mathcal{G}_{\dot{r},s}  \left(\dt \Jba \right) 
        + \mathcal{H}_{\dot{r},s}  \left(\dt \Jbd \right)  \Big\} \left[\Jba\right]^{r} \left[\Ja\right]^{s} 
        \left[\Jbd\right]^{1-r}  \left[\Jd\right]^{2-s}     \\
        &+  \sum_{r=0}^{2} \sum_{s=0}^{1}            \Big\{  \mathcal{G}_{r,\dot{s}}  \left(\dt \Ja \right)  + \mathcal{H}_{r,\dot{s}}  \left(\dt \Jd \right)  \Big\}  \left[\Jba\right]^{r} \left[\Ja\right]^{s} \left[\Jbd\right]^{2-r}  \left[\Jd\right]^{1-s} \ ,
    \end{split}
\end{equation}
where all of the above-mentioned terms/coefficients are given in Appendix~\ref{app:explicitL0L1} (see Eqs.~\eqref{eq:explicitGrs}, \eqref{eq:explicitGrdots} and \eqref{eq:explicitHrs}). These terms/coefficients are not all independent. Their interrelations arise from:
\begin{itemize}
    \item the \emph{reality} of the on-shell action ,

    \item the \emph{underlying KMS symmetry}, which gives rise to non-linear fluctuation-dissipation relations.
\end{itemize}
\noindent
Importantly, these relations connect terms with and without derivatives (i.e., dissipation vs. fluctuation), and terms with different combinations of average ($\Ja, \Jba$) and difference ($\Jd, \Jbd$) sources. This structure becomes more transparent in the Witten diagram representation, where different diagram topologies naturally encode these dependencies.

Without further ado, we now write down the non-linear FDRs among the quartic coefficients:
\begin{equation}\label{eq:FDRsmu0not}
\begin{split}
 &\mG_{\dot{0},0} - \mathcal{H}_{\dot{1},0} - i\b \mG_{0,0}+\frac{i\b}{24} \left( \mG_{2,0}  + \mG_{1,1} + \mG_{0,2} \right) = 0 \ ,\\
-\frac{1}{3}&\left( \mG_{\dot{0},2} +\mG_{\dot{1},1}  \right)+  \mathcal{H}_{\dot{1},2} +\frac{i\b}{6} \left( \mG_{2,0}  + \mG_{1,1} + \mG_{0,2} \right) = 0 \ , \\
-\frac{1}{3}&\left( \mG_{\dot{1},0} +\mG_{\dot{0},1}  \right)+ \frac{1}{5} \mathcal{H}_{\dot{0},2} +\frac{2}{5} \mathcal{H}_{\dot{1},1} -\frac{i\b}{16} \left( \mG_{1,2}  + \mG_{2,1} \right) + \frac{i\b}{4} \left( \mG_{0,1}  + \mG_{1,0} \right) = 0 \ .
\end{split}
\end{equation}
where the terms containing time derivatives correspond to dissipation coefficients, while those without derivatives represent the fluctuation strength and coefficient $\beta$ (inverse temperature) acts as the thermal weighting factor. Hence, these are the non-linear FDRs at finite density, derived from the structure of the on-shell gravitational action. It is important to emphasize, however, that these relations have been obtained by working only to linear order in the small chemical potential. Going beyond this approximation may modify these relations, a question we leave for future work.

These relations can be further understood using the inverse MSR formalism \cite{Martin:1973zz}, where one transforms the influence functional into a stochastic field theory. In that language:
\begin{itemize}
\item The \textit{average source} $\Ja$ is identified with the classical stochastic field.

\item The \textit{difference source} $i\Jd$ corresponds to the stochastic noise.
\end{itemize}
For the purposes of this work, we focus on the on-shell action and postpone a detailed derivation of the corresponding stochastic field equations to future work. 

\medskip

The non-linear relations we have derived represent a new class of holographic FDRs valid at finite density. Unlike the well-studied two-point correlators with linear FDT, these relations govern the dynamics of higher-order correlators in strongly coupled systems with chemical potential and describe how non-linear fluctuations and dissipative responses are interrelated. The fact that such relations are encoded in the dynamics of bulk gravity further underscores holography as a powerful organising principle for understanding fluctuations in strongly coupled systems.

The holographic FDRs at finite density presented here constitute one of the novel contributions of this work. To our knowledge, this is the first time such relations have been explicitly derived. Naturally, this raises the question of whether they can be independently verified through other means. Unfortunately, strongly coupled field theories at finite density generally lack non-holographic tools that allow for computing such relations. This very limitation highlights the power and utility of holography in accessing non-perturbative regimes.

That said, an important consistency check is to consider the zero-density limit ($\mu \to 0$), in which the dual black hole background becomes neutral. One would expect to recover previously known results in this limit. However, the system we study here differs from the one examined in earlier zero-density analyses (e.g, the system we reviewed in section~$\S$\ref{sec:realatmuzeror}). In particular, our zero-density limit involves two coupled scalar fields with cross-interactions, as opposed to a single self-interacting scalar field considered in \cite{Jana:2020vyx}. Thus, the check is less straightforward than it may appear.

A meaningful first step is to derive the holographic FDRs for two cross-interacting scalar fields at zero density. This itself constitutes a new result and is worth pursuing independently. Once obtained, these FDRs can be compared to those in \cite{Jana:2020vyx} (or system presented in section~$\S$\ref{sec:realatmuzeror}). However, the outcome of such a comparison is not obvious \emph{a priori}, given the structural differences between the two setups. As such, the question is exploratory in nature and could lead to new insights into how FDRs behave in the presence of multiple interacting fields.

\subsection{Zero-density limit}

The zero-density limit of our setup corresponds to a self-interacting complex scalar field propagating in the grSK geometry instead of the RNSK geometry, with the action given by,
\begin{equation}
    S = -\oint_{\rm grSK} \left( |\partial_{M} \Phi |^2 + \frac{\lambda}{2! 2!} |\Phi|^{4} \right) \ .
\end{equation}
This new setup can equivalently be described in terms of real degrees of freedom by replacing the complex field $\Phi$ with a pair of real scalar fields $\{\phi, \varphi\}$ defined as,
\begin{equation}
    \Phi \equiv \frac{\phi + i \varphi}{\sqrt{2}} \ .
\end{equation}
The action in terms of the $\phi$--$\varphi$ variables then becomes,
\begin{equation}
    S = -\oint_{\rm grSK} \left( \frac{1}{2}(\partial \phi)^2 + \frac{1}{2}(\partial \varphi)^2 + \frac{\lambda}{16} \phi^{4} + \frac{\lambda}{16} \varphi^{4} + \frac{\lambda}{8} \phi^{2} \varphi^{2} \right) \ .
\end{equation}

Inserting the perturbative solution\footnote{The notation for the perturbative expansion of $\phi - \varphi$ system is similar to that of $\Phi$, as given in Eq.~(\ref{eq:Phipert}).} into the bare action yields the \emph{on-shell action} $S_{\rm os}$ in the following form:
\begin{equation}
\begin{split}
    S_{\rm os} &= -\frac{1}{2} \int_{\partial \mathcal{M}} \phi_{(0)} \left( \partial_{A} \phi_{(0)} \right) - \frac{1}{2} \int_{\partial \mathcal{M}} \varphi_{(0)} \left( \partial_{A} \varphi_{(0)} \right) \\
    &\quad + \lambda \oint_{\mathcal{M}} \left[ \frac{1}{8} \left( (\phi - \phi_{(0)})\phi + i \varphi_{(0)} \phi - i \phi_{(0)} \varphi + (\varphi - \varphi_{(0)})\varphi \right)(\phi^2 + \varphi^2) \right. \\
    &\qquad\quad \left. - \frac{1}{16} \phi^{4} - \frac{1}{16} \varphi^{4} - \frac{1}{8} \phi^{2} \varphi^{2} \right] \ ,
\end{split}
\end{equation}
where $\partial \mathcal{M}$ denotes the boundary of the bulk manifold $\mathcal{M}$ (i.e., grSK geometry in this setup). In deriving the above expression, we have used the field equations of the $\phi$--$\varphi$ system and imposed appropriate boundary conditions to simplify the on-shell action.

Expanding the on-shell action $S_{\rm os}$ in powers of $\lambda$, we obtain:
\begin{equation}
    S_{\rm os} = S_{(2)} + S_{(4)} + \ldots \ ,
\end{equation}
where the subscripts indicate the number of boundary sources in each term. In particular, the quadratic and quartic contributions are given by:
\begin{equation} \label{eq:S2S4S6muzero}
\begin{split}
    S_{(2)} &= -\frac{1}{2} \int_{\partial \mathcal{M}} \phi_{(0)} \left( \partial_{A} \phi_{(0)} \right) - \frac{1}{2} \int_{\partial \mathcal{M}} \varphi_{(0)} \left( \partial_{A} \varphi_{(0)} \right) \ , \\
    S_{(4)} &= -\oint_{\rm grSK} \left( \frac{\lambda}{16} \phi_{(0)}^4 + \frac{\lambda}{16} \varphi_{(0)}^4 + \frac{\lambda}{8} \phi_{(0)}^2 \varphi_{(0)}^2 \right) \ .
\end{split}
\end{equation}
Again, these terms correspond to the free and contact contributions at tree-level in the Feynman diagram expansion.

\medskip

Coming back to the complex field viewpoint, since the chemical potential vanishes in the zero-density case, we can directly write down the gradient-expanded solution for the zeroth-order complex field $\Phi$ (with $\mathbf{k} = 0$) by setting $\mu = 0$ (see Eq.~\eqref{eq:Ginexp} for reference). As previously noted, the only non-trivial component we require is the ingoing solution $\Gin$, which, after setting $\mu = 0$, takes the form:
\begin{equation}
    \Gin = \sum_{i} \Gin_{i}(\z) \left( \frac{\beta k^0}{2} \right)^i \ ,
\end{equation}
with coefficients explicitly given by:
\begin{equation} \label{eq:Ginpart1muzero}
\begin{split}
    \Gin_0 &= 1 \ , \quad \Gin_1 = \mathbb{F}_1(r) \ , \\
    \Gin_2 &= \int_{\z_c}^{\z} \frac{d \bar{\z}}{\bar{r}^{d-1}} \int_{\z_h}^{\bar{\z}} d \hat{\z} \left[\hat{r}^{d-1}\right]' \mathbb{F}_1(\hat{r})
    - 2 \int_{\z_c}^{\z} d \bar{\z} \left( \mathbb{F}_1(r) - \frac{r_+^{d-1}}{r^{d-1}} \mathbb{F}_1(r_+) \right) \ ,
\end{split}
\end{equation}
where we again define,
\begin{equation}
    \mathbb{F}_n(r) \equiv \int_{\z_c}^{\z} d \bar{\z} \left[ \left( \frac{r_+}{\bar{r}} \right)^{d-n} - 1 \right] \ , \qquad
    \z_h \equiv \z(r_+), \qquad \z_c \equiv \z(r_c) \ ,
\end{equation}
where $r_+$ now denotes the horizon size of the grSK geometry. The outgoing solution is obtained via \emph{CPT} conjugation discussed in section \ref{sec:phiinRNSK} (see Eq.~(\ref{eq:GintoGout}) for the mapping) by setting $\muq=0$.

Using this, the full solution for the zeroth-order complex field at zero density is:
\begin{equation}
    \Phi_{(0)} \Big|_{\mu =0} = \Gin_{q}(k) \left[ \Ja(k) + \left(n_k + \frac{1}{2}\right) \Jd(k) \right] - e^{-\beta k^0 (\z - 1)} \Gin_{-q}(-k) n_k \Jd(k) \ ,
\end{equation}
where we have suppressed the radial dependence of the Green's functions.

To conclude the discussion on the gradient expansion, we present the solution up to linear order in derivatives, which will be useful when we evaluate the non-linear FDRs in the upcoming text.
\begin{equation} \label{Phigrdexpposmuzero}
\small
\begin{split}
    \Phi_{(0)}\Big|_{\mu=0} &= \left[ \Ja + \left( \z - \frac{1}{2} + \mathbb{F}_1 \right) \Jd \right] + \frac{i \beta}{2} \left[ \mathbb{F}_1 \, \partial_t \Ja - \left( \z(\z - 1) + \left( \z - \frac{1}{2} \right) \mathbb{F}_1 \right) \partial_t \Jd \right]  \ ,
\end{split}
\end{equation}
where we have replaced $k^0 \mapsto i \partial_t$.

\medskip

As far as the quadratic influence phase $S_{(2)}$ is concerned, the calculation is straightforward. One can simply take the result from the non-zero chemical potential case and set $\mu = 0$. The two-point KMS condition is satisfied and gives rise to two familiar linear FDTs --- one for each real scalar field in the complex field.

However, the interesting structure emerges in the quartic influence phase $S_{(4)}$, where cross-interactions --- i.e., couplings between the two real scalar fields --- first appear. By taking the $\mu \to 0$ limit of Eq.~(\ref{eq:IFquartic}), we obtain,
\begin{equation}
\begin{split}
    S_{(4)}\big|_{\mu=0} = -\lambda \int d^d x \int_{\rm ext}  \mathcal{L}_{0} -\lambda \int d^d x \int_{\rm ext} \mathcal{L}_{1} +\mathcal{O}\left( (\beta \dt)^2\right)\ ,
\end{split}    
\end{equation}
where $\mathcal{L}_0$ and $\mathcal{L}_1$ denote the non-derivative and first-derivative contributions in the zero chemical potential limit, respectively. From here on, we will omit the explicit notation indicating $\mu = 0$, as it will be clear from the context.

The non-derivative part is given by:
\begin{equation}
    \mathcal{L}_0 = \sum_{r,s=0}^{2} \mathcal{G}_{r,s} \left[ \Jba \right]^r \left[ \Ja \right]^s \left[ \Jbd \right]^{2 - r} \left[ \Jd \right]^{2 - s} \ ,
\end{equation}
with independent components:
\begin{equation}\label{eq:Grsind}
\begin{split}
    \mathcal{G}_{0,0} &= \frac{1}{4} \left( \z + \mathbb{F}_1 \right) + \left( \z + \mathbb{F}_1 \right)^3 \ , \qquad \qquad 
    \mathcal{G}_{0,1} = \frac{1}{8} + \frac{3}{2} \left( \z + \mathbb{F}_1 \right)^2 \ , \\
    \mathcal{G}_{0,2} &= \frac{1}{2} \left( \z + \mathbb{F}_1 \right) \ , \quad 
    \mathcal{G}_{1,1} = 2 \left( \z + \mathbb{F}_1 \right) \ , \quad 
    \mathcal{G}_{1,2} = \frac{1}{2} \ , \quad 
    \mathcal{G}_{2,2} = 0 \ .
\end{split}
\end{equation}
Here, all the remaining coefficients are related by the following relation:
\begin{equation}
    \mathcal{G}_{r,s} = \mathcal{G}_{s,r} \  .
\end{equation}
The first-derivative contribution, $\mathcal{L}_1$, remains unchanged from the $\mu \neq 0$ case (See Eqs.~\eqref{eq:L0L1terms} and Eqs.~\eqref{eq:explicitGrdots} and \eqref{eq:explicitHrs}), as it does not contain any $\mu$-dependent terms to begin with. Note that this holds only because we are working up to linear order in the small chemical potential expansion.

Thus, it can be readily checked that these coefficients are related by what we call FDRs at zero density for the system of complex field with quartic interactions (similar to the $\phi-\varphi$ system described above).  The relations which directly follow from taking the $\mu \to 0$ limit are as follows:
\begin{equation}\label{eq:FDRsmu0}
\begin{split}
 &\mG_{\dot{0},0}  - \mathcal{H}_{\dot{1},0} -i\b \mG_{0,0} + \frac{i\b}{24}\mG_{1,1} +\frac{i\b}{12}\mG_{0,2}= 0 \ , \\
-\frac{1}{3}&\left( \mG_{\dot{0},2} +\mG_{\dot{1},1}  \right)+  \mathcal{H}_{\dot{1},2} + \frac{i\b}{6}\mG_{1,1} +\frac{i\b}{3}\mG_{0,2}= 0 \ , \\
-\frac{2}{3}&\left( \mG_{\dot{1},0} +\mG_{\dot{0},1}  \right)+ \frac{2}{5}\mathcal{H}_{\dot{0},2} +\frac{4}{5}  \mathcal{H}_{\dot{1},1} -\frac{i\b}{4}  \mG_{1,2}  + i\b \, \mG_{0,1}   = 0 \ .
\end{split}
\end{equation}
Apart from these relations, we expect additional relations to emerge in the zero-density limit. We leave a systematic exploration of these to future work. For now, our goal is to compare the fluctuation and dissipation coefficients $\mathcal{G}_{r,s}$, $\mathcal{G}_{\dot{r},s}$, and $\mathcal{H}_{\dot{r},s}$ introduced in this section with the coefficients $(\Theta_k, \bar{\Theta}_k)$ that appeared earlier in section~$\S$\ref{sec:realatmuzeror}.

Since both sets of coefficients arise in the zero-density case, it is natural to ask: how are they related? We will shortly demonstrate this connection explicitly, but let us first introduce the idea at a general level by examining how the coefficients $\mathcal{G}_{r,s}$ and related quantities can be expressed in terms of $(\Theta_k, \bar{\Theta}_k)$.

To begin, we consider a prototypical example by taking a coefficient $\mathcal{G}_{1,1}$ defined in Eq.~\eqref{eq:Grsind}. We now convert the exterior integral to the grSK integral, as shown below--
\begin{equation}\label{eq:protocomparison}
\begin{split}
        \int_{\rm ext}\mG_{1,1} &= 2\int_{\rm ext} \left(\z+\mathbb{F}_1 \right)\\
        &=  \int_{\rm ext}\left[ \left(\z+\frac{1}{2}+\mathbb{F}_1\right)^2- \left(\z-\frac{1}{2}+\mathbb{F}_1\right)^2 \right]\\
        &=  \oint_{\rm grSK} \left(\z-\frac{1}{2}+\mathbb{F}_1\right)^2 = i^2 \Theta_{2} \ ,
\end{split}
\end{equation}
where in the last line we used the definition of $\Theta_2$ from Eq.~\eqref{eq:Thetadef}. This computation illustrates that the coefficient $\mathcal{G}_{1,1}$, once integrated over the exterior region, directly relates to $\Theta_2$.
where we have used the definition given in Eq.~\eqref{eq:Thetadef}. 

To streamline notation going forward, we define,
\begin{equation}
    \widetilde{F} \equiv \int_{\rm ext} F(\z) \ ,
\end{equation}
where the `tilde' denotes integration over the exterior region. Then, the result from Eq.~\eqref{eq:protocomparison} simply becomes,
\begin{equation}
    \widetilde{\mathcal{G}}_{1,1} = i^2 \Theta_{2} \ . 
\end{equation}

Using this logic and notation, we can now similarly compare other coefficients. Without going into the details, we just quote the results of the comparison for the zero-derivative terms as,
\begin{equation}\label{eq:Grscompare}
\begin{split}
    \widetilde{\mathcal{G}}_{0,0} &= \frac{i^4}{4}\Theta_4 \ , \qquad \qquad 
    \widetilde{\mathcal{G}}_{0,1} = \frac{i^3}{2}\Theta_3\ , \\
    \widetilde{\mathcal{G}}_{0,2} &= \frac{i^2}{4}\Theta_2 \ , \qquad 
    \widetilde{\mathcal{G}}_{1,1} = i^2 \, \Theta_2 \ , \qquad 
    \widetilde{\mathcal{G}}_{1,2} = \frac{i}{2}\Theta_1 \  ,
\end{split}
\end{equation}
where $\Theta_k$ is defined in Eq.~\eqref{eq:Thetadef} and explicit expressions are given in Eq.~\eqref{eq:Thetaexplicit}.

Next, we quote the comparison of first-derivative terms as,
\begin{equation}\label{eq:Grdotscompare}
\begin{aligned}
    & \widetilde{\mathcal{G}}_{\dot{0},2} =  \frac{i}{2} \bar{\Theta}^{(a)}_{1} \ ,  \quad &&\widetilde{\mathcal{G}}_{\dot{1},1}  = i \bar{\Theta}^{(a)}_{1}  \ , \\
    &\widetilde{\mathcal{G}}_{\dot{1},0}  = \frac{i^2}{2} \bar{\Theta}^{(a)}_{2} \ ,
    &&\widetilde{\mathcal{G}}_{\dot{0},1}  = i^2 \bar{\Theta}^{(a)}_{2} \ , &&&\widetilde{\mathcal{G}}_{\dot{0},0}  = \frac{i^3}{2} \bar{\Theta}^{(a)}_{3} \ ,
\end{aligned}
\end{equation}
and
\begin{equation}\label{eq:Hrdotscompare}
\begin{aligned}
    & \widetilde{\mathcal{H}}_{\dot{1},2} =  \frac{i}{2} \bar{\Theta}^{(d)}_{1} \ , 
    && \widetilde{\mathcal{H}}_{\dot{0},2} =  \frac{i^2}{2} \bar{\Theta}^{(d)}_{2} \ , 
    &&& \widetilde{\mathcal{H}}_{\dot{1},1} =  i^2 \bar{\Theta}^{(d)}_{2} \ , \\
    &\widetilde{\mathcal{H}}_{\dot{0},1} = i^3 \bar{\Theta}^{(d)}_{3} \ , 
    &&\widetilde{\mathcal{H}}_{\dot{1},0} = \frac{i^3}{2} \bar{\Theta}^{(d)}_{3} \ , 
    &&& \widetilde{\mathcal{H}}_{\dot{0},0} = \frac{i^4}{2} \bar{\Theta}^{(d)}_{4}  \ ,
\end{aligned}
\end{equation}
where we have used the expressions for coefficients $\bar{\Theta}^{(a)}_k$ and $\bar{\Theta}^{(d)}_k$ are defined in Eq.~\eqref{eq:Thetabaddef}. These relations provide a precise dictionary between the two sets of coefficients --- those arising in the complex scalar field setup and those in the real scalar field setup --- in the zero-density limit.

\paragraph{Consistency Check:}

To validate this dictionary, let us consider a particular fluctuation-dissipation relation in the $\mu \to 0$ limit. For example, consider the first equation in Eq.~\eqref{eq:FDRsmu0}:
\begin{equation}
\mG_{\dot{0},0}  - \mathcal{H}_{\dot{1},0} -i\b \mG_{0,0} + \frac{i\b}{24}\mG_{1,1} +\frac{i\b}{12}\mG_{0,2}= 0 \ .
\end{equation}
Integrating over the exterior region and plugging in the relations from Eqs.~\eqref{eq:Grscompare}, \eqref{eq:Grdotscompare}, and \eqref{eq:Hrdotscompare}, we find that this becomes the following $\Theta$-based expression:
\begin{equation}
    \frac{2}{\beta}\bar{\Theta}_{3}+ \Theta_{4}+\frac{1}{4}\Theta_2 = 0 \ ,
\end{equation}
where we have used the identity $\bar{\Theta}_{k} = \bar{\Theta}^{(a)}_{k}-  \bar{\Theta}^{(d)}_{k}$. As noted in Eq.~\eqref{eq:FDRTheta}, this is nothing but FDRs for a real scalar setup at zero density.

One can perform similar checks for other FDRs and confirm that they are consistent with the non-linear FDRs expressed in Eq.~\eqref{eq:FDRTheta}. This establishes the equivalence between the two descriptions at zero density, at least for the set of relations we have obtained, namely those in Eq.~\eqref{eq:FDRsmu0}.

\medskip

We now conclude our discussion of the comparison between the zero-density FDRs derived here and those obtained in the earlier work~\cite{Jana:2020vyx}, and proceed to summarise the main insights of this section.

To summarise, we have shown that:
\begin{itemize}
\item 
Non-linear fluctuation-dissipation relations exist in holographic systems at small but finite density.
\item 
These relations are encoded in the structure of the on-shell gravitational action.
\item 
They reduce to known results at zero density, thereby passing a non-trivial consistency check,
\item 
And they represent a step forward in connecting bulk gravity with non-linear stochastic dynamics in strongly coupled quantum systems at finite density.
\end{itemize}
Having resolved the question posed at the beginning of this work, we now conclude with a summary of results and a discussion of future directions.

\section{Discussion}\label{sec:Discussion}

In this work, we have focused on the study of (non)-linear fluctuation-dissipation relations (FDRs) for holographic systems at finite density. This was achieved by employing the RNSK geometry -- the gravitational dual of the Schwinger-Keldysh formalism at finite density. While RNSK geometry has been previously studied for free fields \cite{Loganayagam:2020iol}, our work is the first to apply it to interacting fields, e.g. complex scalars. We derived the non-linear FDRs up to linear order in the small charge (or small density) limit and reproduced the expected zero density results. Although a complete understanding of FDRs at arbitrary charge remains a challenge, we have made notable progress in the small charge regime.

\medskip

In the bulk, our approach constructs an \emph{exterior field theory} to describe interactions in a charged black hole background, confined outside the outermost horizon. The existence of such an exterior field theory in this setting is non-trivial, underscoring the novelty of our framework. Building on the results of \cite{Loganayagam:2024mnj}, we extend the construction of an exterior EFT for Hawking radiation to the case of charged black holes. Within this theory, we develop a tree-level Witten diagrammatic formalism to compute boundary Schwinger–Keldysh correlators at finite density, obtaining explicit results for up to six-point functions. The same diagrammatic framework also enables the computation of the influence phase to arbitrary orders in the bulk coupling constant $\lambda$, thereby allowing the determination of real-time correlators with an arbitrary number of insertions for the holographic CFT at finite temperature and chemical potential.

\medskip

There are several directions in which this investigation can be extended. A natural step forward would be to extend the analysis to systems with arbitrary values of charge. Such an extension would solidify the existence of holographic fluctuation-dissipation relations for large density systems. However, the primary challenge lies in finding solutions at an arbitrary charge, while the rest of the analysis is expected to proceed relatively smoothly.

\medskip

From a technical standpoint, an important next step is the explicit evaluation of the radial integrals appearing in the exterior field theory. In higher dimensions, this is hindered by the fact that bulk propagators are only known perturbatively in a boundary derivative expansion. However, for the BTZ black hole ($d=2$), the bulk-to-boundary propagators are known exactly in terms of hypergeometric functions \cite{Birmingham:2001hc, Jana:2020vyx}. This makes the BTZ geometry an ideal testbed to validate our methods more precisely and perform explicit computations beyond the derivative expansion.

\medskip

Another interesting path involves understanding the effects of loop corrections on the exterior theory. In collaboration with R. Loganayagam and G. Martin, we have obtained preliminary results in this area, which we plan to present soon. These corrections, corresponding to $\frac{1}{N}$ effects in the boundary theory, raise important questions about how fluctuation-dissipation relations transform when such corrections are accounted for. Addressing this would help clarify the distinction between the statistical and quantum origins of fluctuations.\footnote{I would like to thank Hong Liu and R. Loganayagam for a discussion on this question.}

\medskip

A further extension would be to explore fermionic systems, using the framework of open EFTs for interacting fermions \cite{Martin:2024mdm}. Since fermions are ubiquitous in physical systems, establishing fluctuation–dissipation relations in this setting would greatly broaden our understanding of strongly coupled systems.

\medskip

Finally, it would be interesting to investigate possible cosmological analogues of our results. In particular, the methods developed here could be applied to the study of correlators in de Sitter spacetime using the Schwinger–Keldysh formalism, in the spirit of recent works such as \cite{Sleight:2021plv, Schaub:2023scu, Bhattacharya:2023twz}. Such an analysis could yield new insights into the structure of cosmological correlators and clarify the role of fluctuations during the early universe.

\bigskip

\section*{Acknowledgements}

I would like to thank R. Loganayagam for his advice throughout this project and Godwin Martin for many insightful discussions. I would also like to thank Subhro Bhattacharjee, Joydeep Chakravarty, Hong Liu, Gautam Mandal, Suvrat Raju, Ashoke Sen, Omkar Shetye, Akhil Sivakumar and the ICTS string group for valuable discussions. My thanks extend to the organisers of the National Strings Meeting (NSM) 2024 at IIT Ropar and Strings 2025 at NYU Abu Dhabi, where some preliminary results of this work were presented. I would like to acknowledge the support of the Department of Atomic Energy, Government of India, under project no. RTI4001, and the unwavering and generous support provided by the people of India towards research in the fundamental sciences.

\appendix

\section{CPT action}\label{app:CPT}
In this appendix,  we will explicitly see how \emph{CPT} acts on the field equation of the complex field in the RNSK geometry. As we have seen in section~$\S$\ref{sec:RNSK}, it is important for obtaining the outgoing solution from the ingoing solution. Here, we will explicitly show how the mapping given in Eq.~(\ref{eq:intooutmap}) is correct.

We begin with the field equations for a free complex scalar field $\Phi$,
\begin{equation}
    D_{M}D^{M} \Phi  = 0 \ , \qquad \text{where} \quad D_{M} =\partial_{M}-i q \mathcal{A}_{M} \ .
\end{equation}
Using the momentum space representation
\begin{equation}
    \Phi(\z,v,\mathbf{x}) =  \int \frac{d^d k}{(2 \pi)^d} \Phi(\z,k^0 ,\mathbf{k}) e^{-i k^0v+i \mathbf{k.x}} \ .
\end{equation}
Then, the field equation for the complex field can be written explicitly as,
\begin{equation}\label{eq:EOM}
    \begin{split}
        &\Bigg\{\left(\frac{d}{d \z}+\frac{\b k^0 }{2}\right)\left[ r^{d-1} \left(\frac{d}{d \z}+\frac{\b k^0}{2}\right) \right] -r^{d-1} \left[  \left( \frac{\b k^0 }{2}\right)^2 - f \left(\frac{\b \mathbf{k}}{2}\right)^2 \right]\\
        &\hspace{2cm}+\b q \, r^{d-1} \mathcal{A}_{v} \left(\frac{d}{d \z}+\frac{i \b}{4} r f\right)\Bigg\} \Phi_{\rm q} (\z,k^0,\mathbf{k})= 0
    \end{split} 
\end{equation}
where we have restored the $q$ dependence on the field. Using the substitution motivated by \emph{CPT} action (\ref{CFT_CPT}) in the above equation,
\begin{equation}\label{eq:defphitilde}
\Phi_{\rm q}(\z, k^0 ,\mathbf{k} )  = e^{-\beta k^0 \zeta  + \beta \muq \bTh }\ \tilde{\Phi}_{\rm q}(\z, k^0 ,\mathbf{k} )  \ ,  \qquad \text{with} \quad \frac{d \bTh}{d \z} = - \frac{\mathcal{A}_{v}}{\mu} \ ,
\end{equation}
we obtain
\begin{equation}
    \begin{split}
        &\Bigg\{\left(\frac{d}{d \z}-\frac{\b k^0 }{2}-\b q \mathcal{A}_{v} \right)\left[ r^{d-1} \left(\frac{d}{d \z}-\frac{\b k^0 }{2}-\b q \mathcal{A}_{v}\right) \right] -r^{d-1} \left[  \left( \frac{\b k^0 }{2}\right)^2 - f \left(\frac{\b \mathbf{k}}{2}\right)^2 \right]\\
        &\hspace{2cm}+\b q \, r^{d-1} \mathcal{A}_{v} \left(\frac{d}{d \z} -\b k^0 -\b q \mathcal{A}_{v}+\frac{i \b}{4} r f\right)\Bigg\} \tilde{\Phi}_{\rm q}(\z,k^0 ,\mathbf{k})= 0 \ ,
    \end{split} 
\end{equation}
which, when simplified, becomes
\begin{equation}\label{eq:EOMrev}
    \begin{split}
        &\Bigg\{\left(\frac{d}{d \z}-\frac{\b k^0}{2}\right)\left[ r^{d-1} \left(\frac{d}{d \z}-\frac{\b k^0}{2}\right) \right] -r^{d-1} \left[  \left( \frac{\b k^0}{2}\right)^2 - f \left(\frac{\b \mathbf{k}}{2}\right)^2 \right]\\
        &\hspace{2cm}-\b q \, r^{d-1} \mathcal{A}_{v} \left(\frac{d}{d \z}+\frac{i \b}{4} r f\right)\Bigg\} \tilde{\Phi}_{\rm q}(\z,k^0 ,\mathbf{k})= 0
    \end{split} 
\end{equation}

Comparing equations (\ref{eq:EOM}) and (\ref{eq:EOMrev}), we can easily deduce: Given $\Phi_{\rm q} (\z,k)$ is a solution of the field equation, the above algebraic manipulation produces another solution, given by  $\tilde{\Phi}_{\rm -q} (\z,-k)$. Keeping the prefactor in the RHS of Eq.~(\ref{eq:defphitilde}) in mind, we can write the ingoing-to-outgoing map in spacetimes coordinates as:
\begin{equation}
    \Phi^{\rm out}_{\rm q}(\z,v, \mathbf{x}) =  e^{\beta \muq \bTh  } \  \Phi^{\rm in}_{\rm -q}(\z, i \b \z -v,- \mathbf{x})
\end{equation}
We can perform a similar analysis for the conjugate field, with the only major difference being $q\to -q$ in the prefactor.

\section{Bulk-to-Bulk Green's functions in the RNSK geometry }\label{app:bbGRNSK}

In this appendix, we will discuss the derivation of bulk-to-bulk Green's functions for a complex field in the RNSK geometry. Our derivation will closely follow the techniques used in \cite{Arnold:2011hp, Faulkner:2013bna} and recently implemented in the grSK geometry \cite{Loganayagam:2024mnj}. This we want to emphasize because we are using the same notation but a different convention for Green's function.

The bulk-to-bulk Green's function solves the following equation (by writing Eq.~\eqref{eq:bbGdef} in boundary momentum space),\footnote{Comparing it to the equation present in \cite{Loganayagam:2024mnj} (see Eq.~(2.28) in  \cite{Loganayagam:2024mnj}), there is an overall sign difference while taking the zero density limit $q \to 0$.}
\begin{equation}\label{eq:bbGPDE}
    \begin{split}
        &\Bigg\{\left(\frac{d}{d \z}+\frac{\b p^0}{2}\right)\left[ r^{d-1} \left(\frac{d}{d \z}+\frac{\b p^0}{2}\right) \right] -r^{d-1} \left[  \left( \frac{\b p^0}{2}\right)^2 - f \left(\frac{\b \mathbf{p}}{2}\right)^2 \right]\\
        &\hspace{2cm}+\b q \, r^{d-1} \mathcal{A}_{v} \left(\frac{d}{d \z}+\frac{i \b}{4} r f\right)\Bigg\} \bbG_{\rm q}(\z|\z_0,p)= \frac{i \beta}{2}\delta (\z -\z_0) \ ,
    \end{split} 
\end{equation}
with the following bi-normalisable boundary conditions,
\begin{equation}
    \lim_{\z \to 0} \bbG_{\rm q}(\z|\z_0,p) = \lim_{\z \to 1} \bbG_{\rm q}(\z|\z_0,p) = 0 \ .
\end{equation}
Using the notation used by the authors \cite{Loganayagam:2024mnj} and bi-normalisability of the Green's function, our ansatz for the bulk-bulk Green's function is given below,
\begin{equation} \label{eq:ansatz}
     \bbG_{\rm q}(\z|\z_0,p)  = \frac{1}{W_{\rm q}(\z_0,p)} g_{\sL}(\z_>,p) g_{\sR}(\z_<,p) \ .
\end{equation}
where $g_{\sL}$ and $g_{\sR}$ represent boundary-bulk Green's functions from the left and the right boundary, respectively, while $W_{\rm q}$ denotes the undetermined Wronskian. The symbols $\zeta_<$ and $\zeta_>$ are defined by,
\begin{equation}
    \zeta_{< (>)} \equiv \Bigl\{ 
        \begin{array}{cc}
            \zeta & \text{if $\zeta$ comes before (after) $\zeta_0$ on the RNSK contour}  \ , \\
            \zeta_0 & \text{if $\zeta_0$ comes before (after) $\zeta$ on the RNSK contour} \ .
        \end{array}
    \label{zetalessergreaterDef}
\end{equation}

Since the bulk-to-bulk Green's function has to solve the free field equation away from the sources, it is clear that it has to be proportional to the boundary-to-bulk Green's functions. Left and right-normalisability then fixes this combination of the boundary-to-bulk Green's functions to be $g_{\sL}(\z_>,k) g_{\sR}(\z_<,k)$. The other factors in the above expression must be fixed using the jump condition coming from the delta function source in the field equation. The jump condition is obtained by integrating the Eq.~(\ref{eq:bbGPDE}), given as follows,
\begin{equation}\label{eq:jump}
    \left[r^{d-1}\frac{d}{d \z}\bbG_{\rm q}(\z|\z_0,p)\right]^{\z^+_0}_{\z_0^-} = \frac{i \b}{2} \ .
\end{equation}
Now, we put our ansatz (\ref{eq:ansatz}) into the jump condition (\ref{eq:jump}) to obtain the Wronskian $W_{\rm q}$ as,\footnote{Again, note the sign difference in the Wronskian as compared to \cite{Loganayagam:2024mnj}.}
\begin{equation}
    \begin{split}
    W_{\rm q}(\z,p) &= \frac{2}{i \b}r^{d-1}\left[ g_{\sR}(\z,p) \frac{d g_{\sL}(\z,p)}{d \z}- g_{\sL}(\z,p)  \frac{d g_{\sR}(\z,p)}{d \z} \right]\\
    &= g_{\sL}(\z,p)g^{\Pi^{\ast}}_{\sR}(\z,p)-g_{\sR}(\z,p) g^{\Pi^{\ast}}_{\sL}(\z,p) \ ,
    \end{split}
\end{equation}
where we have defined left and right conjugate Green's functions\footnote{By conjugate Green's function, we mean the Green's function corresponding to the conjugate field in question.} as, 
\begin{equation}
    g^{\Pi^{\ast}}_{\sR (\sL)} \equiv   -r^{d-1}\mathbb{D}_{+}g_{\sR (\sL)} = -r^{d-1}\frac{2}{i \b}\left(\frac{d}{d \z}+\frac{\b p^0}{2} +\frac{\b q }{2}\mathcal{A}_{v} \right) g_{\sR (\sL)} \ .
\end{equation}
Let's now apply the above differential operator to the Wronskian given above to get,
\begin{equation}
    \begin{split} 
        \frac{2}{i \b} &\left(\frac{d}{d \z}+\b p^0 +\b q \mathcal{A}_{v} \right) W_{\rm q}(\z,p)= g_{\sR} \mathbb{D}_+ \left[r^{d-1} \mathbb{D}_+ g_{\sL}\right] - g_{\sL} \mathbb{D}_+ \left[r^{d-1} \mathbb{D}_+ g_{\sR}\right]=0 \ .
    \end{split}
\end{equation}
The second equality holds upon using EOM obeyed by $g_{\sR}$ and $g_{\sL}$. Hence, it implies that,
\begin{equation}\label{eq:constantgrSK}
    e^{\b p^0 \z -\beta \muq \bTh } W_{\rm q}(\z,p) = \text{constant along the RNSK radial contour} \ .
\end{equation}

\subsubsection*{Determining the Wronskian}

The retarded 2-point boundary correlator $\Kin$ (or $K_{\rm ret}$ in Eq.~\eqref{eq:Kretdef}) is obtained by taking the boundary limit of the ingoing boundary-bulk conjugate Green's function \cite{Ghosh:2020lel}, as shown below--
\begin{equation}\label{eq:Kin}
    \lim_{\z \to 0,1} r^{d-1} \mathbb{D}_{+} \Gin_{\rm q}(\z,p) = \lim_{r \to \infty} r^{d-1} \mathbb{D}_{+} \Gin_{\rm q}(\z,p) \Big|_{\rm ren}= K^{\rm in}_{\rm q}(p)  \ ,
\end{equation}
where subscript `ren' denotes the standard holographic renormalisation \cite{Skenderis:2008dg}. Henceforth, we will stop writing this subscript as it is to be assumed in every boundary limit. 

Similarly, the boundary limits of the outgoing boundary-bulk conjugate Green's function give,
\begin{equation}\label{eq:Kout}
\begin{split}
    &\lim_{\z \to 0} r^{d-1} \mathbb{D}_{+} \Gout_{\rm q}(\z,p) = \lim_{r \to \infty} r^{d-1} \mathbb{D}_{+} \Gin_{\rm -q}(\z,-p) = K^{\rm in}_{\rm -q}(-p)  \ ,\\
    &\lim_{\z \to 1} r^{d-1} \mathbb{D}_{+} \Gout_{\rm q}(\z,p) =  e^{-\b (p^0 -q \mu )} \lim_{r \to \infty} r^{d-1} \mathbb{D}_{+} \Gin_{\rm -q}(\z,-p) =  e^{-\b (p^0 -\mu q )} K^{\rm in}_{\rm -q}(-p) \ ,
\end{split}
\end{equation}
where we have used the \emph{CPT} isometry: $\Gout_{\rm q}(\z,k) = e^{-\beta k^0 \z} e^{ \beta \muq \bTh } \Gin_{\rm -q}(\z,-k)$. Note that the difference between the left and right boundary limit of the outgoing conjugate Green's function as opposed to the same limit of the ingoing conjugate Green's function.

We now turn to the left and right conjugate Green's functions and their boundary limits. Using above-mentioned boundary limits, i.e. equations (\ref{eq:Kin}) and (\ref{eq:Kout}), we obtain the left boundary limits as,
\begin{equation}\label{eq:LBCgPRgPL}
    \begin{split}
        \lim_{\z \to 0}g^{\Pi^{\ast}}_{\sR} &\equiv   -\lim_{\z \to 0}r^{d-1}\mathbb{D}_{+} g_{\sR}= -  (1+\nbe_{p})\left[ K^{\rm in}_{\rm q}(p) -K^{\rm in}_{\rm -q}(-p)\right] \ ,\\
        \lim_{\z \to 0}g^{\Pi^{\ast}}_{\sL} &\equiv   -\lim_{\z \to 0}r^{d-1}\mathbb{D}_{+} g_{\sL}= -  \nbe_{p} \left[ K^{\rm in}_{\rm q}(p) - e^{\b (p^0-\mu q)}K^{\rm in}_{\rm -q}(-p)\right] \ ,
    \end{split}
\end{equation}
and the corresponding right boundary limits as,
\begin{equation}\label{eq:RBCgPRgPL}
    \begin{split}
        \lim_{\z \to 1}g^{\Pi^{\ast}}_{\sR} &\equiv   -\lim_{\z \to 0}r^{d-1}\mathbb{D}_{+} g_{\sR}= -  (1+\nbe_{p})\left[ K^{\rm in}_{\rm q}(p) -e^{-\b (p^0 -\mu q)}K^{\rm in}_{\rm -q}(-p)\right]\\
        \lim_{\z \to 1}g^{\Pi^{\ast}}_{\sL} &\equiv   -\lim_{\z \to 0}r^{d-1}\mathbb{D}_{+} g_{\sL}= -  \nbe_{p} \left[ K^{\rm in}_{\rm q}(p) - K^{\rm in}_{\rm -q}(-p)\right] \ .
    \end{split}
\end{equation}
Using these boundary limits and boundary limits of the right and the left boundary-bulk Green's functions Eqs.~(\ref{eq:LBCgRgL}) and (\ref{eq:RBCgRgL}), we obtain the constant in Eq. ~(\ref{eq:constantgrSK}) at the left boundary, 
\begin{equation}
    \begin{split}
            \lim_{\z \to 0} e^{\b p^0 \z - \b \muq \bTh } W_{\rm q}(\z,p) &= \lim_{\z \to 0}\left[  g_{\sL}(\z,p)g^{\Pi^{\ast}}_{\sR}(\z,p)-g_{\sR}(\z,p) g^{\Pi^{\ast}}_{\sL}(\z,p) \right] \\
            &=  (1+\nbe_{p})\left[ K^{\rm in}_{\rm q}(p) -K^{\rm in}_{\rm -q}(-p)\right] \ ,
    \end{split}
\end{equation}
and at the right boundary also gives the same result as expected from Eq. (\ref{eq:constantgrSK}). Since the above quantity is the same everywhere in the bulk, the expression of the Wronskian can be easily found to be given by,
\begin{equation}\label{eq:Wronskian}
      W_{\rm q}(\z,p)  =  e^{-\b p^0 \z + \b \muq \bTh}(1+\nbe_{p})\left[ K^{\rm in}_{\rm q}(p) -K^{\rm in}_{\rm -q}(-p)\right] \ .
\end{equation}

\subsubsection*{Binormalisable bulk-to-bulk Green's function}

Now that we have expression for the Wronskian (\ref{eq:Wronskian}), the bi-normalisable bulk-to-bulk Green's function (\ref{eq:ansatz}) can be explicitly written as,
\begin{equation}\label{eq:bbGRLbasis}
    \begin{split}
        \bbG_{\rm q}(\z|\z_0,p)  &= \frac{e^{\b p^0 \z_0 - \b \muq \bTh(\z_0) }}{(1+\nbe_{p})\left[ K^{\rm in}_{\rm q}(p) -K^{\rm in}_{\rm -q}(-p)\right]} g_{\sL}(\z_>,p) g_{\sR}(\z_<,p)\\
         &= \frac{e^{\b p^0 \z_0 - \b \muq \bTh(\z_0) }}{(1+\nbe_{p})\left[ K^{\rm in}_{\rm q}(p) -K^{\rm in}_{\rm -q}(-p)\right]} \\
         & \times \Big\{  \Theta_{\rm SK}(\z>\z_0) g_{\sL}(\z,p) g_{\sR}(\z_0,p)+ \Theta_{\rm SK}(\z<\z_0) g_{\sR}(\z,p) g_{\sL}(\z_0,p)\Big\} \ .
    \end{split}
\end{equation}
After $q \to 0$ limit, when we compare this binormalisable bulk-to-bulk Green's function to the one given in \cite{Loganayagam:2020eue} (see Eq.~(A.31) in \cite{Loganayagam:2020eue}), we can see a clear overall sign difference as expected. The bulk-to-bulk Green's function, when written ingoing-outgoing basis using the boundary-to-bulk Green's functions given in Eq.~(\ref{gRgLmain}), we obtain,
\begin{equation}\label{eq:bbGq}
    \begin{split}
        \bbG_{\rm q}(\z|\z_0,p)  &= \frac{1}{\left[ K^{\rm in}_{\rm q}(p) -K^{\rm in}_{\rm -q}(-p)\right]}  \Bigg\{  \ \nbe_{p} \, e^{\b p^0 \z_0 - \b \muq \bTh(\z_0) } \Gin_{\rm q}(\z,p) \Gin_{\rm q}(\z_0,p)\\
         &\hspace{4cm}+(1+\nbe_{p}) e^{-\b p^0 \z + \b \muq \bTh(\z) } \Gin_{\rm -q}(\z,-p) \Gin_{\rm -q}(\z_0,-p)\\
         &\hspace{0.2cm} - \left[ \Theta_{\rm SK}(\z <\z_0)+\nbe_{p} \right] \Gin_{\rm q}(\z,p) \Gin_{\rm -q}(\z_0,-p)\\
         &\hspace{0.2cm} - \left[ \Theta_{\rm SK}(\z >\z_0)+\nbe_{p} \right] e^{\b p^0(\z_0-\z)} e^{ -\b \muq [ \bTh(\z_0)- \bTh(\z) ]} \Gin_{\rm -q}(\z,-p) \Gin_{\rm q}(\z_0,p) \Bigg\} \ .
    \end{split}
\end{equation}
As expected, it resembles the bi-normalisable bulk-to-bulk Green's function for a real scalar field in the grSK geometry \cite{Martin:2024mdm}. Once we have constructed the bi-normalisable Green's function, we now move to retarded and advanced Green's functions useful in describing the causal scattering processes in the bulk. Later, we will see that they are not completely independent but related by reciprocity.

\subsection{Retarded and Advanced bulk-to-bulk Green's functions}\label{app:bbGRA}
After constructing a bi-normalisable bulk-to-bulk Green's function, we can also construct bulk-to-bulk Green's functions that have specific causal properties, i.e., the retarded and the advanced bulk-to-bulk Green's functions. These Green's functions are normalisable at the left boundary and analytic in the upper half and the lower half of the frequency plane, respectively. Here, we will not provide a derivation of these Green's functions, as it is similar to the computation of the bi-normalisable bulk-to-bulk Green's function given above.\footnote{Again, readers are encouraged to look at \cite{Loganayagam:2024mnj} for detailed analysis.} We just quote the end results for the retarded bulk-to-bulk Green's function here as,
\begin{equation}\label{eq:bbGRq}
    \begin{split}
        \bbGR(\z|\z_0,& p) = \frac{1}{\left[ K^{\rm in}_{\rm q}(p) -K^{\rm in}_{\rm -q}(-p)\right]} \\
         &\times \Bigg\{ -e^{\b p^0 \z_0 - \b  \muq \bTh (\z_0)} \Gin_{\rm q}(\z,p) \Gin_{\rm q}(\z_0,p)   +\Theta_{\rm SK}(\z >\z_0) \Gin_{\rm q}(\z,p) \Gin_{\rm -q}(\z_0,-p)\\
         & \quad + \Theta_{\rm SK}(\z <\z_0) e^{\b p^0(\z_0-\z)} e^{ - \b \muq [ \bTh(\z_0)- \bTh(\z) ]} \Gin_{\rm -q}(\z,-p) \Gin_{\rm q}(\z_0,p) \Bigg\}
    \end{split}
\end{equation}
and the advanced bulk-bulk Green's function,
\begin{equation}\label{eq:bbGAq}
    \begin{split}
        \bbGA(\z|\z_0,& p)  = \frac{1}{\left[ K^{\rm in}_{\rm q}(p) -K^{\rm in}_{\rm -q}(-p)\right]} \\
         &\times \Bigg\{ e^{-\b p^0 \z +\b  \muq \bTh(\z)} \Gin_{\rm -q}(\z,-p) \Gin_{\rm -q}(\z_0,-p)   -\Theta_{\rm SK}(\z <\z_0) \Gin_{\rm q}(\z,p) \Gin_{\rm -q}(\z_0,-p)\\
         & \quad - \Theta_{\rm SK}(\z >\z_0) e^{\b p^0(\z_0-\z)} e^{ - \b \muq [ \bTh(\z_0)- \bTh(\z) ]} \Gin_{\rm -q}(\z,-p) \Gin_{\rm q}(\z_0,p) \Bigg\}
    \end{split}
\end{equation}
where there is again, but due to the same reason, an overall sign difference in $\bbGR$ and $\bbGA$ as compared to \cite{Loganayagam:2024mnj}. Note that we have suppressed the label $q$ for the retarded and advanced Green's functions.

The reciprocity relation of the bi-normalisable bulk-bulk Green's function can be easily found from Eq.~(\ref{eq:bbGq}) as follows,
\begin{equation}
    \bbG_{\rm q}(\z|\z_0,p) =    \bbG_{\rm -q}(\z_0 |\z, - p) \equiv \bar{\bbG}_{\rm q} (\z_0 |\z, - p) \ ,
\end{equation}
where we have defined the bar as an action which changes the sign of charge: $q \mapsto -q$.

The bi-normalisable bulk-bulk Green's function can be written as a linear combination of retarded and advanced bulk-bulk Green's functions, given as,
\begin{equation}\label{eq:bbGARbbG}
     \bbG(\z|\z_0,p) = - \nbe_{p} \bbG_{\rm ret}(\z|\z_0,p) +(1+\nbe_{p} )\bbG_{\rm adv}(\z|\z_0,p)  \ ,
\end{equation}
and then the reciprocity at the level of retarded and advanced Green's functions is given by,
\begin{equation}\label{eq:bbGARreci}
      \bbG_{\rm adv}(\z|\z_0,p) = \bar{\bbG}_{\rm ret}(\z_0|\z,-p) \ .
\end{equation}

\section{Discontinuities on the RNSK geometry }\label{app:Disc}

In this appendix, we examine the integration over the RNSK geometry, relevant to the on-shell action calculation. Specifically, the calculation involves a contour integral over the complexified radius, as depicted in figure~(\ref{fig:fixedv}). This integral can be simplified into a single exterior integral, defined in Eq.~(\ref{eq:extint}), with the discontinuities arising from the horizon cap. Here, we will outline these discontinuities and demonstrate how they are computed.

These discontinuity arises due to the non-analytic nature of the integrand, which originates from the outgoing propagators. These non-analyticities are introduced through terms such as $\zeta$ and $\bTh(\z)$. Notably, $e^{\b \muq \bTh}$ experiences a jump by $e^{\beta \muq}$ across contour branches, and $\zeta$ and $\mathbb{\Theta}$ jumps by unity across the horizon cap. These factors collectively explain the origin of the discontinuity. 

To understand this, consider the evaluation of the following integral,
\begin{equation}\label{RNSKtoExt1}
    \oint_{\zeta} \left[e^{\b \mu q + \b  \muq \bTh (\z)}\right]^{\alpha} \prod_{i=1}^{\alpha}e^{\beta k^0_i (1-\zeta)}  \mathcal{F}(\zeta) = \frac{-1}{\nq_{\sum k_i, \alpha}} \int_{\rm ext}\left[e^{ \b  \muq \bTh(\z)}\right]^{\alpha}  \prod_{i=1}^{\alpha} e^{-\beta k^0_i \zeta}  \mathcal{F}(\zeta) \ ,
\end{equation}
where $\mathcal{F}(\z)$ is an analytic function and recall the definition of  $\int_{\rm ext}$ and $\nq_{k, \alpha}$
\begin{equation}
    \int_{\rm ext} \equiv \int_{r_+}^{r_c} \diff r \ r^{d-1} \ , \qquad  \qquad \nq_{k, \alpha} = \frac{1}{e^{\b (k^0- \alpha \muq)}-1}  \ .
    \label{eq:DefExteriorIntegral}
\end{equation}
This is the only type of discontinuity that arises in a single integral, representing the most general form of a single discontinuity. It can be explicitly verified that this result yields the correct expressions for contact diagrams. However, as we move beyond contact diagrams -- for example, to exchange diagrams -- a greater number of bulk integrals appear, often accompanied by multiple bulk-to-bulk propagators. Therefore, it becomes necessary to extend this result to multi-discontinuities, starting with double discontinuities.

Now, we will come to the computation of exchange diagrams that contain bulk-to-bulk propagators. The simplest exchange diagrams consist of double integrals over the RNSK contour. The most general form of such an integral is given by,
\begin{equation}\label{RNSKtoExt2}
\begin{split}
    \oint_{\z_1,\z_2}  \prod_{i=1}^{2}e^{-\b \muq \alpha_i[1-\mathbb{\Theta}(\z_i)] }& e^{\beta \kappa_i (1-\z_i)}  \mathcal{F}(\z_1,\z_2) \bbG(\z_2|\z_1,p)\\
    &= \int_{\rm ext}  \prod_{i=1}^{2}e^{-\b \muq \alpha_i[1-\mathbb{\Theta}(\z_i)] } e^{\beta \kappa_i (1-\z_i)}  \mathcal{F}(\z_1,\z_2) \bbG_{\rm dd}(\z_2|\z_1,p)  \ ,
\end{split}    
\end{equation}
where $\mathcal{F}(\z_1,\z_2)$ is again an analytic function in both variables and the double-discontinuity $ \bbG_{\rm dd}$ is given by,
\begin{equation}
\begin{split}
     \bbG_{\rm dd}(\z_2|\z_1,p) \equiv &  \bbG(\z_2|\z_1,p)  -e^{\b(\muq \alpha_1 -\kappa_1) } \bbG(\z_2|\z_1+1,p)\\
     &-e^{\b(\muq \alpha_2 -\kappa_2) } \bbG(\z_2+1|\z_1,p)+e^{\b(\muq \alpha_1+ \muq \alpha_2 -\kappa_1-\kappa_2) } \bbG(\z_2+1|\z_1+1,p) \ .
\end{split}     
\end{equation}
The exponential factors in the above expression are simply the monodromy factors due to the branch cut extending from the horizon radius $r_+$ to the cut-off radius $r_c$ on the complex $r$ plane. We can now use the expression for the bulk-to-bulk propagator given in Eq.~(\ref{eq:bbGq}) to evaluate the integrand above explicitly. We will use the theta function identities. Using the following properties of radial step functions on the grSK/RNSK contour, we find,
\begin{equation}
    \Theta_{\rm SK}(\z+1>\z'+1) =   \Theta_{\rm SK}(\z < \z')  \ ,   
\end{equation}
and,
\begin{equation}
    \Theta_{\rm SK}(\z+1>\z') = 1  \ , \qquad   \Theta_{\rm SK}(\z >\z'+1) = 0   \ ,     
\end{equation}
where $\z,\z' < \z_h$ or both $\z$ and $\z'$ are on the left exterior contour.

Substituting the expressions for the bulk-to-bulk propagator and using relations of Heavyside step functions, the double-discontinuity obtained is as follows, 
\begin{equation}
\begin{split}
    \bbG_{\rm dd}(\z|\z',p) &= \frac{-n_{p,1}}{(1+n_{\kappa_1 -1, \alpha_1 -1})(1+n_{ \kappa_2,\alpha_2})} \bbGR(\z|\z',p) \\
    &\ + \frac{(1+n_{p,1})}{(1+n_{\kappa_1, \alpha_1})(1+n_{ \kappa_2+p, \alpha_2 +1})} \bbGA(\z|\z',p) \ .
\end{split}    
\end{equation}
We observe that the double-discontinuity formula reduces to the result in \cite{Loganayagam:2024mnj} in the $q \to 0$ limit. Though a generalisation to multi-discontinuities is possible, we only focus on the double-discontinuity in this work.

\section{On-shell action for $|\Phi|^4$ theory}\label{app:onshell}

In this appendix, we present the on-shell action for the $|\Phi|^4$ theory using a method similar to \cite{Loganayagam:2024mnj}. This approach circumvents the need to evaluate the boundary terms in the on-shell action explicitly. It leverages the fact that the boundary value of the higher-order solution vanishes, thereby simplifying the computation.

The bare action is given by
\begin{equation}
    S_{\rm bare} = \ -\oint d^{d+1}x \sqrt{-g} \left( |D_{M} \Phi |^2 + \frac{\lambda}{2!\, 2!}\left(\Phib \Phi\right)^{2} \right) \  .
\end{equation}
Inserting the perturbative solution into the bare action yields the on-shell action in the following form.
\begin{equation}
    S_{\rm os} = \ -\int_{\partial\mathcal{M}} d \Sigma^{A} \ \Phi_{(0)}  \overline{\left( D_{A} \Phi_{(0)} \right)} + \lambda \oint_{\mathcal{M}}  \Big[ \frac{ 1 }{2!}   \left(\Phib -\Phib_{(0)}\right)\Phi  \left(\Phib \Phi\right)- \frac{ 1 }{2! 2!}   \left(\Phib \Phi\right)^{2} \Big]  \ ,   
\end{equation}
where we have used the equations of motion and imposed the appropriate boundary conditions to significantly simplify the expression for the on-shell action. When expanded in terms of the coupling constant, the on-shell action is explicitly expressed as follows:
\begin{equation}
    \begin{split}
        S_{(2)} &= -\int_{\partial\mathcal{M}} d \Sigma^{A} \ \Phi_{(0)}  \overline{\left( D_{A} \Phi_{(0)} \right)} \ , \\
        S_{(4)} &=  -\frac{\lambda}{2!\, 2!} \oint d^{d+1}x \sqrt{-g}  \left(\Phib_{(0)} \Phi_{(0)}\right)^{2} \ , \\
        S_{(6)} &= -\frac{\lambda^2}{2} \oint d^{d+1}x \sqrt{-g}    \    \Phib_{(0)} \Phi_{(1)}  \Phib_{(0)} \Phi_{(0)}  \ ,
    \end{split}
\end{equation}
where the above terms represent both free and contact contributions at the tree level in Feynman diagrams.

We find it convenient to express the on-shell action in the Past-Future (PF) basis, i.e. Eqs.~(\ref{FPbasis}) and (\ref{FbPbbasis}). In this basis, SK collapse (unitarity) and KMS condition (thermality) are reflected in the vanishing of coefficients of all Future and the Past terms, respectively. Moreover, this basis exclusively involves retarded/advanced propagators in the correlators, ensuring the description of causal scattering processes. 

Now, we turn to the explicit calculation of the on-shell action, starting from the contact term $S_{(4)}$ of the $|\Phi|^4$ theory.

\paragraph{Contact term: }
The contact term  $S_{(4)}$ is given in the second line of the eq.~(\ref{eq:S2S4S6}). Passing into the boundary momentum variables, it can be written as,
\begin{equation}\label{eq:S4PFgrSKexp}
    \begin{split}
        S_{(4)} &=   -\frac{\lambda}{4}   \int_{k_{1,2,3,4} } \oint_{\z} \ \Phib_{(0)}(\z,k_1)  \Phi_{(0)}(\z,k_2)  \Phib_{(0)}(\z,k_3)  \Phi_{(0)}(\z,k_4)\\
        &=  -\frac{\lambda}{4}  \int_{k_{1,2,3,4} }  \oint_{\z} \ \Big\{  - \Ginb(\z,k_1) \Jb_{\Fb}(k_1)+e^{\beta(k^0_1 +\muq)} \Goutb(\z,k_1) \Jb_{\Pb}(k_1) \Big\}\\
        &\hspace{3cm} \times \Big\{ - \Gin(\z,k_2) J_{\Fb}(k_2)+e^{\beta(k^0_2 -\muq)} \Gout(\z,k_2) J_{\Pb}(k_2) \Big\}\\
        &\hspace{3cm} \times \Big\{  - \Ginb(\z,k_3) \Jb_{\Fb}(k_3)+e^{\beta(k^0_3 +\muq)} \Goutb(\z,k_3) \Jb_{\Pb}(k_3) \Big\}\\
        &\hspace{3cm} \times \Big\{ - \Gin(\z,k_4) J_{\Fb}(k_4)+e^{\beta(k^0_4 -\muq)} \Gout(\z,k_4) J_{\Pb}(k_4) \Big\} \ ,
    \end{split}
\end{equation}
where we have written the solution in terms of ingoing and outgoing solutions, as given in Eqs.~(\ref{Phi0PF}) and (\ref{Phib0PF}).

Taking into account the analyticity of $\Gin$ and $\Ginb$ in the $\z$ coordinate and using momentum conservation, the RNSK radial integral leads us to the following observations: 
\begin{enumerate}
    \item \textbf{SK collapse holds:} The term with only future sources, i.e. $\Jb_{\Fb}$ and $J_{\Fb}$, vanishes.
    \item \textbf{KMS condition holds:} The term with only past sources, i.e. $\Jb_{\Pb}$ and $J_{\Pb}$, vanishes.
\end{enumerate}
As a result, seven of the original nine terms (expand Eq.~\eqref{eq:S4PFgrSKexp} and relabel the momentum to see nine independent terms) remain. The next step involves performing the RNSK radial integral to obtain a single exterior integral, using Eq.~(\ref{RNSKtoExt1}) from the previous appendix. Thus, we obtain the following,
\begin{equation}\label{eq:IFcontact}
    \begin{split}
        S_{(4)} 
        &=  -\lambda \int_{k_{1,2,3,4} }  \int_{\rm ext}  \Bigg\{ \frac{1}{2\nq_{k_4, 1}}  \Jb_{\Fb}(k_1) J_{\Fb}(k_2) \Jb_{\Fb}(k_3) J_{\Pb}(k_4) \Ginb(\z,k_1) \Gin(\z,k_2)   \Ginb(\z,k_3) \Gout(\z,k_4)\\
        &+ \frac{1}{2\nqb_{k_3,1}}  \Jb_{\Fb}(k_1) J_{\Fb}(k_2) \Jb_{\Pb}(k_3) J_{\Fb}(k_4) \Ginb(\z,k_1) \Gin(\z,k_2) \Goutb(\z,k_3)  \Gin(\z,k_4)\\
        &- \frac{1}{\nq_{k_{34},0}}  \Jb_{\Fb}(k_1) J_{\Fb}(k_2) \Jb_{\Pb}(k_3) J_{\Pb}(k_4) \Ginb(\z,k_1) \Gin(\z,k_2)   \Goutb(\z,k_3) \Gout(\z,k_4)\\
        &- \frac{1}{\nq_{k_{24},2}}\Jb_{\Fb}(k_1) J_{\Pb}(k_2) \Jb_{\Fb}(k_3) J_{\Pb}(k_4) \Ginb(\z,k_1)   \Gout(\z,k_2)   \Ginb(\z,k_3)  \Gout(\z,k_4)\\
        &- \frac{1}{\nqb_{k_{13},2}}\Jb_{\Pb}(k_1) J_{\Fb}(k_2) \Jb_{\Pb}(k_3) J_{\Fb}(k_4)  \Goutb(\z,k_1) \Gin(\z,k_2)   \Goutb(\z,k_3) \Gin(\z,k_4)\\
        &+\frac{1}{2\nq_{k_{234},1}} \Jb_{\Fb}(k_1) J_{\Pb}(k_2) \Jb_{\Pb}(k_3) J_{\Pb}(k_4) \Ginb(\z,k_1) \Gout(\z,k_2)   \Goutb(\z,k_3) \Gout(\z,k_4)\\
        &+ \frac{1}{2\nqb_{k_{134},1}}  \Jb_{\Pb}(k_1) J_{\Fb}(k_2) \Jb_{\Pb}(k_3) J_{\Pb}(k_4) \Goutb(\z,k_1) \Gin(\z,k_2)    \Goutb(\z,k_3) \Gout(\z,k_4)
        \Bigg\} \ .
    \end{split}
\end{equation}
As noted previously, these exterior integrals naturally exhibit Bose-Einstein factors at different `effective' chemical potentials. These seven terms correspond to the seven diagrams shown in figure~(\ref{fig:fourpointcontact}).

\paragraph{Exchange terms: }
Having addressed the contact term, the next step is to examine the exchange terms. In the case of $|\Phi|^4$ theory, the first exchange term originates from the six-point influence phase $S_{(6)}$, which appears at quadratic order in $\lambda$. The $S_{(6)}$ is given in the third line of Eq.~(\ref{eq:S2S4S6}) and is expressed in momentum space as,
\begin{equation}\label{eq:appDS6one}
    \begin{split}
        S_{(6)} 
        &=  -\frac{\lambda^2}{2} \int_{k_{1,2,3,4}}  \oint_{\z,\z'}\   \bar{\Phi}_{(0)}(\z,k_1)\Phi_{(0)}(\z,k_2) \bar{\Phi}_{(0)}(\z,k_3)\bbG(\z|\z',k_4) \mathbb{J}_{(1)}(\z',k_4)  \ .
    \end{split}
\end{equation}
Upon using the expression of bulk source $\mathbb{J}_{(1)}$ from Eq.~(\ref{bulksources}), we find,
\begin{equation}
    \begin{split}
        S_{(6)} &=  -\frac{\lambda^2}{4} \int_{ k_{1,2,...,6}} \oint_{\z,\z'} \\
        &\times \Phib_{(0)}(\z,k_1) \Phi_{(0)}(\z,k_2) \Phib_{(0)}(\z,k_3) \bbG(\z|\z',k_{456}) \Phi_{(0)}(\z',k_4) \Phib_{(0)}(\z',k_5) \Phi_{(0)}(\z',k_6) \ .
    \end{split}
\end{equation}
Using the expressions of zeroth-order solution equations (\ref{Phi0PF}) and (\ref{Phib0PF}), we get,
\begin{equation}
\footnotesize
    \begin{split}
        &S_{(6)} =  -\frac{\lambda^2}{4} \int_{ k_{1,2,...,6}} \oint_{\z,\z'} \left[  - \Ginb(\z,k_1) \Jb_{\Fb}(k_1)+e^{\beta(k^0_1 +\mu q)} \Goutb(\z,k_1) \Jb_{\Pb}(k_1) \right] \\
        & \times \left[ - \Gin(\z,k_2) J_{\Fb}(k_2)+e^{\beta(k^0_2 -\mu q)} \Gout(\z,k_2) J_{\Pb}(k_2) \right] \left[  - \Ginb(\z,k_3) \Jb_{\Fb}(k_3)+e^{\beta(k^0_3 +\mu q)} \Goutb(\z,k_3) \Jb_{\Pb}(k_3) \right] \\
        & \times \bbG(\z|\z',k_{456} ) \left[ - \Gin(\z,k_4) J_{\Fb}(k_4)+e^{\beta(k^0_4 -\mu q)} \Gout(\z,k_4) J_{\Pb}(k_4) \right]\\
        &\times \left[  - \Ginb(\z,k_5) \Jb_{\Fb}(k_5)+e^{\beta(k^0_5 +\mu q)} \Goutb(\z,k_5) \Jb_{\Pb}(k_5) \right]  \left[ - \Gin(\z,k_6) J_{\Fb}(k_6)+e^{\beta(k^0_6 -\mu q)} \Gout(\z,k_6) J_{\Pb}(k_6) \right] \ .
    \end{split}
\end{equation}
Once again, the SK collapse and the KMS condition are evident, as the terms with all future and past sources vanish in the six-point influence phase $S_{(6)}$, respectively.

We proceed with the RNSK radial integral to express $S_{(6)}$ in terms of the exterior radial integrals. Here, we present the result for a single term of $S_{(6)}$, specifically the component with five future and one past sources, denoted as $S_{5\Fb, 1\Pb}$. This term comprises two independent contributions, as shown below--
\begin{equation}
 S_{5\Fb, 1\Pb} =   S[\Fb \F \Fb \F \Fb \Psmall]+  S[\Fb \F \Fb \F \Pb \F]  \ ,
\end{equation}
where the notation is designed to clearly indicate whether the past source is from the field or its conjugate. Now, let us analyse the case where the past source is from the field, written as,
\begin{equation}
    \begin{split}
        &S[\Fb \F \Fb \F \Fb \Psmall] =  \frac{\lambda^2}{4} \int_{ k_{1,2,...,6}}  \Jb_{\Fb}(k_1) J_{\Fb}(k_2)  \Jb_{\Fb}(k_3) J_{\Fb}(k_4) \Jb_{\Fb}(k_5) J_{\Pb}(k_6) \oint_{\z,\z'} \bbG(\z|\z',k_{456})\\
        &\times\left[ \Ginb(\z,k_1) \Gin(\z,k_2) \Ginb(\z,k_3) \Gin(\z',k_4) \Ginb(\z',k_5)e^{\beta(k^0_6 -\mu q)} \Gout(\z',k_6)  \right] \ .
    \end{split}
\end{equation}
Using double-discontinuity integral given in the Eq. (\ref{RNSKtoExt2}) from the previous Appendix, we find,
\begin{equation}
    \begin{split}
        &S[\Fb \F \Fb \F \Fb \Psmall] =  \frac{\lambda^2}{2} \int_{ k_{1,2,...,6}} \frac{1}{\nq_{k_6, 1}} \Jb_{\Fb}(k_1) J_{\Fb}(k_2)  \Jb_{\Fb}(k_3) J_{\Fb}(k_4) \Jb_{\Fb}(k_5) J_{\Pb}(k_6) \\
        &\times \int_{\rm ext} \bbGR(\z|\z',k_{456}) \left[ \Ginb(\z,k_1) \Gin(\z,k_2) \Ginb(\z,k_3) \Gin(\z',k_4) \Ginb(\z',k_5) \Gout(\z',k_6)  \right] \ .
    \end{split}
\end{equation}
It is worth emphasizing that the result above was derived by leveraging the reciprocity between the retarded and advanced bulk-to-bulk propagators.

Similarly, one could extend this approach to compute additional terms in the six-point influence phase. However, since the same principles apply, we restrict our focus to this single term in this appendix. By employing a similar analysis and multi-discontinuity integrals, exchange terms contributing to the higher-point influence phase can also be determined.

\section{Explicit terms of quartic on-shell action $S_{(4)}$}\label{app:explicitL0L1}

In this section, we will explicitly compute the four-point influence phase at linear order in the gradient expansion. In the time domain, we start with the solution itself given in Eq.~\eqref{Phigrdexppos}, as,
\begin{equation}
\begin{split}
    \Phi &= \left[\Ja + \left( \z-\frac{1}{2}+ \mathbb{F}_{1} \right) \Jd \right]+ \frac{ i \b}{2} \left[ \mathbb{F}_{1}  \dt \Ja -\left( \z (\z-1)+ \left[ \z -\frac{1}{2} \right] \mathbb{F}_{1} \right) \dt \Jd \right]  \\
    &+\frac{\b \mu_{q}}{2}\left[  \left( \mathbb{F}_{2} -\mathbb{F}_{1} \right) \Ja +\left( \z (\z-1)+ \left[ \z -\frac{1}{2} \right] \mathbb{F}_{2} +  \left( \z-\frac{1}{2}+ \mathbb{F}_{2} \right) \mathbb{F}_{1} \right)\Jd \right] \ .
\end{split}
\end{equation}
Since we are only working up to linear order in derivatives while discarding the mixed terms (see the discussion around Eq.~\eqref{eq:S4linear}), we can divide the following terms into two parts:
\begin{equation}
\begin{split}
    S_{(4)} &= -\lambda\int d^d x\oint d\z \sqrt{-g}\frac{|\Phi_{(0)}|^4}{4} \\
    &= -\lambda \int d^d x \int_{\rm ext}  \mathcal{L}_{0} -\lambda \int d^d x \int_{\rm ext} \mathcal{L}_{1}+ \mathcal{O}((\beta \dt)^2)+ \mathcal{O}((\beta \muq)^2)+ \mathcal{O}( \beta^2 \muq \dt) \ ,
\end{split}    
\end{equation}
where $\mathcal{L}_{0}$ and $\mathcal{L}_{1}$ are non-derivative and first derivative term.

As we can see now, we can make our lives easier by moving into position space. It will remove unnecessary notation and declutter the expressions.

\paragraph{Zeroth-derivative term $\mathcal{L}_{0}$: } The $\mathcal{L}_{0}$ can be explicitly written as,
\begin{equation}
    \begin{split}
        \mathcal{L}_{0} &=  \sum_{r,s=0}^{2} \mathcal{G}_{r,s} \  \left[\Jba\right]^{r} \left[\Ja\right]^{s} 
        \left[\Jbd\right]^{2-r}  \left[\Jd\right]^{2-s}     \ ,
    \end{split}
\end{equation}
where the coefficients of these terms are as follows:
\begin{equation}\label{eq:explicitGrs}
\begin{aligned}
    &\mathcal{G}_{2,2} =  0  \ , \qquad \qquad \mathcal{G}_{1,1} =  2 \left( \z+ \mathbb{F}_1 \right)\ ,  
    &&\mathcal{G}_{1,2} =  \frac{1}{2} -\frac{\b \mu_{q}}{2} \left( \z+ \mathbb{F}_1 \right)\ , \\
    &\mathcal{G}_{2,1} =  \frac{1}{2} +\frac{\b \mu_{q}}{2} \left( \z+ \mathbb{F}_1 \right)\ , 
    &&\mathcal{G}_{0,0} =  \frac{1}{4}\left( \z+ \mathbb{F}_1 \right) +\left( \z+ \mathbb{F}_1 \right)^3 \ ,\\
    &\mathcal{G}_{0,2} =  \frac{1}{2}\left( \z+ \mathbb{F}_1 \right) -\frac{ 3\, \b \mu_{q}}{4} \left( \z+ \mathbb{F}_1 \right)^2 \ ,    &&\mathcal{G}_{2,0} =  \frac{1}{2}\left( \z+ \mathbb{F}_1 \right) +\frac{ 3\, \b \mu_{q}}{4} \left( \z+ \mathbb{F}_1 \right)^2 \ , \\
    &\mathcal{G}_{0,1} = \frac{1}{8}+ \frac{ 3}{2} \left( \z+ \mathbb{F}_1 \right)^2 -\frac{\b \mu_{q}}{8} \left( \z+ \mathbb{F}_1 \right) -\b \mu_{q} \left( \z+ \mathbb{F}_1 \right)^3 \ , \\
    &\mathcal{G}_{1,0} = \frac{1}{8}+ \frac{ 3}{2} \left( \z+ \mathbb{F}_1 \right)^2 +\frac{\b \mu_{q}}{8} \left( \z+ \mathbb{F}_1 \right) +\b \mu_{q} \left( \z+ \mathbb{F}_1 \right)^3 \  .
\end{aligned}
\end{equation}
Note that the following is true,
\begin{equation}
    \mG_{r,s} = \mG_{s,r}\Big|_{q \to -q} \ . 
\end{equation}

\medskip

\paragraph{First-derivative term $\mathcal{L}_{1}$: }
In the first derivative term, the time derivative can appear on any of the four sources. Thus, we can explicitly write $\mathcal{L}_{1}$ as,
\begin{equation}
\footnotesize
    \begin{split}
        \mathcal{L}_{1} &=  \sum_{r=0}^{1} \sum_{s=0}^{2} \mathcal{G}_{\dot{r},s}  \left(\dt \Jba \right) \left[\Jba\right]^{r} \left[\Ja\right]^{s} 
        \left[\Jbd\right]^{1-r}  \left[\Jd\right]^{2-s}  +     \sum_{r=0}^{2} \sum_{s=0}^{1} \mathcal{G}_{r,\dot{s}}  \left(\dt \Ja \right)  \left[\Jba\right]^{r} \left[\Ja\right]^{s} \left[\Jbd\right]^{2-r}  \left[\Jd\right]^{1-s}    \\
        &+  \sum_{r=0}^{1} \sum_{s=0}^{2} \mathcal{H}_{\dot{r},s}  \left(\dt \Jbd \right) \left[\Jba\right]^{r} \left[\Ja\right]^{s} 
        \left[\Jbd\right]^{1-r}  \left[\Jd\right]^{2-s} + \sum_{r=0}^{2} \sum_{s=0}^{1} \mathcal{H}_{r,\dot{s}}  \left(\dt \Jd \right) \left[\Jba\right]^{r} \left[\Ja\right]^{s} \left[\Jbd\right]^{2-r}  \left[\Jd\right]^{1-s} \ ,
    \end{split}
\end{equation}
which in concise form is given by,
\begin{equation}
    \begin{split}
        \mathcal{L}_{1} &=  \sum_{r=0}^{1} \sum_{s=0}^{2}  \Big\{ \mathcal{G}_{\dot{r},s}  \left(\dt \Jba \right) 
        + \mathcal{H}_{\dot{r},s}  \left(\dt \Jbd \right)  \Big\} \left[\Jba\right]^{r} \left[\Ja\right]^{s} 
        \left[\Jbd\right]^{1-r}  \left[\Jd\right]^{2-s}     \\
        &+  \sum_{r=0}^{2} \sum_{s=0}^{1}                  
        \Big\{  \mathcal{G}_{r,\dot{s}}  \left(\dt \Ja \right)  + \mathcal{H}_{r,\dot{s}}  \left(\dt \Jd \right)  \Big\}  \left[\Jba\right]^{r} \left[\Ja\right]^{s} \left[\Jbd\right]^{2-r}  \left[\Jd\right]^{1-s}   \ ,
    \end{split}
\end{equation}
where an independent set of the terms involving $(\dt \Jba)$ is as follows:
\begin{equation}\label{eq:explicitGrdots}
\begin{aligned}
    &\mathcal{G}_{\dot{1},2} =  0  \ , \qquad  \mathcal{G}_{\dot{0},2} =  \frac{i\b}{4} \mathbb{F}_1 \ ,  \quad &&\mathcal{G}_{\dot{1},1}  = \frac{i\b}{2} \mathbb{F}_1  \ , \\
    &\mathcal{G}_{\dot{1},0}  = \frac{i\b}{2} \mathbb{F}_1 (\mathbb{F}_1 +\z) \ ,
    &&\mathcal{G}_{\dot{0},1}  = i\b \mathbb{F}_1 (\mathbb{F}_1 +\z) \ ,\\
    &\mathcal{G}_{\dot{0},0}  = \frac{i\b}{4} \mathbb{F}_1 \left( \frac{1}{4}+3(\mathbb{F}_1 +\z)^2 \right) \ .
\end{aligned}
\end{equation}
Now, we turn to the independent terms involving $(\dt \Jbd)$ and they are explicitly given as,
\begin{equation}\label{eq:explicitHrs}
\begin{aligned}
    & \mathcal{H}_{\dot{1},2} =  -\frac{i \b}{4} \left( \mathbb{F}_1 +2 \z  \right)\ , \\
    & \mathcal{H}_{\dot{0},2} =  \frac{i \b }{8} \Bigg\{ -2 (\mathbb{F}_1 + \z) (\mathbb{F}_1 + 3\z)  \Bigg\} , \\
    & \mathcal{H}_{\dot{1},1} =  \frac{i \b }{4} \Bigg\{ -2 (\mathbb{F}_1 + \z) (\mathbb{F}_1 + 3\z)  \Bigg\} , \\
    & \mathcal{H}_{\dot{0},1} = \frac{i \b}{8} \Bigg\{ (-\mathbb{F}_1 - 4 \mathbb{F}_1^3 - 2 \z - 24 \mathbb{F}_1^2 \z - 36 \mathbb{F}_1 \z^2 - 
   16 \z^3 \Bigg\} , \\
    & \mathcal{H}_{\dot{1},0} = \frac{i \b}{16} \Bigg\{ (-\mathbb{F}_1 - 4 \mathbb{F}_1^3 - 2 \z - 24 \mathbb{F}_1^2 \z - 36 \mathbb{F}_1 \z^2 - 
   16 \z^3 \Bigg\} , \\
    & \mathcal{H}_{\dot{0},0} = \frac{i \b}{32}   \Bigg\{ -2 (\mathbb{F}_1 + \z) (3 \mathbb{F}_1 + 4 \mathbb{F}_1^3 + 7 \z + 28 \mathbb{F}_1^2 \z + 44 \mathbb{F}_1 \z^2 + 20 \z^3)  \Bigg\} \ ,
\end{aligned}
\end{equation}
where all the remaining terms can be obtained from the following relation,
\begin{equation}
    \mG_{r,\dot{s}} = \mG_{\dot{s},r}\ ,  \qquad \mathcal{H}_{r,\dot{s}} = \mathcal{H}_{\dot{s},r} \ .
\end{equation}

\bibliography{HolFDRs}

\end{document}